\documentclass[12pt]{article}

\usepackage[utf8]{inputenc}

\usepackage[parfill]{parskip}

\usepackage[a4paper, margin=2.5cm]{geometry}

\usepackage{amsmath}
\usepackage{amssymb}
\usepackage[affil-it]{authblk}
\usepackage{bm}
\usepackage{graphicx}

\usepackage{xcolor}
\usepackage{tikz}
\usetikzlibrary{3d}

\usepackage[textwidth=8em,textsize=small]{todonotes}
\usepackage{url}

\usepackage{subcaption}

\DeclareCaptionFormat{subfig}{~#1#2#3}
\renewcommand\Authfont{\fontsize{12}{14.4}\selectfont}
\renewcommand\Affilfont{\fontsize{9}{10.8}\itshape}

\title{Modelling the impact of social mixing and behaviour on infectious disease transmission:\\ application to SARS-CoV-2
}

\author[1]{Alison C Hale\thanks{Alison Hale: a.c.hale@lancaster.ac.uk}}
\author[2]{Jonathan M Read}
\author[3]{Christopher P Jewell}
\affil[1, 3]{Department of Mathematics and Statistics, Lancaster University, UK.}
\affil[2]{Centre for Health Informatics Computing and Statistics, Lancaster Medical School, Lancaster University, UK.}

\date{\today}

\begin{document}

\maketitle

\section*{Abstract}

In regard to infectious diseases socioeconomic determinants are strongly associated with differential exposure and susceptibility however they are seldom accounted for by standard compartmental infectious disease models.

These associations are explored with a novel compartmental infectious disease model which, stratified by deprivation and age, accounts for population-level behaviour including social mixing patterns.

As an exemplar using a fully Bayesian approach our model is fitted, in real-time, to the UKHSA COVID-19 community testing case data from England. Metrics including reproduction number and forecasts of daily case incidence are estimated from the posterior.

From this dataset it is observed that during the initial period of the pandemic the most deprived groups reported the most cases however this trend reversed after the summer of 2021. Forward simulation experiments based on the fitted model demonstrate that this reversal can be accounted for by differential changes in population level behaviours including social mixing and testing behaviour, but it is not explained by the depletion of susceptible individuals.

In future epidemics, with a focus on socioeconomic factors the approach outlined here provides the possibility of identifying those groups most at risk with a view to helping policy-makers better target their support.

\bigskip

\textbf{keywords words:} \textit{Infectious diseases, Epidemic model, Socioeconomic factors, Social mixing patterns}

\bigskip

\section{Introduction} \label{sec:introduction}

Disparities in disease morbidity and mortality are commonly observed within many countries, though the underlying causes are often complex or unclear \cite{Rasanathan2018}.
In Europe, lower socioeconomic status and deprivation are associated with higher levels of morbidity and mortality \cite{Marmot2006}. 
For instance a large study in the UK \cite{Charlton2013} found that the prevalence of individuals with three or more health conditions was 29\% in the most deprived quintile of the population compared with 14\% in the least deprived quintile.
Unequal burdens of morbidity and mortality are further exacerbated by infectious diseases \cite{Quinn2014}.

At a population level, the transmission of infectious diseases are thought to be linked to inequality, poverty and social determinants of health.
In this context differential exposure and susceptibility
have been highlighted by the COVID\nobreakdash-19 pandemic: globally \cite{Jensen2021}, across Europe \cite{WHO2020HIE} and within England \cite{PHEGW14472020}.
For example, a large community cohort study based in England and Wales \cite{Beale2022} concluded that participants in the most deprived areas had an increased risk of exposure relative to the least deprived due a greater number of non-household contacts associated with attending workplaces or schools, a greater reliance on public transport and vehicle sharing, and a greater frequency in visiting shops.
Additionally, household level factors relating to the highest risk of transmission included poverty, multi-generational households where vulnerable or elderly members live, and inability to isolate due to overcrowding or financial reasons \cite{EMG2021}.

Deprivation is known to be correlated with age, and is quantified by the Index of Multiple Deprivation (IMD) \cite{GOVIMD2019}.
For instance the 2019 population estimates from the Office for National Statistics \cite{ONSIMD2019} show that $55\%$ of the under forties age group are in the most deprived IMD deciles 1 to 5, whereas this drops to $45\%$ for the over fifties. 
Furthermore, differential risk of disease due to age may also have played a significant role during the COVID\nobreakdash-19 pandemic in the UK due to: the vaccination programme being rolled-out in the most part by age (oldest age-groups first); age-related decline in immune system function \cite{Bajaj2021}; and behavioural or life-style differences across age groups e.g. children often have a large number of close contacts from attending school and may also be less likely to adopt good personal hygiene practices.

Standard compartmental infectious disease models are commonly used to explore contagion dynamics \cite{Brauer2017} although they rarely account for the heterogeneity in infection risk arising from socioeconomic determinants \cite{Galanis2021}.
In this regard the effect of deprivation on infection incidence could, for example, be explored from the perspective of differential depletion of susceptible individuals, and also differential social mixing patterns or other relevant population-level behaviours (e.g. testing behaviour).
Consequently, the inclusion of socioeconomic determinants has the potential to enhance the model's effectiveness to make inference, to improve disease incidence forecast, and to more reliably inform policy decisions.

In this paper, a novel stochastic compartmental model is implemented where the population is stratified into deprivation-age groups.
As an exemplar the model is fitted to the UKHSA English community testing COVID\nobreakdash-19 positive case data.
Using a fully Bayesian approach the model sequentially learns as the epidemic evolves, as such it captures the changing epidemic dynamics across both strata and time.
Using transmission metrics, such as the reproduction number, it is  possible to determine from the fitted model which stratum are most responsible for driving the epidemic in real-time.
Furthermore it was observed from this COVID-19 data that during the first year and a half of the pandemic the most deprived groups reported the highest incidence, but beyond spring 2021 this switches such that highest incidence appears to be in some of the least deprived strata.
The mechanisms behind this phenomena, referred to here as \textit{deprivation-switching}, have not been fully explored or explained previously in the literature.
To address this, we constructed a number of forward simulation experiments, based on the fitted model's posterior samples, to explore factors which may underlie the phenomena. Specifically we assess whether the observed deprivation-switching could have been generated by heterogeneous depletion of susceptible individuals, changes in social mixing, and changes in population-level behaviour.
The focus here is on assessing the effect of each factor, it is not model selection using statistical hypothesis testing.

The remainder of this paper is structured as follows: section \ref{sec:dataset} describes the motivating dataset; section \ref{sec:description} gives the mathematical details of our model; in section \ref{sec:application} our model is applied and fitted to the UKHSA COVID-19 dataset which leads to results relating to model selection, model predictions and also forward simulations designed to explore deprivation-switching; and finally the results are discussed in section \ref{sec:discussion}.

\section{Motivating dataset} \label{sec:dataset}

This study was motivated by the COVID\nobreakdash-19 pandemic in England.
Between March 2020 and March 2022 the unfolding pandemic exhibited complex dynamics with respect to both deprivation and age.
The variations in the observed case incidence with respect to age are usually attributed to age-specific mixing behaviour \cite{Mossong2008, Klepac2020, Danon2013, Davies2020}, such as attending school, however the rates relating to deprivation are not well understood so are the main focus in this paper.

The UK Health Security Agency (UKHSA) COVID\nobreakdash-19 positive case data was collected through community testing in England, often referred to as Pillar 2 testing, from February 2020.
The data were recorded using polymerase chain reaction (PCR) tests or lateral flow tests (LFTs).
The data only included an individual's first reported positive test which may or may not have been confirmed by a PCR test; positive tests from an individual were not counted more than once.
Reinfections were not included in this data; they were relatively rare compared to first episodes. For example, on 1 January 2022 there had been $0.29$ million recorded reinfections against $11.6$ million first episodes.
For each positive test recorded explanatory variables included specimen date of test, age in years, and a deprivation index determined by the individual's current place of residence.

By 31 March 2022 over 17 million first cases had been recorded in England, this corresponds to approximately $30\%$ of the population.
There are two important caveats.
First, during the early stages of the pandemic mass testing for the general public was unavailable but availability was increased during 2020, and widely available from April 2021 until the end of March 2022.
Secondly, the case data was not a random sample from the population, as community testing was voluntary and individuals could choose whether to test and choose whether to report a positive result.
Factors including asymptomatic disease and conceivably personal factors (e.g. not financially viable to take time off work due to illness) may have resulted in some individuals being unable to report their SARS-CoV-2 infection.
Note that it is beyond the scope of this work to directly account for missing data.

The 2019 English Index of Multiple Deprivation (IMD) \cite{GOVIMD2019} was used to determine the deprivation index of each reported case.
IMD is categorised into deciles where $1$ refers to the most deprived $10\%$ of the population and $10$ the least deprived $10\%$.
Where needed, aggregated case data was normalised by deprivation and/or age group using 2019 Office for National Statistics population estimates \cite{ONSIMD2019}.

\begin{figure}[!ht]
  \begin{subfigure}[]{.95\linewidth}
    \centering
    \subcaption{Case incidence per 100k population with 14-day moving average window. Left: aggregated by age. Right: aggregated by IMD (1 refers to most deprived).}
    \includegraphics[width=11cm]{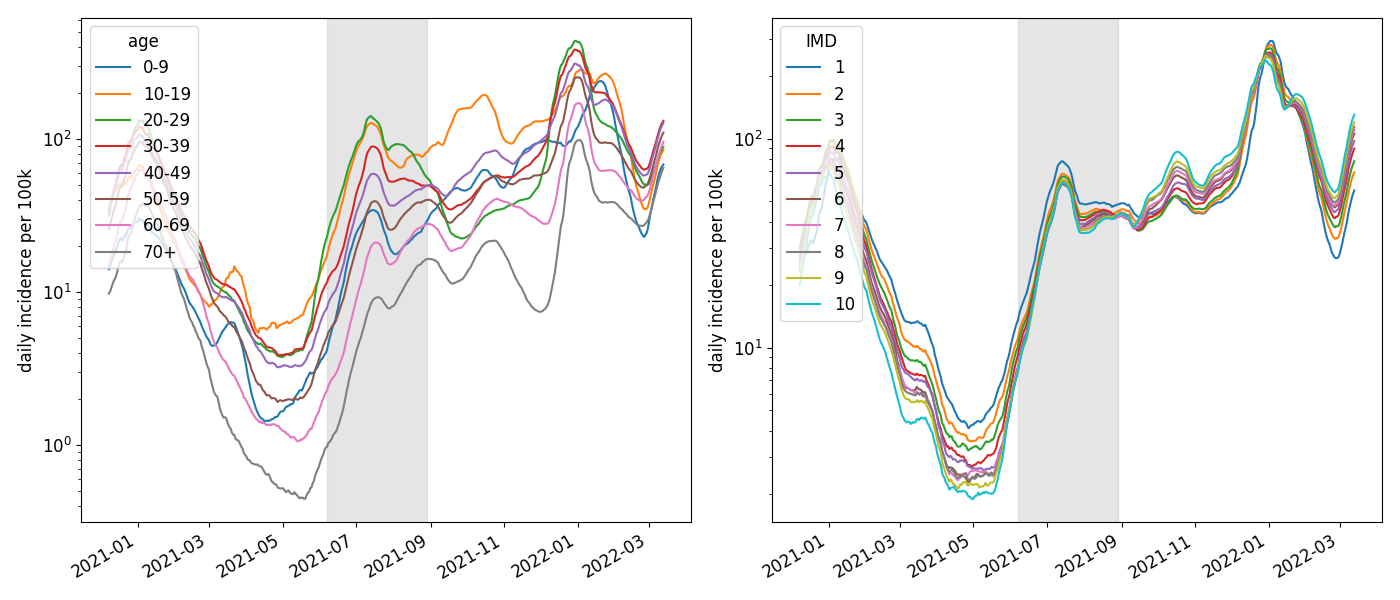}
    \label{fig:raw_data_all_counts}
  \end{subfigure}
  \begin{subfigure}[]{.95\linewidth}
    \centering
    \subcaption{Case incidence per 100k population over training and forecast time period. Left: aggregated by age, showing weekday-weekend effect; less cases recorded at weekend. Right: aggregated by IMD and smoothed with 14-day moving average window, showing deprivation-switching during September/October 2021.}
    \includegraphics[width=11cm]{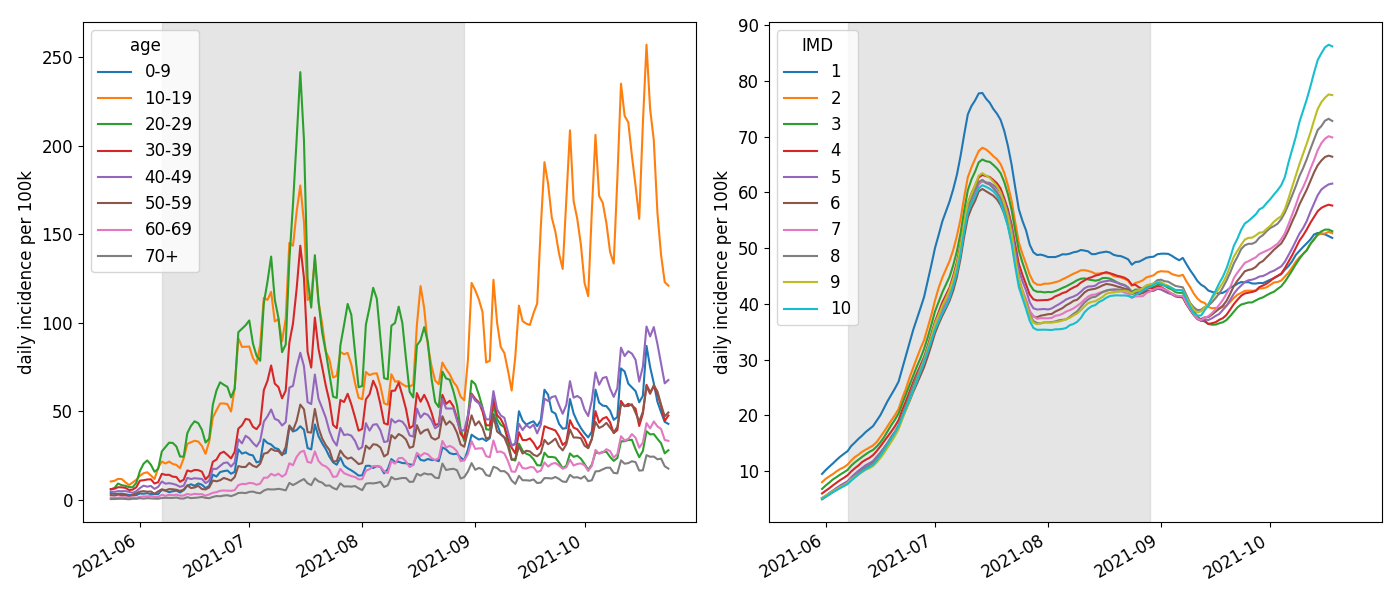}
    \label{fig:plot_raw_counts_per_pop_short_timeseries}
  \end{subfigure}
  \begin{subfigure}[]{.95\linewidth}
    \centering
    \subcaption{Cumulative sum as a proportion of each age (left) or IMD (right) group's population size.}
    \includegraphics[width=11cm]{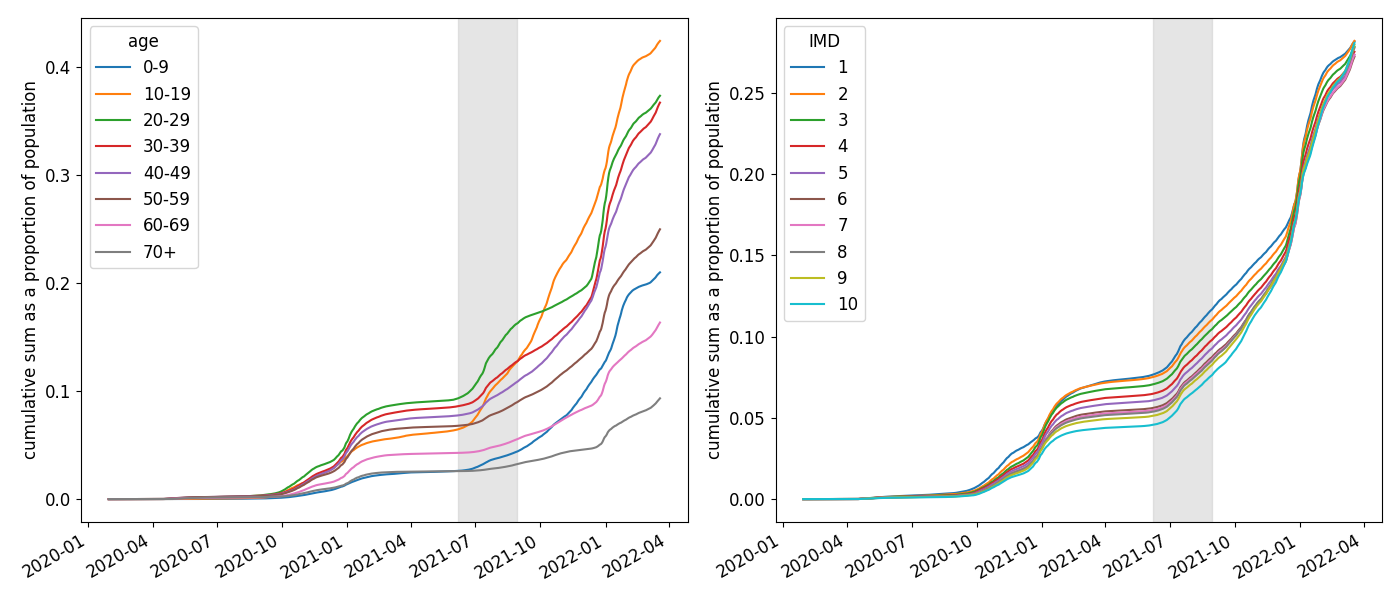}
    \label{fig:raw_data_all_sumsums}
  \end{subfigure}  
  \caption{UKHSA COVID-19 community testing case data for England aggregated by age group and deprivation (IMD) decile. Grey bands from 7 June to 29 August 2021 inclusive indicate the time period of the training data.}
  \label{fig:UKHSAdata}
\end{figure}

Until the spring of 2021, the highest incidence was in the most deprived groups while the lowest incidence was in least deprived groups: the intermediate deprivation groups are ordered similarly; Figures \ref{fig:raw_data_all_counts} and \ref{fig:plot_raw_counts_per_pop_short_timeseries}.
However beyond the spring of 2021, in the epoch of the delta and omicron variants, this pattern reverses and furthermore the inversion is complete with all the deprivation groups appearing in reverse order: in other words the highest case incidences were reported by the least deprived groups.
This phenomena of deprivation-switching is further complicated by sometimes occurring during different time periods depending on the individual’s age: further detail is given in Figure S1 in Supplementary Material.
When the case data is aggregated by deprivation-age groups, the age-deprivation structure is clear.
For example, Figure \ref{fig:raw_data_choropleth} shows that cases aggregated over a 12-week period exhibit a trend across IMD deciles; a similar pattern is observed on a monthly time scale between April 2020 and March 2022 as shown in Figure S1.
For reference the cumulative sum of cases is shown in Figure \ref{fig:raw_data_all_sumsums}, it is of interest to note that by early 2022 there is very little difference between the deprivation groups although there is still a considerable difference between age groups.


Ethical approval: the UKHSA COVID-19 data were supplied after anonymisation under strict data protection protocols agreed between the Lancaster University and Public Health England. The ethics of the use of these data for these purposes was agreed by Public Health England with the UK government SPI-M(O)/SAGE committees.


\clearpage
\newpage

\section{Model description} \label{sec:description}


\textit{Notation}, we adopt the following conventions by denoting objects as follows:
\begin{itemize}
    \item matrices - bold uppercase symbols e.g. $\mathbf{X}$
    \item vectors i.e. row/column matrices - bold lower case symbols e.g. $\mathbf{x}$
    \item matrix elements - italics lower case e.g. $x_{i,j}$
    \item $\mathbf{1}_{I \times J}$ - matrix of ones with size $I \times J$
    \item a tilde over a symbol - object is centered e.g. $\mathbf{\tilde{x}} = \mathbf{x} - \frac{1}{N} \sum^{N}_{n=1} x_n$
\end{itemize}

\bigskip

Consider a system where individuals are distributed between one of four states: susceptible to infection (S); exposed but not yet infectious (E); infectious (I); and removed due to recovery or death (R).
These states form a directed graph such that over time individuals may transition between successive states in the direction of the sink state R.
Let the population be stratified, as such each individual is assigned a stratum in which they remain over all time.
In terms of column matrices the system's state at time $t$ is fully described by the state matrices $\mathbf{x}^{\mathrm{S}}(t)$, $\mathbf{x}^{\mathrm{E}}(t)$, $\mathbf{x}^{\mathrm{I}}(t)$ and $\mathbf{x}^{\mathrm{R}}(t)$ whose length equals to the number of strata.
As such $\mathbf{x}^{\mathrm{S}}(t)$ represents the number of individuals in state S at time $t$ and likewise for the other three states.
The system dynamics are defined by a compartmental model \cite{Brauer2019} of the form
\begin{subequations}
\label{eq:odes}
\begin{align}
    \dot{\mathbf{x}}^{\mathrm{S}}(t) &= -\mathbf{h}^{\mathrm{SE}}(t) \, \mathbf{x}^{\mathrm{S}}(t)\label{eq:ode1}\\
    \dot{\mathbf{x}}^{\mathrm{E}}(t) &= \mathbf{h}^{\mathrm{SE}}(t) \, \mathbf{x}^{\mathrm{S}}(t) - \mathbf{h}^{\mathrm{EI}}(t) \, \mathbf{x}^{\mathrm{E}}(t)\label{eq:ode2}\\
    \dot{\mathbf{x}}^{\mathrm{I}}(t) &= \mathbf{h}^{\mathrm{EI}}(t) \, \mathbf{x}^{\mathrm{E}}(t) - \mathbf{h}^{\mathrm{IR}}(t) \, \mathbf{x}^{\mathrm{I}}(t)\label{eq:ode3}\\
    \dot{\mathbf{x}}
    ^{\mathrm{R}}(t) &= \mathbf{h}^{\mathrm{IR}}(t) \, \mathbf{x}^{\mathrm{I}}(t)\label{eq:ode4}
\end{align}
\end{subequations}
where dot denotes time derivative and the $\mathbf{h}(t)$ terms describe the hazard rates of transitions between successive states: for example $\mathbf{h}^{\mathrm{SE}}(t)$ is the hazard rate for a transition from the susceptible to exposed state for each stratum.
This rate will be defined such that a susceptible individual experiences a force of infection from all infectious individuals within their stratum and also a force weighted by contact rates from infectious individuals in all other strata.
Note that $\dot{\mathbf{x}}^{\mathrm{S}}(t) + \dot{\mathbf{x}}^{\mathrm{E}}(t) + \dot{\mathbf{x}}^{\mathrm{I}}(t) +\dot{\mathbf{x}}^{\mathrm{R}}(t) = \bm{0}$ thus the number of individuals in each stratum is conserved throughout time.

We require that our model is complex enough to represent the dynamics observed in the UKHSA COVID-19 data and also flexible enough to account for observed processes at stratum level.
To this end we extend the above model such that the hazard rate for the transition from state $\mathrm{S}$ to $\mathrm{E}$, i.e. $\mathbf{h}^{\mathrm{SE}}(t)$, includes the following:
age-structured social mixing matrix;
deprivation-structured social mixing matrix;
relative hazard of each age group;
relative hazard of each imd group, per age group;
and a random walk to account for the inherent stochasticity in the data.
In terms of interpretation: `social mixing patterns' are accounted for by the mixing matrices, to be denoted as $\bm{\mathcal{C}}$; and `behavioural patterns' aside from mixing patterns are accounted for by the relative hazard across age and deprivation strata, these will be incorporated in a single term $\bm{\chi}$. 

The full mathematical details of the model are described below.
The population is stratified by deprivation and age giving $L$ strata.
Time is discretised into units of days such that $t \in \{1,2,...,T\}$.
To formulate all hazard rates the following definitions are required:
\begin{enumerate}
    \item Constants:
        \begin{itemize}
            \item $J$ total number of deprivation strata
            \item $K$ total number of age strata
            \item $L$ total number of strata i.e. $L = J \, K$
            \item $T$ total number of time points
            \item Fixed model parameters: $\eta$, $\phi$, $\nu$, $\gamma_0$
        \end{itemize}
    \item Parameters to be estimated along with their corresponding priors:
        \begin{itemize}
            \item $\bm{\psi} \overset{\mathrm{iid}}\sim \mathrm{Beta}(1, 1)$, with components $[\psi_1,\psi_2, ...\psi_K]^\mathrm{T}$
            \item $\bm{\rho} \overset{\mathrm{iid}}\sim \mathrm{Beta}(1, 1)$, with components $[\rho_1,\rho_2, ...\rho_K]^\mathrm{T}$
            \item $\gamma_1 \sim \mathcal{N}(0, \sigma_{\gamma_1})$
            \item $\alpha_0 \sim \mathcal{N}(0, \sigma_{\alpha_0})$
            \item $\bm{\alpha_t} \sim \mathcal{N}(\bm{0}, \bm{\Sigma}_{\alpha_t})$ with size $T$
        \end{itemize}

    \item Covariates:
        \begin{itemize}
            \item The day of the week effect, e.g. as observed in left panel of Figure \ref{fig:plot_raw_counts_per_pop_short_timeseries}, is described by $\mathbf{w}=[w_1, w_2, ...,w_T]^\mathrm{T}$ with each element taking one of two values: $2/7$ for weekdays otherwise $-5/7$.
            \item Each element of $\mathbf{d}=[d_1, d_2, ...,d_J]^\mathrm{T}$ represents a unique deprivation group.
            \item The contact rates between age groups are defined by $\bm{\mathcal{C}}^A$, size $K \times K$, an age-structured social mixing matrix.  Similarly contact rates among deprivation groups are represented by $\bm{\mathcal{C}}^D$, size $J \times J$. 
            The Kronecker product of age and deprivation contact rates, denoted as $\mathcal{K}(\bm{\mathcal{C}}^A , \bm{\mathcal{C}}^D)$, describes the overall contact rate structure.
            \item A matrix of population sizes is denoted by $\mathbf{n}=[n_1, n_2, .., n_{L}]^\mathrm{T}$ where each element relates to a given stratum, hence $\mathbf{n}$ has length $L=JK$.  Let $1/\mathbf{n}$ denote $ [1/n_1, 1/n_2, ..., 1/n_L]^\mathrm{T}$.
        \end{itemize}
    \item To account for differential behaviour (excluding social mixing) across stratum we introduce a behavioural adaptation vector $\bm{\chi} = [\bm{\varkappa}_1, \bm{\varkappa}_2, ..., \bm{\varkappa}_K]^\mathrm{T}$.
    In its simplest form $\bm{\varkappa}_k$ is an intercept and slope model for each age stratum, for reference its structure is depicted in Figure \ref{fig:visualofChi}.
    With mean quantities $\bar{\psi} = \frac{1}{K} \sum\nolimits^K_{k=1} \psi_k$ and $\bar{d} = \frac{1}{J} \sum\nolimits^J_{j=1} d_j$, then  $\bm{\varkappa}_k$ is defined as:
                \begin{subequations}
                \label{eq:chis}
                \begin{align}
                    \bm{\tilde{\psi}} &= \bm{\psi} - \bar{\psi}\\
                    \bm{\tilde{\rho}} &= \bm{\rho} - \frac{1}{2}\\
                    \mathbf{\tilde{d}} &= \mathbf{d} - \bar{d}\\
                    \bm{\varkappa}_k  &= \phi \left(1  + \tilde{\psi}_k \right)  +  \eta \left( \frac{1}{2} + \tilde{\rho}_k  f( \mathbf{\tilde{d}}) \right) \label{eq:chi}
                \end{align}
                \end{subequations}
                where vector $\bm{\varkappa}_k$ has length $J$, $\phi>0$ and $\eta>0$.
                Under the given parameterisation the function $f(\mathbf{\tilde{d}})$ should be chosen to span $-1$ to $1$ so that all elements of $\bm{\chi}$ are greater than zero.
                By design the area under $\bm{\chi}$ is invariant to changes in $\bm{\psi}$ and $\bm{\rho}$, in this sense $\bm{\chi}$ accounts for the relative effect by age and deprivation strata.
                Vector $\bm{\tilde{\psi}}$ is interpreted as the relative force of infection among the age groups.
                For a given age group $k$, function $f(\mathbf{\tilde{d}})$ describes the force as a function of deprivation; for example
                the function could be a slope which is parameterised by $\tilde{\rho}_k$.
                The precise form of $\bm{\chi}$ will be chosen in the Section \ref{sec:model_selection} to reflect the observed structure seen in the UKHSA COVID-19 data, as observed in Figures \ref{fig:raw_data_choropleth} or S1.
                Note that, the form of $\bm{\varkappa}_k$, particularly the second term, is application specific so it may need to be adapted or generalised in future applications however the key feature of preserving the area under $\bm{\chi}$ is recommended as this aids fitting when using an MCMC.
        \item 
        A temporally-correlated random effect, specifically a random walk, is included to account for the temporal variability in the global transmission rate as follows
        \begin{equation}
            a(t) = \alpha_0 + \sum_{q=1}^{Q_t} \bm{\alpha}_{\bm{t}_q}
        \end{equation}
        where $Q_t$ is the number of jumps in the interval $(0, t)$.

\end{enumerate}

With these definitions the hazard rate vectors of length $L$ are:
\begin{subequations}
    \label{eq:hazardrates}
    \begin{align} 
        \mathbf{h}^{\mathrm{SE}}(t) &= \exp(a(t)) \; \bm{\chi} \odot \frac{1}{\mathbf{n}} \odot \left[   \mathcal{K}(\bm{\mathcal{C}}^A , \bm{\mathcal{C}}^D)   \cdot \left(\mathbf{x}^{\mathrm{I}}(t) \odot \frac{1}{\mathbf{n}}\right)\right] \label{eq:hazardrates1}\\ 
        \mathbf{h}^{\mathrm{EI}}(t) &= \nu \mathbf{1}_{L \times 1} \label{eq:hazardrates2}\\
        \mathbf{h}^{\mathrm{IR}}(t) &= \exp(\gamma_0+\gamma_1 w_t) \mathbf{1}_{L \times 1} \label{eq:hazardrates3}
    \end{align}
\end{subequations}
where $\odot$ denotes element-wise multiplication and $\cdot$ matrix multiplication.
It follows that on average individuals spend $1/\nu$ days in the exposed state followed by approximately $1 / \exp(\gamma_0)$ days in the infectious state before entering the sink state.

Given the state at time $t$ the approximate reproduction number $r_{j}(t)$ is defined as the expected number of further individuals that one individual in stratum $j$ will go on to infect.
Consequently using the force of infection exerted by an individual in $j$ on a susceptible individual in $i$ then with reference to Equation \ref{eq:hazardrates1}:
\begin{equation}
    r_{j}(t) \approx \frac{1 - \exp \left( - \exp(a(t)) \delta_t \sum_{i=1}^L x_i^{\mathrm{S}}(t) \frac{\chi_i}{N_i} \frac{\kappa_{ij}}{ N_j} \right)}{1-\exp( - \gamma_0 \delta_t)}
\label{eq:Rjt}
\end{equation}
where 
$x_i^{\mathrm{I}}(t)=1$ individual $\forall \, i$, 
time step length $\delta_t=1$ (day) and $\kappa_{ij}$ denotes elements of $\mathcal{K}(\bm{\mathcal{C}}^A , \bm{\mathcal{C}}^D)$. 
Over the course of an individual's infectious period both $x_i^{\mathrm{S}}(t)$ and $\mathbf{h}^{\mathrm{SE}}(t)$ are assumed constant.

\section{Application to COVID-19 in England} \label{sec:application}

\subsection{Model specification} \label{sec:model_spec}

The model was specified with the following sources of publicly available data.
The 2019 English Index of Multiple Deprivation (IMD) \cite{GOVIMD2019} was used to categorised each case into a deprivation decile to this end $\mathbf{d}=[1,2,...,10]^\mathrm{T}$ where $1$ refers to the most deprived group and $10$ the least.
The 2019 Office for National Statistics population estimates \cite{ONSIMD2019} were used to construct vector $\mathbf{n}$.
The empirical survey data, \texttt{POLYMOD} \cite{polymod_data} \cite{Mossong2008}, was used to estimate the age-structured social mixing matrix $\bm{\mathcal{C}}^A$ with the R-package \texttt{socialmixr} \cite{socialmxr}.
To the authors' knowledge there is not an empirical study which estimates contact rates between deprivation groups in England, or a comparable population, therefore homogeneous mixing was assumed as such $\bm{\mathcal{C}}^D = \mathbf{1}_{J \times J}$.

Observations from the UKHSA COVID\nobreakdash-19 data, e.g. Figures \ref{fig:raw_data_choropleth} and S1, suggest that an intercept and slope model is a reasonable first order approximation for $\bm{\varkappa}_k$.
With reference to Equation \ref{eq:chis} let $f(\mathbf{\tilde{d}})=  \tanh(-\xi \mathbf{\tilde{d}})$, with $\xi > 0$, be sufficiently large to produce an approximately linear slope across the deciles.
It then follows that each component of $\bm{\tilde{\rho}}$ determines, for a given age group, the magnitude and direction of the monotonic slope across the deprivation groups.
Investigations into more complex forms of $f(\mathbf{\tilde{d}})$ and therefore $\bm{\varkappa}_k$, e.g. polynomials of degree greater than one, are beyond the scope of this work.

Age groups were chosen to span ten years from $0\text{-}9$ to $60\text{-}69$, with the final category defined as $70$ or older.
Given these age groups and the deprivation deciles accordingly there were $L=80$ strata.

The training data was the UKHSA COVID-19 positive case data from 7 June to 29 August 2021 inclusive; during this epoch the delta variant was dominant.
The daily case incidence, aggregated by strata, represents the number of events for the I to R transition, that is $| \mathbf{x}^{\mathrm{I}}(t) - \mathbf{x}^{\mathrm{R}}(t)|$.
The constants and hyperparameters were defined as:
        $\eta = 2$,
        $\phi = 2$,
        $\xi = 0.3$, 
        $\nu = 0.28$,
        $\gamma_0=\log_e(0.25)$,
        $\sigma_{\gamma_1} = 100$,
        $\sigma_{\alpha_0} = 10$,
        $\bm{\Sigma}_{\alpha_t} = \mathrm{diag}(\bm{5})/1000$.
This model was fitted using a fully Bayesian approach
based on discrete-time Markov chain methodology where the differential equations were modelled with a stochastic Euler scheme based on the Chain Binomial method \cite{Abbey1952} \cite{Fine1977}.
A brief outline of this methodology given in the Appendix.

The code used in relation to this paper is publicly available: the \texttt{imd-age-covid19uk} repository \cite{imdage} contains the model specification; and the \texttt{gemlib} repository \cite{gemlib} holds the MCMC algorithms.
A straightforward example of using these repositories is given on the home page of the \texttt{imd-age-covid19uk} repository.

\subsection{Model selection} \label{sec:model_selection}

The behavioural adaptation term and contact network structure between strata are both unobserved, and were optimised using predictive model selection.  Here, the continuous rank probability score (CRPS) \cite{Matheson1976} was used to compare model predictions with the observed data: 
the posterior predictive CRPS was calculated for each stratum $i$ at each time point $t$, denoted by $\mathrm{cprs}_{i,t}$ where smaller values indicate a better fit.



\begin{figure}[!ht]
  \begin{subfigure}[]{.95\linewidth}
    \centering
    \subcaption{CRPS per stratum $i$, specifically the median over time of all respective $\mathrm{cprs}_{i,t}$: lower values indicate a better fit. Left: least complex model, A,  (Table \ref{tab:RSP_imdage}, top row). Right: full model, D, (Table \ref{tab:RSP_imdage}, bottom row, i.e. Equations \ref{eq:chis} and \ref{eq:hazardrates}).}
    \includegraphics[width=11cm]{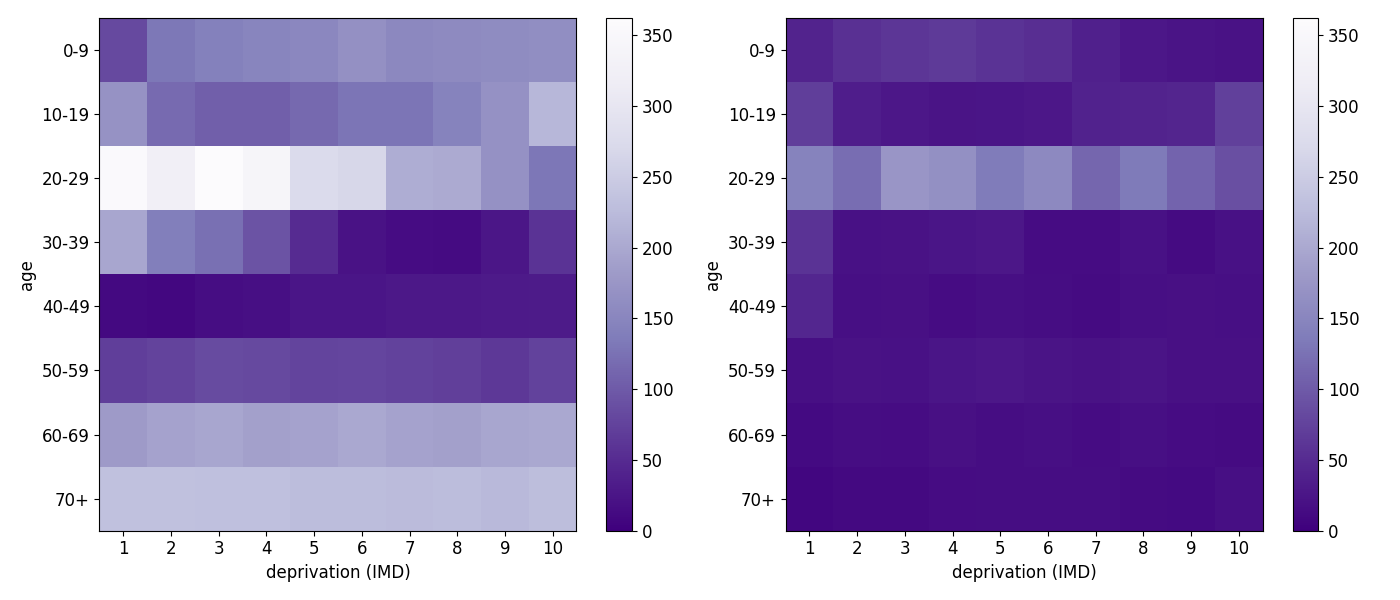}
    \label{fig:fitted_data_RPS}
  \end{subfigure}
  \begin{subfigure}[]{.95\linewidth}
    \centering
    \subcaption{Expected values of the elements of the behavioural adaptation vector, $\bm{\chi}$. Left: without normalisation. Right: normalised by population size, specifically elements of vector $\bm{\chi} \odot \frac{1}{\mathbf{n}}$.}
    \includegraphics[width=11cm]{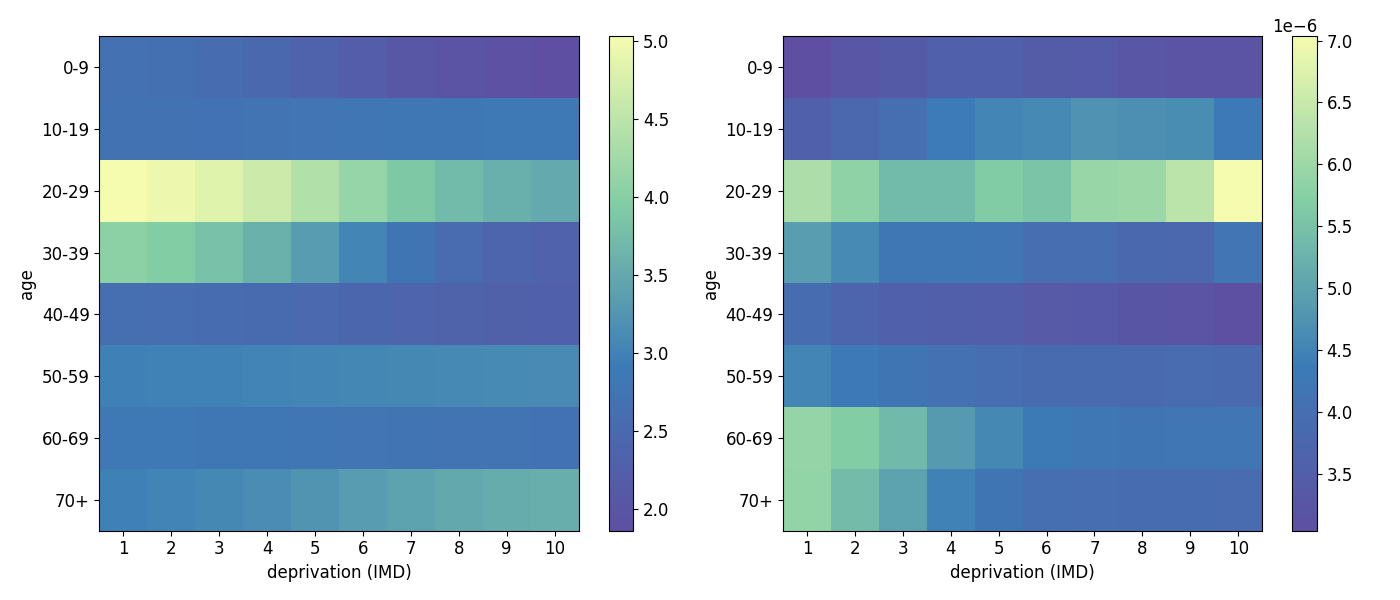}
    \label{fig:fitted_data_chi}
  \end{subfigure}
   \begin{subfigure}[]{.95\linewidth}
    \centering
    \subcaption{Case incidence aggregated by deprivation and age over the training data period. Left: case count. Right: case count per 100k population.}
    \includegraphics[width=11cm]{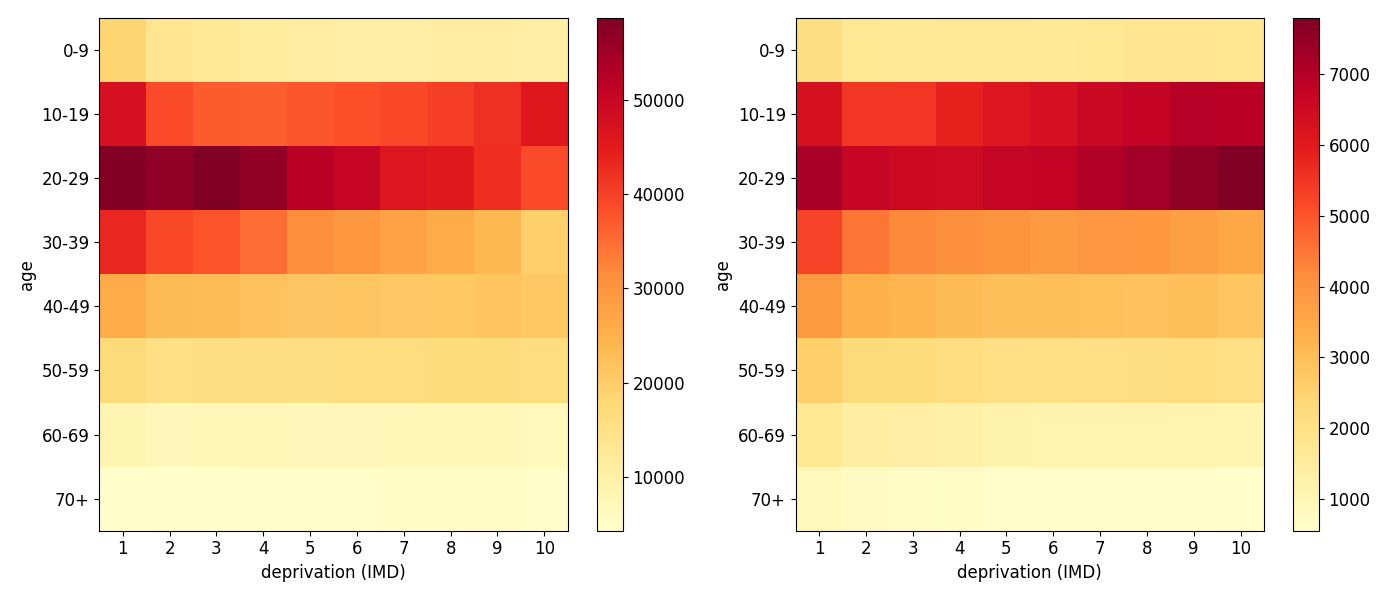}
    \label{fig:raw_data_choropleth}
  \end{subfigure}
  \caption{Top two panels: CRPS and behavioural adaptation term estimates for model fitted to training data (07/06/2021 - 29/08/2021). Bottom panel: UKHSA COVID-19 positive case data aggregated over the training data period.}
  \label{fig:RPS_CHI_choropeth}
\end{figure}

Using the training data, four models are fitted each with increasing complexity as detailed in Table \ref{tab:RSP_imdage}.
With reference to this table the simplest model, A, has both constant hazard/susceptibility and homogeneous mixing across all age and deprivation groups.
Model B is more complex in that it includes inhomogeneous mixing across age groups.
Model C builds on B in that it also includes relative hazard/susceptibility across deprivation groups per age stratum.
Finally model D, the full model, adds complexity to model C by additionally including relative hazard/susceptibility across age groups; note that model D is given by Equations \ref{eq:chis} and \ref{eq:hazardrates}. 
The CRPS estimates summarised in Table \ref{tab:RSP_imdage} show that the median CRPS of the full model, D, is about one fifth lower than that of the least complex model, A.
Visually this can be seen in Figure \ref{fig:fitted_data_RPS} which shows that the full model, D, has on average a substantially better fit across the strata when compared with the least complex model, A.

Furthermore given full model fitted to the training data then $\bm{\chi} = [\bm{\varkappa}_1, \bm{\varkappa}_2, ..., \bm{\varkappa}_K]^\mathrm{T}$ reflects the general structure of the UKHSA COVID-19 data. 
This can be seen by comparing the expected value of the elements of $\bm{\chi}$ in Figure \ref{fig:fitted_data_chi} left panel with the case data shown in Figure \ref{fig:raw_data_choropleth} left panel: this is similarly true when comparing the right panels which have been adjusted by population size.
For instance the similarity between the $20\text{-}29$ age group in the left panels of Figures \ref{fig:fitted_data_chi} and \ref{fig:raw_data_choropleth} is particularly striking.
This confirms that the model fit appears to be consistent with the data.

In the interests of parsimony additional model complexity is not considered here.
The full model, D, has sixteen parameters which need to be fitted i.e. $2K=16$ however in principle the most complex form of $\bm{\chi}$ (assuming it is not time dependent) would have $L=80$ parameters that is one parameter per stratum.
For reference Figures S2 to S5 in the Supplementary Material show the posterior samples of all estimated parameters for the full model, D, fitted to the training data.

\begin{table}
\caption{\label{tab:RSP_imdage}CRPS for different models fitted to the training data where successive rows relate to models of increasing complexity, the bottom row is the full model as defined in Equations \ref{eq:chis} and \ref{eq:hazardrates}. The CRPS is summarised using the quartiles ($Q_{1,2,3}$) of all $\mathrm{cprs}_{i,t}$.  Increasing model complexity is associated with decreasing CRPS and hence a better fit.}
\centering
\begin{tabular}{lll|lll}
\multicolumn{3}{c|}{Model} & \multicolumn{3}{c}{$\mathrm{cprs}_{i,t}$} \\
type & $\bm{\varkappa}_k$ & $\bm{\mathcal{C}}^A$ & $Q_1$ & $Q_2$ & $Q_3$\\ \hline
A & $\mathbf{1}_{J \times 1}$ & $\mathbf{1}_{K \times K}$ &  41.5 & 122.5 & 217.2\\
B & $\mathbf{1}_{J \times 1}$ & $\bm{\mathcal{C}}^A$ &  13.0 & 41.9 & 138.3\\
C & $\eta \left( \frac{1}{2} + \tilde{\rho}_k  f( \mathbf{\tilde{d}}) \right)$ & $\bm{\mathcal{C}}^A$ &  11.6 & 35.5 & 115.7 \\
D & $\phi \left(1  + \tilde{\psi}_k \right)  +  \eta \left( \frac{1}{2} + \tilde{\rho}_k  f( \mathbf{\tilde{d}}) \right)$ & $\bm{\mathcal{C}}^A$ & 9.1 & 25.4 & 63.2
\end{tabular}
\end{table}


\clearpage
\newpage

\subsection{Model predictions} \label{sec:predictions}

To gain insight into the unfolding epidemic, metrics are estimated using the posterior samples of the model fitted to the training data (7 June to 29 August 2021 inclusive).

\begin{figure}[!ht]
  \begin{subfigure}[]{1.0\linewidth}
    \centering
    \subcaption{Predicted reproduction number per stratum $j$ where $\tau$ is the last day in the training data. Left: expected values of $r_j(\tau)$. Right: probability that $r_j(\tau)>1$.}
    \includegraphics[width=11cm]{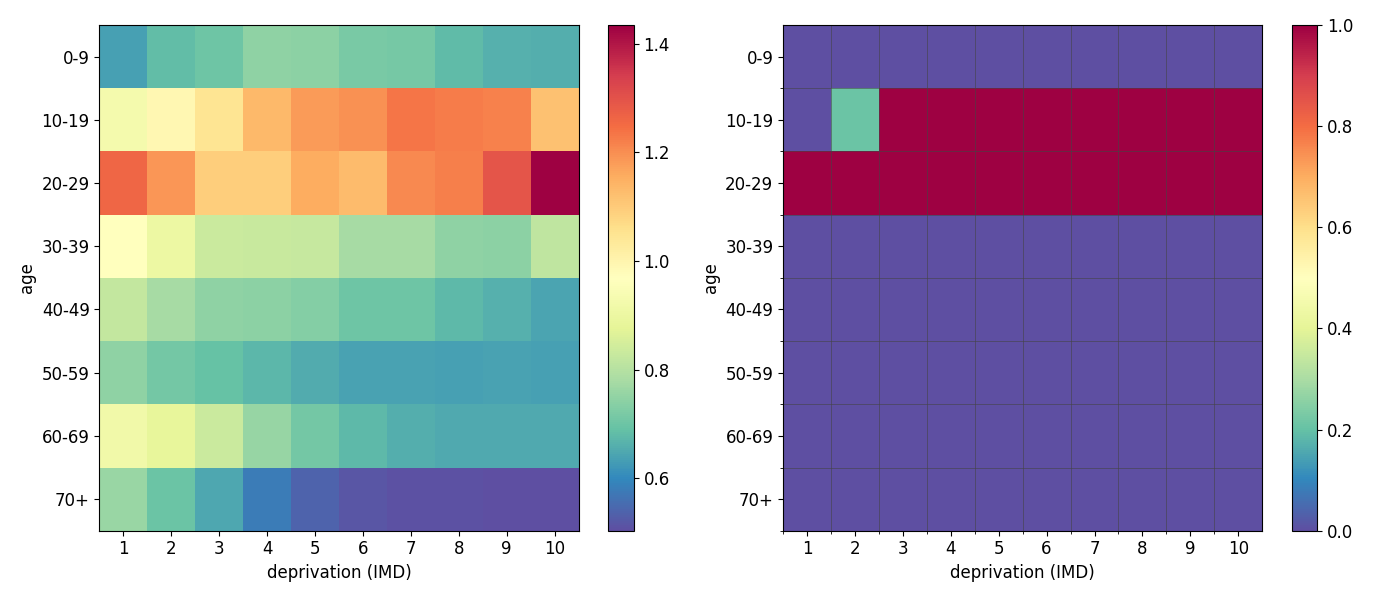}
    \label{fig:fitted_data_choropleth_Rt}
  \end{subfigure}
  \begin{subfigure}[]{1.0\linewidth}
    \centering
    \subcaption{Model forecast for first day beyond the training data. Left: expected case incidence. Right: expected case incidence per 100k.}
    \includegraphics[width=11cm]{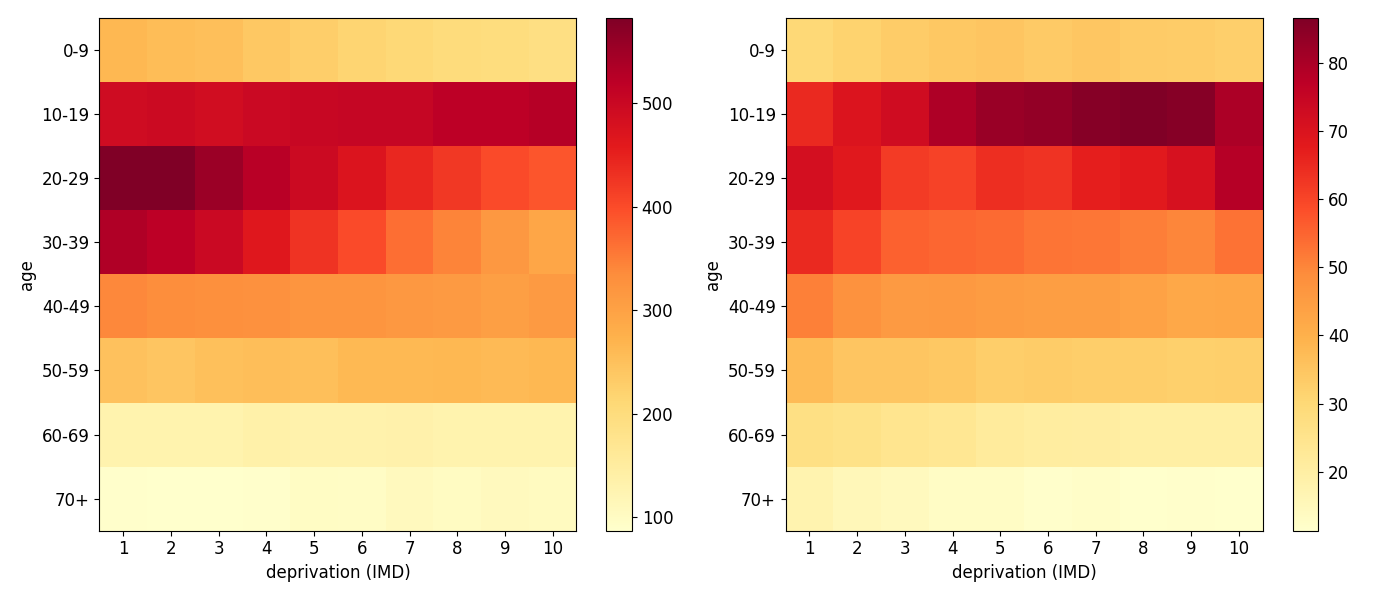}
    \label{fig:fitted_data_choropleth_incidence}
  \end{subfigure}
  \begin{subfigure}[]{1.0\linewidth}
    \centering
    \subcaption{8-week forecast of expected case incidence (solid lines) alongside their respective $90\%$ credible intervals, and observed case incidence (dotted lines). 
    CRPS quartiles over all $\mathrm{cprs}_{i,t}$ are $(Q_1,Q_2,Q_3) = (41.5, 97.2, 197.3)$.
    Left: aggregated by age group. Right: aggregated by deprivation (IMD) decile.}
    \includegraphics[width=11cm]{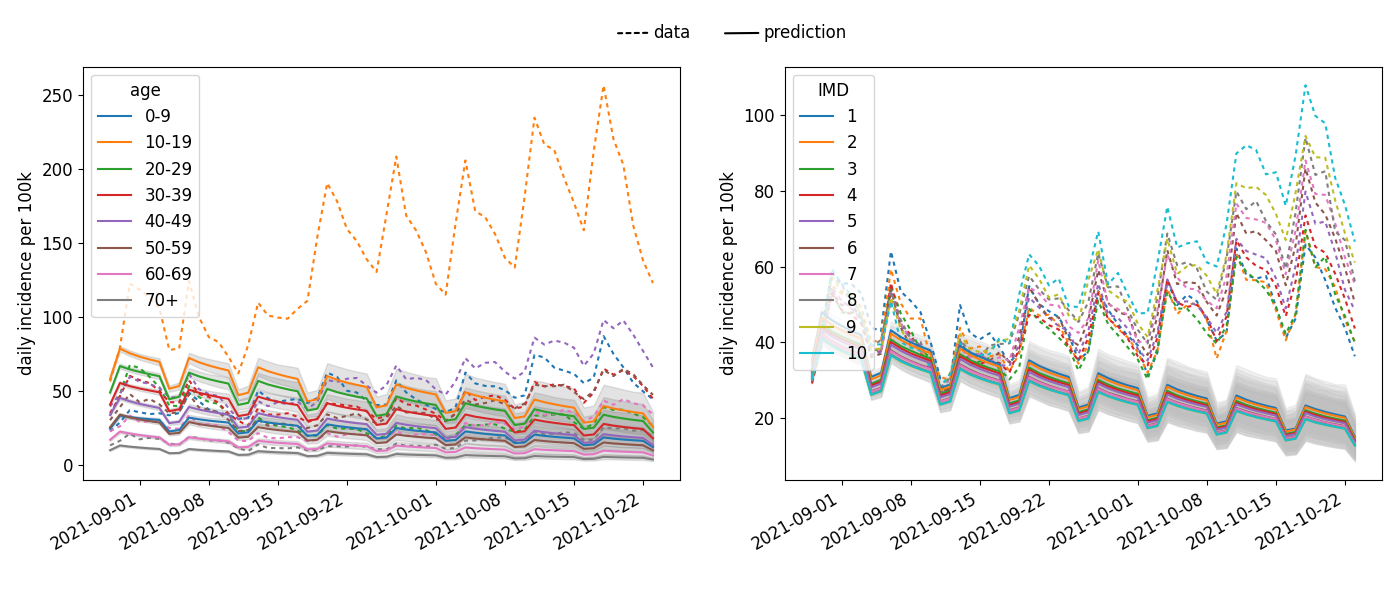}
    \label{fig:fitted_data_ts_incidence_per_100k}
  \end{subfigure}
  \caption{Predictions from model fitted to training data (07/06/2021 - 29/08/2021). Deprivation (IMD) 1 decile refers to the most deprived.}
  \label{fig:inference}
\end{figure}

First, the expected value of the reproduction number is estimated for each stratum using Equation \ref{eq:Rjt}.
An example is given in Figure \ref{fig:fitted_data_choropleth_Rt} which shows that on 29 August 2021 the $20\text{-}29$ age group were predicted to have a reproduction number greater than one irrespective of deprivation.
The highest reproduction number was predicted to be in the least deprived $20\text{-}29$ stratum, in contrast the lowest reproduction numbers were associated with the least deprived in the $70+$ age group.

Secondly, case incidence forecasts for each stratum are estimated using the posterior samples and simulating forward in time beyond the training data epoch.
As an example see the expected case incidence forecast for the 30 August 2021 in Figure \ref{fig:fitted_data_choropleth_incidence}: note that this figure has similar characteristics to the UKHSA COVID-19 data shown in Figure \ref{fig:raw_data_choropleth}.
Aggregating by age and deprivation, the time course over an 8-week forecast period is shown in Figure \ref{fig:fitted_data_ts_incidence_per_100k}, in the left panel the weekday-weekend effect is clearly visible however in contrast to the UKHSA COVID-19 case data (e.g. Figure \ref{fig:plot_raw_counts_per_pop_short_timeseries}) the right panel does not show any sign of the deprivation-switching.
Furthermore the dynamics of this simulated forecast contradict those observed, in that the case incidence forecast (solid lines) has a downward trend over time whereas the observed case incidence (dotted lines) has an upward trend.
This anomaly arises because the simulated forecast is with respect to the fitted model which cannot account for future yet-to-be-observed external changes, hence an unobserved external factor not present during the training period is resulting in the observed epidemic growing.
Note that this anomaly is not due to a new more infectious variant as the dominant variant during this epoch is Delta, this does not change until mid to late December 2021 when Omicron begins to dominate.

\begin{figure}
    \centering
    \includegraphics[width=0.95\textwidth]{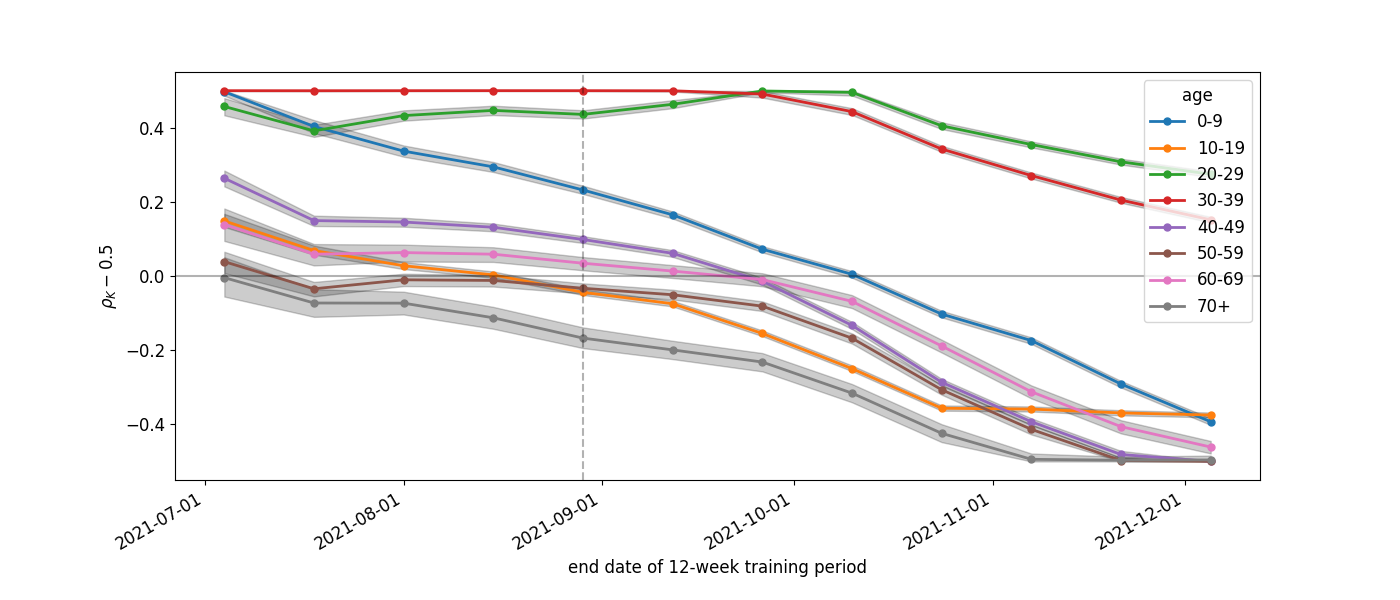}
    \caption{Expected value and 98\% credible interval of posterior samples for model parameters $\bm{\tilde{\rho}} = \bm{\rho} - 0.5$ where the model is fitted to a number of different training datasets, each with a 12-week window.  The circles are at fortnightly intervals and located at the final date of the estimate's respective training data period.  The vertical dashed line relates to posterior estimates from the model fitted to the training data, 07/06/2021 - 29/08/2021. Given age group $k$, then if $(\rho_k-0.5) >0$ the most deprived groups report the most cases, $(\rho_k-0.5) <0$ the least deprived groups report the most cases, otherwise there is no difference between the groups or the trend is non-linear and symmetric e.g. a second-degree polynomial. For completeness $\bm{\tilde{\psi}}$ is depicted in Figure S14.}
    \label{fig:plot_rho_from_multiple_MCMC_runs}
\end{figure}

Finally, to investigate further changing patterns in recorded incidence the model is fitted, at fortnightly intervals, using training data with a 12-week window throughout the last part of 2021: note that everywhere else in this paper the training data is always from 7 June to 29 August 2021 inclusive.
The results are given in Figure \ref{fig:plot_rho_from_multiple_MCMC_runs} in terms of the elements of $\bm{\tilde{\rho}}$, values below zero indicate that the least deprived individuals report the most cases.
As time increases there is a downward trend across all age groups which is consistent with the UKHSA COVID-19 case data in-so-much as increasing numbers of cases are reported by the least deprived groups. 
The model fit over multiple training data time periods therefore reflects the trends observed in the case data.
The changes in $\bm{\tilde{\rho}}$ suggest there is an underlying process with respect to deprivation which is not directly identifiable from the UKHSA COVID-19 case data, such as population-level changes in behaviour including testing patterns and social mixing.

\clearpage
\newpage

\subsection{Simulation of deprivation-switching} \label{sec:sim_switching}

In Section \ref{sec:dataset} deprivation-switching and a generally increasing trend in reported case incidence was observed in the UKHSA COVID-19 positive case data during the autumn of 2021, e.g. Figure \ref{fig:plot_raw_counts_per_pop_short_timeseries}, however these dynamics were not reproduced by the forward simulations of Section \ref{sec:predictions}, Figure \ref{fig:fitted_data_ts_incidence_per_100k}.
The focus of this section is these anomalies. 
The model is fitted to the training data (7 June to 29 August 2021 inclusive) and then forward simulation experiments, based on the posterior samples, are used to investigate possible hypotheses regarding deprivation-switching, specifically: depletion of susceptible individuals; differential increase in social mixing across deprivation deciles; and differential changes in testing patterns across deprivation deciles.
The objective here is to gauge the plausibility of each hypothesis since to go beyond this would require additional data which is not available to this study, for example: the true case counts; empirical time-varying social mixing data; and a time-varying measure of population level behaviour in terms of deprivation-age strata e.g. testing behaviour.
In the absence of such data there are too many degrees of freedom for the model parameters to be uniquely identified.
To reduce the degrees of freedom age-related changes are not considered in this section as deprivation is the main interest, in this regard changes in deprivation are assumed to be independent of age.
It is therefore left for future work, with additional data sources, to undertake rigorous model selection using standard statistical hypothesis testing approaches.

\subsubsection{Depletion of susceptible individuals} \label{sec:sim_depletion}

The true COVID-19 case incidence in England was under reported, largely due to mass testing being unavailable during the initial months of the pandemic.
Additionally due to the relatively small case incidences hospital and care setting are not included in the data used in this work.
To investigate the depletion of susceptible individuals the model is first fitted to the training data assuming there is no depletion and secondly the cumulative sum from the beginning of the pandemic is computed from the UKHSA COVID-19 case data (Pillar 2 only), as shown in Figure \ref{fig:raw_data_all_sumsums}.
For a given forward simulation the cumulative sum is subtracted from the initial susceptible state and added to the initial recovered state.
To simulate deprivation-switching it was found that the depletion of susceptible individuals needed to be at least five times higher than that observed, this is shown in Figure \ref{fig:plot_simulated_timeseries_incidence_100k_depletion} where  deprivation-switching is visible.
Although the UKHSA COVID-19 case data will not have recorded all cases it is an unrealistic to presume such a large proportion of case data is missing.
Furthermore during September 2021, this simulated case incidence exhibited a downward trend, but the reverse is observed in the UKHSA COVID-19 case data: for example see the dotted lines in either Figure \ref{fig:plot_simulated_timeseries_incidence_100k_depletion} or Figure S6 in the supplementary material.
Note that during the autumn of 2021 the dominant variant, Delta, remains unchanged therefore a new more infectious variant is not the reason for the observed increase in case incidence during this epoch.

\textit{\textbf{In summary}}, depletion of susceptible individuals is not a plausible explanation for the observed deprivation-switching as the simulated dynamics exhibit a rapidly shrinking epidemic which disagrees with the observed case data in which the epidemic is growing.

\subsubsection{Differential increase in social mixing across
deprivation deciles}\label{sec:sim_social_mixing}

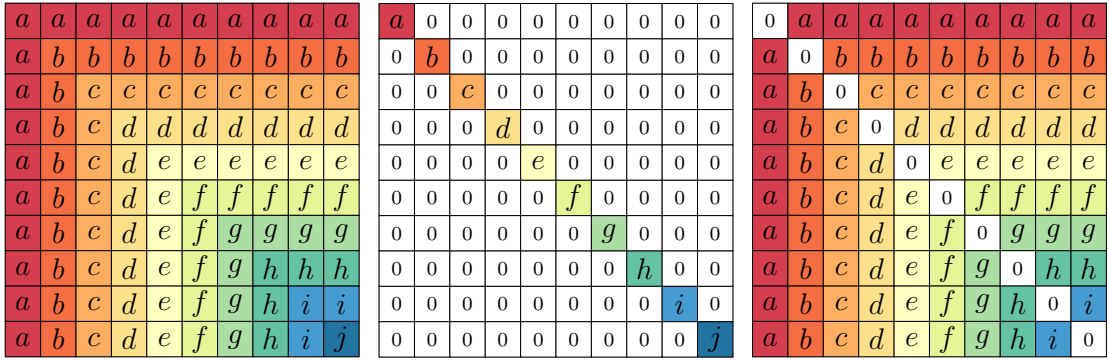
\begin{figure}
\centering

\begin{subfigure}{.33\textwidth}
\centering
\begin{tikzpicture}[x=5mm, y=5mm]
\draw[step=1cm,gray,very thin] (0,0) grid (10,10);

// 1
\filldraw[fill={rgb,255:red,213; green,62; blue,79}, draw=black] (0,9) rectangle (1,10);
\node at (0.5,9.5) {$a$};
\filldraw[fill={rgb,255:red,213; green,62; blue,79}, draw=black] (1,9) rectangle (2,10);
\node at (1.5,9.5) {$a$};
\filldraw[fill={rgb,255:red,213; green,62; blue,79}, draw=black] (2,9) rectangle (3,10);
\node at (2.5,9.5) {$a$};
\filldraw[fill={rgb,255:red,213; green,62; blue,79}, draw=black] (3,9) rectangle (4,10);
\node at (3.5,9.5) {$a$};
\filldraw[fill={rgb,255:red,213; green,62; blue,79}, draw=black] (4,9) rectangle (5,10);
\node at (4.5,9.5) {$a$};
\filldraw[fill={rgb,255:red,213; green,62; blue,79}, draw=black] (5,9) rectangle (6,10);
\node at (5.5,9.5) {$a$};
\filldraw[fill={rgb,255:red,213; green,62; blue,79}, draw=black] (6,9) rectangle (7,10);
\node at (6.5,9.5) {$a$};
\filldraw[fill={rgb,255:red,213; green,62; blue,79}, draw=black] (7,9) rectangle (8,10);
\node at (7.5,9.5) {$a$};
\filldraw[fill={rgb,255:red,213; green,62; blue,79}, draw=black] (8,9) rectangle (9,10);
\node at (8.5,9.5) {$a$};
\filldraw[fill={rgb,255:red,213; green,62; blue,79}, draw=black] (9,9) rectangle (10,10);
\node at (9.5,9.5) {$a$};

// 2
\filldraw[fill={rgb,255:red,213; green,62; blue,79}, draw=black] (0,8) rectangle (1,9);
\node at (0.5,8.5) {$a$};
\filldraw[fill={rgb,255:red,244; green,109; blue,67}, draw=black] (1,8) rectangle (2,9);
\node at (1.5,8.5) {$b$};
\filldraw[fill={rgb,255:red,244; green,109; blue,67}, draw=black] (2,8) rectangle (3,9);
\node at (2.5,8.5) {$b$};
\filldraw[fill={rgb,255:red,244; green,109; blue,67}, draw=black] (3,8) rectangle (4,9);
\node at (3.5,8.5) {$b$};
\filldraw[fill={rgb,255:red,244; green,109; blue,67}, draw=black] (4,8) rectangle (5,9);
\node at (4.5,8.5) {$b$};
\filldraw[fill={rgb,255:red,244; green,109; blue,67}, draw=black] (5,8) rectangle (6,9);
\node at (5.5,8.5) {$b$};
\filldraw[fill={rgb,255:red,244; green,109; blue,67}, draw=black] (6,8) rectangle (7,9);
\node at (6.5,8.5) {$b$};
\filldraw[fill={rgb,255:red,244; green,109; blue,67}, draw=black] (7,8) rectangle (8,9);
\node at (7.5,8.5) {$b$};
\filldraw[fill={rgb,255:red,244; green,109; blue,67}, draw=black] (8,8) rectangle (9,9);
\node at (8.5,8.5) {$b$};
\filldraw[fill={rgb,255:red,244; green,109; blue,67}, draw=black] (9,8) rectangle (10,9);
\node at (9.5,8.5) {$b$};

// 3
\filldraw[fill={rgb,255:red,213; green,62; blue,79}, draw=black] (0,7) rectangle (1,8);
\node at (0.5,7.5) {$a$};
\filldraw[fill={rgb,255:red,244; green,109; blue,67}, draw=black] (1,7) rectangle (2,8);
\node at (1.5,7.5) {$b$};
\filldraw[fill={rgb,255:red,253; green,174; blue,97}, draw=black] (2,7) rectangle (3,8);
\node at (2.5,7.5) {$c$};
\filldraw[fill={rgb,255:red,253; green,174; blue,97}, draw=black] (3,7) rectangle (4,8);
\node at (3.5,7.5) {$c$};
\filldraw[fill={rgb,255:red,253; green,174; blue,97}, draw=black] (4,7) rectangle (5,8);
\node at (4.5,7.5) {$c$};
\filldraw[fill={rgb,255:red,253; green,174; blue,97}, draw=black] (5,7) rectangle (6,8);
\node at (5.5,7.5) {$c$};
\filldraw[fill={rgb,255:red,253; green,174; blue,97}, draw=black] (6,7) rectangle (7,8);
\node at (6.5,7.5) {$c$};
\filldraw[fill={rgb,255:red,253; green,174; blue,97}, draw=black] (7,7) rectangle (8,8);
\node at (7.5,7.5) {$c$};
\filldraw[fill={rgb,255:red,253; green,174; blue,97}, draw=black] (8,7) rectangle (9,8);
\node at (8.5,7.5) {$c$};
\filldraw[fill={rgb,255:red,253; green,174; blue,97}, draw=black] (9,7) rectangle (10,8);
\node at (9.5,7.5) {$c$};

//4
\filldraw[fill={rgb,255:red,213; green,62; blue,79}, draw=black] (0,6) rectangle (1,7);
\node at (0.5,6.5) {$a$};
\filldraw[fill={rgb,255:red,244; green,109; blue,67}, draw=black] (1,6) rectangle (2,7);
\node at (1.5,6.5) {$b$};
\filldraw[fill={rgb,255:red,253; green,174; blue,97}, draw=black] (2,6) rectangle (3,7);
\node at (2.5,6.5) {$c$};
\filldraw[fill={rgb,255:red,254; green,224; blue,139}, draw=black] (3,6) rectangle (4,7);
\node at (3.5,6.5) {$d$};
\filldraw[fill={rgb,255:red,254; green,224; blue,139}, draw=black] (4,6) rectangle (5,7);
\node at (4.5,6.5) {$d$};
\filldraw[fill={rgb,255:red,254; green,224; blue,139}, draw=black] (5,6) rectangle (6,7);
\node at (5.5,6.5) {$d$};
\filldraw[fill={rgb,255:red,254; green,224; blue,139}, draw=black] (6,6) rectangle (7,7);
\node at (6.5,6.5) {$d$};
\filldraw[fill={rgb,255:red,254; green,224; blue,139}, draw=black] (7,6) rectangle (8,7);
\node at (7.5,6.5) {$d$};
\filldraw[fill={rgb,255:red,254; green,224; blue,139}, draw=black] (8,6) rectangle (9,7);
\node at (8.5,6.5) {$d$};
\filldraw[fill={rgb,255:red,254; green,224; blue,139}, draw=black] (9,6) rectangle (10,7);
\node at (9.5,6.5) {$d$};

// 5
\filldraw[fill={rgb,255:red,213; green,62; blue,79}, draw=black] (0,5) rectangle (1,6);
\node at (0.5,5.5) {$a$};
\filldraw[fill={rgb,255:red,244; green,109; blue,67}, draw=black] (1,5) rectangle (2,6);
\node at (1.5,5.5) {$b$};
\filldraw[fill={rgb,255:red,253; green,174; blue,97}, draw=black] (2,5) rectangle (3,6);
\node at (2.5,5.5) {$c$};
\filldraw[fill={rgb,255:red,254; green,224; blue,139}, draw=black] (3,5) rectangle (4,6);
\node at (3.5,5.5) {$d$};
\filldraw[fill={rgb,255:red,255; green,255; blue,191}, draw=black] (4,5) rectangle (5,6);
\node at (4.5,5.5) {$e$};
\filldraw[fill={rgb,255:red,255; green,255; blue,191}, draw=black] (5,5) rectangle (6,6);
\node at (5.5,5.5) {$e$};
\filldraw[fill={rgb,255:red,255; green,255; blue,191}, draw=black] (6,5) rectangle (7,6);
\node at (6.5,5.5) {$e$};
\filldraw[fill={rgb,255:red,255; green,255; blue,191}, draw=black] (7,5) rectangle (8,6);
\node at (7.5,5.5) {$e$};
\filldraw[fill={rgb,255:red,255; green,255; blue,191}, draw=black] (8,5) rectangle (9,6);
\node at (8.5,5.5) {$e$};
\filldraw[fill={rgb,255:red,255; green,255; blue,191}, draw=black] (9,5) rectangle (10,6);
\node at (9.5,5.5) {$e$};

// 6
\filldraw[fill={rgb,255:red,213; green,62; blue,79}, draw=black] (0,4) rectangle (1,5);
\node at (0.5,4.5) {$a$};
\filldraw[fill={rgb,255:red,244; green,109; blue,67}, draw=black] (1,4) rectangle (2,5);
\node at (1.5,4.5) {$b$};
\filldraw[fill={rgb,255:red,253; green,174; blue,97}, draw=black] (2,4) rectangle (3,5);
\node at (2.5,4.5) {$c$};
\filldraw[fill={rgb,255:red,254; green,224; blue,139}, draw=black] (3,4) rectangle (4,5);
\node at (3.5,4.5) {$d$};
\filldraw[fill={rgb,255:red,255; green,255; blue,191}, draw=black] (4,4) rectangle (5,5);
\node at (4.5,4.5) {$e$};
\filldraw[fill={rgb,255:red,230; green,245; blue,152}, draw=black] (5,4) rectangle (6,5);
\node at (5.5,4.5) {$f$};
\filldraw[fill={rgb,255:red,230; green,245; blue,152}, draw=black] (6,4) rectangle (7,5);
\node at (6.5,4.5) {$f$};
\filldraw[fill={rgb,255:red,230; green,245; blue,152}, draw=black] (7,4) rectangle (8,5);
\node at (7.5,4.5) {$f$};
\filldraw[fill={rgb,255:red,230; green,245; blue,152}, draw=black] (8,4) rectangle (9,5);
\node at (8.5,4.5) {$f$};
\filldraw[fill={rgb,255:red,230; green,245; blue,152}, draw=black] (9,4) rectangle (10,5);
\node at (9.5,4.5) {$f$};

// 7
\filldraw[fill={rgb,255:red,213; green,62; blue,79}, draw=black] (0,3) rectangle (1,4);
\node at (0.5,3.5) {$a$};
\filldraw[fill={rgb,255:red,244; green,109; blue,67}, draw=black] (1,3) rectangle (2,4);
\node at (1.5,3.5) {$b$};
\filldraw[fill={rgb,255:red,253; green,174; blue,97}, draw=black] (2,3) rectangle (3,4);
\node at (2.5,3.5) {$c$};
\filldraw[fill={rgb,255:red,254; green,224; blue,139}, draw=black] (3,3) rectangle (4,4);
\node at (3.5,3.5) {$d$};
\filldraw[fill={rgb,255:red,255; green,255; blue,191}, draw=black] (4,3) rectangle (5,4);
\node at (4.5,3.5) {$e$};
\filldraw[fill={rgb,255:red,230; green,245; blue,152}, draw=black] (5,3) rectangle (6,4);
\node at (5.5,3.5) {$f$};
\filldraw[fill={rgb,255:red,171; green,221; blue,164}, draw=black] (6,3) rectangle (7,4);
\node at (6.5,3.5) {$g$};
\filldraw[fill={rgb,255:red,171; green,221; blue,164}, draw=black] (7,3) rectangle (8,4);
\node at (7.5,3.5) {$g$};
\filldraw[fill={rgb,255:red,171; green,221; blue,164}, draw=black] (8,3) rectangle (9,4);
\node at (8.5,3.5) {$g$};
\filldraw[fill={rgb,255:red,171; green,221; blue,164}, draw=black] (9,3) rectangle (10,4);
\node at (9.5,3.5) {$g$};

// 8
\filldraw[fill={rgb,255:red,213; green,62; blue,79}, draw=black] (0,2) rectangle (1,3);
\node at (0.5,2.5) {$a$};
\filldraw[fill={rgb,255:red,244; green,109; blue,67}, draw=black] (1,2) rectangle (2,3);
\node at (1.5,2.5) {$b$};
\filldraw[fill={rgb,255:red,253; green,174; blue,97}, draw=black] (2,2) rectangle (3,3);
\node at (2.5,2.5) {$c$};
\filldraw[fill={rgb,255:red,254; green,224; blue,139}, draw=black] (3,2) rectangle (4,3);
\node at (3.5,2.5) {$d$};
\filldraw[fill={rgb,255:red,255; green,255; blue,191}, draw=black] (4,2) rectangle (5,3);
\node at (4.5,2.5) {$e$};
\filldraw[fill={rgb,255:red,230; green,245; blue,152}, draw=black] (5,2) rectangle (6,3);
\node at (5.5,2.5) {$f$};
\filldraw[fill={rgb,255:red,171; green,221; blue,164}, draw=black] (6,2) rectangle (7,3);
\node at (6.5,2.5) {$g$};
\filldraw[fill={rgb,255:red,102; green,194; blue,165}, draw=black] (7,2) rectangle (8,3);
\node at (7.5,2.5) {$h$};
\filldraw[fill={rgb,255:red,102; green,194; blue,165}, draw=black] (8,2) rectangle (9,3);
\node at (8.5,2.5) {$h$};
\filldraw[fill={rgb,255:red,102; green,194; blue,165}, draw=black] (9,2) rectangle (10,3);
\node at (9.5,2.5) {$h$};

// 9
\filldraw[fill={rgb,255:red,213; green,62; blue,79}, draw=black] (0,1) rectangle (1,2);
\node at (0.5,1.5) {$a$};
\filldraw[fill={rgb,255:red,244; green,109; blue,67}, draw=black] (1,1) rectangle (2,2);
\node at (1.5,1.5) {$b$};
\filldraw[fill={rgb,255:red,253; green,174; blue,97}, draw=black] (2,1) rectangle (3,2);
\node at (2.5,1.5) {$c$};
\filldraw[fill={rgb,255:red,254; green,224; blue,139}, draw=black] (3,1) rectangle (4,2);
\node at (3.5,1.5) {$d$};
\filldraw[fill={rgb,255:red,255; green,255; blue,191}, draw=black] (4,1) rectangle (5,2);
\node at (4.5,1.5) {$e$};
\filldraw[fill={rgb,255:red,230; green,245; blue,152}, draw=black] (5,1) rectangle (6,2);
\node at (5.5,1.5) {$f$};
\filldraw[fill={rgb,255:red,171; green,221; blue,164}, draw=black] (6,1) rectangle (7,2);
\node at (6.5,1.5) {$g$};
\filldraw[fill={rgb,255:red,102; green,194; blue,165}, draw=black] (7,1) rectangle (8,2);
\node at (7.5,1.5) {$h$};
\filldraw[fill={rgb,255:red,70; green,156; blue,209}, draw=black] (8,1) rectangle (9,2);
\node at (8.5,1.5) {$i$};
\filldraw[fill={rgb,255:red,70; green,156; blue,209}, draw=black] (9,1) rectangle (10,2);
\node at (9.5,1.5) {$i$};

//10
\filldraw[fill={rgb,255:red,213; green,62; blue,79}, draw=black] (0,0) rectangle (1,1);
\node at (0.5,0.5) {$a$};
\filldraw[fill={rgb,255:red,244; green,109; blue,67}, draw=black] (1,0) rectangle (2,1);
\node at (1.5,0.5) {$b$};
\filldraw[fill={rgb,255:red,253; green,174; blue,97}, draw=black] (2,0) rectangle (3,1);
\node at (2.5,0.5) {$c$};
\filldraw[fill={rgb,255:red,254; green,224; blue,139}, draw=black] (3,0) rectangle (4,1);
\node at (3.5,0.5) {$d$};
\filldraw[fill={rgb,255:red,255; green,255; blue,191}, draw=black] (4,0) rectangle (5,1);
\node at (4.5,0.5) {$e$};
\filldraw[fill={rgb,255:red,230; green,245; blue,152}, draw=black] (5,0) rectangle (6,1);
\node at (5.5,0.5) {$f$};
\filldraw[fill={rgb,255:red,171; green,221; blue,164}, draw=black] (6,0) rectangle (7,1);
\node at (6.5,0.5) {$g$};
\filldraw[fill={rgb,255:red,102; green,194; blue,165}, draw=black] (7,0) rectangle (8,1);
\node at (7.5,0.5) {$h$};
\filldraw[fill={rgb,255:red,70; green,156; blue,209}, draw=black] (8,0) rectangle (9,1);
\node at (8.5,0.5) {$i$};
\filldraw[fill={rgb,255:red,35; green,115; blue,160}, draw=black] (9,0) rectangle (10,1);
\node at (9.5,0.5) {$j$};

\end{tikzpicture}
\end{subfigure}%
\begin{subfigure}{.33\textwidth}
\centering
\begin{tikzpicture}[x=5mm, y=5mm]

\draw[step=1cm,gray,very thin] (0,0) grid (10,10);

// 1
\filldraw[fill={rgb,255:red,213; green,62; blue,79}, draw=black] (0,9) rectangle (1,10);
\node at (0.5,9.5) {$a$};
\filldraw[fill={rgb,255:red,255; green,255; blue,255}, draw=black] (1,9) rectangle (2,10);
\node at (1.5,9.5) {${\scriptstyle 0}$};
\filldraw[fill={rgb,255:red,255; green,255; blue,255}, draw=black] (2,9) rectangle (3,10);
\node at (2.5,9.5) {${\scriptstyle 0}$};
\filldraw[fill={rgb,255:red,255; green,255; blue,255}, draw=black] (3,9) rectangle (4,10);
\node at (3.5,9.5) {${\scriptstyle 0}$};
\filldraw[fill={rgb,255:red,255; green,255; blue,255}, draw=black] (4,9) rectangle (5,10);
\node at (4.5,9.5) {${\scriptstyle 0}$};
\filldraw[fill={rgb,255:red,255; green,255; blue,255}, draw=black] (5,9) rectangle (6,10);
\node at (5.5,9.5) {${\scriptstyle 0}$};
\filldraw[fill={rgb,255:red,255; green,255; blue,255}, draw=black] (6,9) rectangle (7,10);
\node at (6.5,9.5) {${\scriptstyle 0}$};
\filldraw[fill={rgb,255:red,255; green,255; blue,255}, draw=black] (7,9) rectangle (8,10);
\node at (7.5,9.5) {${\scriptstyle 0}$};
\filldraw[fill={rgb,255:red,255; green,255; blue,255}, draw=black] (8,9) rectangle (9,10);
\node at (8.5,9.5) {${\scriptstyle 0}$};
\filldraw[fill={rgb,255:red,255; green,255; blue,255}, draw=black] (9,9) rectangle (10,10);
\node at (9.5,9.5) {${\scriptstyle 0}$};

// 2
\filldraw[fill={rgb,255:red,255; green,255; blue,255}, draw=black] (0,8) rectangle (1,9);
\node at (0.5,8.5) {${\scriptstyle 0}$};
\filldraw[fill={rgb,255:red,244; green,109; blue,67}, draw=black] (1,8) rectangle (2,9);
\node at (1.5,8.5) {$b$};
\filldraw[fill={rgb,255:red,255; green,255; blue,255}, draw=black] (2,8) rectangle (3,9);
\node at (2.5,8.5) {${\scriptstyle 0}$};
\filldraw[fill={rgb,255:red,255; green,255; blue,255}, draw=black] (3,8) rectangle (4,9);
\node at (3.5,8.5) {${\scriptstyle 0}$};
\filldraw[fill={rgb,255:red,255; green,255; blue,255}, draw=black] (4,8) rectangle (5,9);
\node at (4.5,8.5) {${\scriptstyle 0}$};
\filldraw[fill={rgb,255:red,255; green,255; blue,255}, draw=black] (5,8) rectangle (6,9);
\node at (5.5,8.5) {${\scriptstyle 0}$};
\filldraw[fill={rgb,255:red,255; green,255; blue,255}, draw=black] (6,8) rectangle (7,9);
\node at (6.5,8.5) {${\scriptstyle 0}$};
\filldraw[fill={rgb,255:red,255; green,255; blue,255}, draw=black] (7,8) rectangle (8,9);
\node at (7.5,8.5) {${\scriptstyle 0}$};
\filldraw[fill={rgb,255:red,255; green,255; blue,255}, draw=black] (8,8) rectangle (9,9);
\node at (8.5,8.5) {${\scriptstyle 0}$};
\filldraw[fill={rgb,255:red,255; green,255; blue,255}, draw=black] (9,8) rectangle (10,9);
\node at (9.5,8.5) {${\scriptstyle 0}$};

// 3
\filldraw[fill={rgb,255:red,255; green,255; blue,255}, draw=black] (0,7) rectangle (1,8);
\node at (0.5,7.5) {${\scriptstyle 0}$};
\filldraw[fill={rgb,255:red,255; green,255; blue,255}, draw=black] (1,7) rectangle (2,8);
\node at (1.5,7.5) {${\scriptstyle 0}$};
\filldraw[fill={rgb,255:red,253; green,174; blue,97}, draw=black] (2,7) rectangle (3,8);
\node at (2.5,7.5) {$c$};
\filldraw[fill={rgb,255:red,255; green,255; blue,255}, draw=black] (3,7) rectangle (4,8);
\node at (3.5,7.5) {${\scriptstyle 0}$};
\filldraw[fill={rgb,255:red,255; green,255; blue,255}, draw=black] (4,7) rectangle (5,8);
\node at (4.5,7.5) {${\scriptstyle 0}$};
\filldraw[fill={rgb,255:red,255; green,255; blue,255}, draw=black] (5,7) rectangle (6,8);
\node at (5.5,7.5) {${\scriptstyle 0}$};
\filldraw[fill={rgb,255:red,255; green,255; blue,255}, draw=black] (6,7) rectangle (7,8);
\node at (6.5,7.5) {${\scriptstyle 0}$};
\filldraw[fill={rgb,255:red,255; green,255; blue,255}, draw=black] (7,7) rectangle (8,8);
\node at (7.5,7.5) {${\scriptstyle 0}$};
\filldraw[fill={rgb,255:red,255; green,255; blue,255}, draw=black] (8,7) rectangle (9,8);
\node at (8.5,7.5) {${\scriptstyle 0}$};
\filldraw[fill={rgb,255:red,255; green,255; blue,255}, draw=black] (9,7) rectangle (10,8);
\node at (9.5,7.5) {${\scriptstyle 0}$};

//4
\filldraw[fill={rgb,255:red,255; green,255; blue,255}, draw=black] (0,6) rectangle (1,7);
\node at (0.5,6.5) {${\scriptstyle 0}$};
\filldraw[fill={rgb,255:red,255; green,255; blue,255}, draw=black] (1,6) rectangle (2,7);
\node at (1.5,6.5) {${\scriptstyle 0}$};
\filldraw[fill={rgb,255:red,255; green,255; blue,255}, draw=black] (2,6) rectangle (3,7);
\node at (2.5,6.5) {${\scriptstyle 0}$};
\filldraw[fill={rgb,255:red,254; green,224; blue,139}, draw=black] (3,6) rectangle (4,7);
\node at (3.5,6.5) {$d$};
\filldraw[fill={rgb,255:red,255; green,255; blue,255}, draw=black] (4,6) rectangle (5,7);
\node at (4.5,6.5) {${\scriptstyle 0}$};
\filldraw[fill={rgb,255:red,255; green,255; blue,255}, draw=black] (5,6) rectangle (6,7);
\node at (5.5,6.5) {${\scriptstyle 0}$};
\filldraw[fill={rgb,255:red,255; green,255; blue,255}, draw=black] (6,6) rectangle (7,7);
\node at (6.5,6.5) {${\scriptstyle 0}$};
\filldraw[fill={rgb,255:red,255; green,255; blue,255}, draw=black] (7,6) rectangle (8,7);
\node at (7.5,6.5) {${\scriptstyle 0}$};
\filldraw[fill={rgb,255:red,255; green,255; blue,255}, draw=black] (8,6) rectangle (9,7);
\node at (8.5,6.5) {${\scriptstyle 0}$};
\filldraw[fill={rgb,255:red,255; green,255; blue,255}, draw=black] (9,6) rectangle (10,7);
\node at (9.5,6.5) {${\scriptstyle 0}$};

// 5
\filldraw[fill={rgb,255:red,255; green,255; blue,255}, draw=black] (0,5) rectangle (1,6);
\node at (0.5,5.5) {${\scriptstyle 0}$};
\filldraw[fill={rgb,255:red,255; green,255; blue,255}, draw=black] (1,5) rectangle (2,6);
\node at (1.5,5.5) {${\scriptstyle 0}$};
\filldraw[fill={rgb,255:red,255; green,255; blue,255}, draw=black] (2,5) rectangle (3,6);
\node at (2.5,5.5) {${\scriptstyle 0}$};
\filldraw[fill={rgb,255:red,255; green,255; blue,255}, draw=black] (3,5) rectangle (4,6);
\node at (3.5,5.5) {${\scriptstyle 0}$};
\filldraw[fill={rgb,255:red,255; green,255; blue,191}, draw=black] (4,5) rectangle (5,6);
\node at (4.5,5.5) {$e$};
\filldraw[fill={rgb,255:red,255; green,255; blue,255}, draw=black] (5,5) rectangle (6,6);
\node at (5.5,5.5) {${\scriptstyle 0}$};
\filldraw[fill={rgb,255:red,255; green,255; blue,255}, draw=black] (6,5) rectangle (7,6);
\node at (6.5,5.5) {${\scriptstyle 0}$};
\filldraw[fill={rgb,255:red,255; green,255; blue,255}, draw=black] (7,5) rectangle (8,6);
\node at (7.5,5.5) {${\scriptstyle 0}$};
\filldraw[fill={rgb,255:red,255; green,255; blue,255}, draw=black] (8,5) rectangle (9,6);
\node at (8.5,5.5) {${\scriptstyle 0}$};
\filldraw[fill={rgb,255:red,255; green,255; blue,255}, draw=black] (9,5) rectangle (10,6);
\node at (9.5,5.5) {${\scriptstyle 0}$};

// 6
\filldraw[fill={rgb,255:red,255; green,255; blue,255}, draw=black] (0,4) rectangle (1,5);
\node at (0.5,4.5) {${\scriptstyle 0}$};
\filldraw[fill={rgb,255:red,255; green,255; blue,255}, draw=black] (1,4) rectangle (2,5);
\node at (1.5,4.5) {${\scriptstyle 0}$};
\filldraw[fill={rgb,255:red,255; green,255; blue,255}, draw=black] (2,4) rectangle (3,5);
\node at (2.5,4.5) {${\scriptstyle 0}$};
\filldraw[fill={rgb,255:red,255; green,255; blue,255}, draw=black] (3,4) rectangle (4,5);
\node at (3.5,4.5) {${\scriptstyle 0}$};
\filldraw[fill={rgb,255:red,255; green,255; blue,255}, draw=black] (4,4) rectangle (5,5);
\node at (4.5,4.5) {${\scriptstyle 0}$};
\filldraw[fill={rgb,255:red,230; green,245; blue,152}, draw=black] (5,4) rectangle (6,5);
\node at (5.5,4.5) {$f$};
\filldraw[fill={rgb,255:red,255; green,255; blue,255}, draw=black] (6,4) rectangle (7,5);
\node at (6.5,4.5) {${\scriptstyle 0}$};
\filldraw[fill={rgb,255:red,255; green,255; blue,255}, draw=black] (7,4) rectangle (8,5);
\node at (7.5,4.5) {${\scriptstyle 0}$};
\filldraw[fill={rgb,255:red,255; green,255; blue,255}, draw=black] (8,4) rectangle (9,5);
\node at (8.5,4.5) {${\scriptstyle 0}$};
\filldraw[fill={rgb,255:red,255; green,255; blue,255}, draw=black] (9,4) rectangle (10,5);
\node at (9.5,4.5) {${\scriptstyle 0}$};

// 7
\filldraw[fill={rgb,255:red,255; green,255; blue,255}, draw=black] (0,3) rectangle (1,4);
\node at (0.5,3.5) {${\scriptstyle 0}$};
\filldraw[fill={rgb,255:red,255; green,255; blue,255}, draw=black] (1,3) rectangle (2,4);
\node at (1.5,3.5) {${\scriptstyle 0}$};
\filldraw[fill={rgb,255:red,255; green,255; blue,255}, draw=black] (2,3) rectangle (3,4);
\node at (2.5,3.5) {${\scriptstyle 0}$};
\filldraw[fill={rgb,255:red,255; green,255; blue,255}, draw=black] (3,3) rectangle (4,4);
\node at (3.5,3.5) {${\scriptstyle 0}$};
\filldraw[fill={rgb,255:red,255; green,255; blue,255}, draw=black] (4,3) rectangle (5,4);
\node at (4.5,3.5) {${\scriptstyle 0}$};
\filldraw[fill={rgb,255:red,255; green,255; blue,255}, draw=black] (5,3) rectangle (6,4);
\node at (5.5,3.5) {${\scriptstyle 0}$};
\filldraw[fill={rgb,255:red,171; green,221; blue,164}, draw=black] (6,3) rectangle (7,4);
\node at (6.5,3.5) {$g$};
\filldraw[fill={rgb,255:red,255; green,255; blue,255}, draw=black] (7,3) rectangle (8,4);
\node at (7.5,3.5) {${\scriptstyle 0}$};
\filldraw[fill={rgb,255:red,255; green,255; blue,255}, draw=black] (8,3) rectangle (9,4);
\node at (8.5,3.5) {${\scriptstyle 0}$};
\filldraw[fill={rgb,255:red,255; green,255; blue,255}, draw=black] (9,3) rectangle (10,4);
\node at (9.5,3.5) {${\scriptstyle 0}$};

// 8
\filldraw[fill={rgb,255:red,255; green,255; blue,255}, draw=black] (0,2) rectangle (1,3);
\node at (0.5,2.5) {${\scriptstyle 0}$};
\filldraw[fill={rgb,255:red,255; green,255; blue,255}, draw=black] (1,2) rectangle (2,3);
\node at (1.5,2.5) {${\scriptstyle 0}$};
\filldraw[fill={rgb,255:red,255; green,255; blue,255}, draw=black] (2,2) rectangle (3,3);
\node at (2.5,2.5) {${\scriptstyle 0}$};
\filldraw[fill={rgb,255:red,255; green,255; blue,255}, draw=black] (3,2) rectangle (4,3);
\node at (3.5,2.5) {${\scriptstyle 0}$};
\filldraw[fill={rgb,255:red,255; green,255; blue,255}, draw=black] (4,2) rectangle (5,3);
\node at (4.5,2.5) {${\scriptstyle 0}$};
\filldraw[fill={rgb,255:red,255; green,255; blue,255}, draw=black] (5,2) rectangle (6,3);
\node at (5.5,2.5) {${\scriptstyle 0}$};
\filldraw[fill={rgb,255:red,255; green,255; blue,255}, draw=black] (6,2) rectangle (7,3);
\node at (6.5,2.5) {${\scriptstyle 0}$};
\filldraw[fill={rgb,255:red,102; green,194; blue,165}, draw=black] (7,2) rectangle (8,3);
\node at (7.5,2.5) {$h$};
\filldraw[fill={rgb,255:red,255; green,255; blue,255}, draw=black] (8,2) rectangle (9,3);
\node at (8.5,2.5) {${\scriptstyle 0}$};
\filldraw[fill={rgb,255:red,255; green,255; blue,255}, draw=black] (9,2) rectangle (10,3);
\node at (9.5,2.5) {${\scriptstyle 0}$};

// 9
\filldraw[fill={rgb,255:red,255; green,255; blue,255}, draw=black] (0,1) rectangle (1,2);
\node at (0.5,1.5) {${\scriptstyle 0}$};
\filldraw[fill={rgb,255:red,255; green,255; blue,255}, draw=black] (1,1) rectangle (2,2);
\node at (1.5,1.5) {${\scriptstyle 0}$};
\filldraw[fill={rgb,255:red,255; green,255; blue,255}, draw=black] (2,1) rectangle (3,2);
\node at (2.5,1.5) {${\scriptstyle 0}$};
\filldraw[fill={rgb,255:red,255; green,255; blue,255}, draw=black] (3,1) rectangle (4,2);
\node at (3.5,1.5) {${\scriptstyle 0}$};
\filldraw[fill={rgb,255:red,255; green,255; blue,255}, draw=black] (4,1) rectangle (5,2);
\node at (4.5,1.5) {${\scriptstyle 0}$};
\filldraw[fill={rgb,255:red,255; green,255; blue,255}, draw=black] (5,1) rectangle (6,2);
\node at (5.5,1.5) {${\scriptstyle 0}$};
\filldraw[fill={rgb,255:red,255; green,255; blue,255}, draw=black] (6,1) rectangle (7,2);
\node at (6.5,1.5) {${\scriptstyle 0}$};
\filldraw[fill={rgb,255:red,255; green,255; blue,255}, draw=black] (7,1) rectangle (8,2);
\node at (7.5,1.5) {${\scriptstyle 0}$};
\filldraw[fill={rgb,255:red,70; green,156; blue,209}, draw=black] (8,1) rectangle (9,2);
\node at (8.5,1.5) {$i$};
\filldraw[fill={rgb,255:red,255; green,255; blue,255}, draw=black] (9,1) rectangle (10,2);
\node at (9.5,1.5) {${\scriptstyle 0}$};

//10
\filldraw[fill={rgb,255:red,255; green,255; blue,255}, draw=black] (0,0) rectangle (1,1);
\node at (0.5,0.5) {${\scriptstyle 0}$};
\filldraw[fill={rgb,255:red,255; green,255; blue,255}, draw=black] (1,0) rectangle (2,1);
\node at (1.5,0.5) {${\scriptstyle 0}$};
\filldraw[fill={rgb,255:red,255; green,255; blue,255}, draw=black] (2,0) rectangle (3,1);
\node at (2.5,0.5) {${\scriptstyle 0}$};
\filldraw[fill={rgb,255:red,255; green,255; blue,255}, draw=black] (3,0) rectangle (4,1);
\node at (3.5,0.5) {${\scriptstyle 0}$};
\filldraw[fill={rgb,255:red,255; green,255; blue,255}, draw=black] (4,0) rectangle (5,1);
\node at (4.5,0.5) {${\scriptstyle 0}$};
\filldraw[fill={rgb,255:red,255; green,255; blue,255}, draw=black] (5,0) rectangle (6,1);
\node at (5.5,0.5) {${\scriptstyle 0}$};
\filldraw[fill={rgb,255:red,255; green,255; blue,255}, draw=black] (6,0) rectangle (7,1);
\node at (6.5,0.5) {${\scriptstyle 0}$};
\filldraw[fill={rgb,255:red,255; green,255; blue,255}, draw=black] (7,0) rectangle (8,1);
\node at (7.5,0.5) {${\scriptstyle 0}$};
\filldraw[fill={rgb,255:red,255; green,255; blue,255}, draw=black] (8,0) rectangle (9,1);
\node at (8.5,0.5) {${\scriptstyle 0}$};
\filldraw[fill={rgb,255:red,35; green,115; blue,160}, draw=black] (9,0) rectangle (10,1);
\node at (9.5,0.5) {$j$};

\end{tikzpicture}
\end{subfigure}%
\begin{subfigure}{.33\textwidth}
\centering
\begin{tikzpicture}[x=5mm, y=5mm]

\draw[step=1cm,gray,very thin] (0,0) grid (10,10);

// 1
\filldraw[fill={rgb,255:red,255; green,255; blue,255}, draw=black] (0,9) rectangle (1,10);
\node at (0.5,9.5) {${\scriptstyle 0}$};
\filldraw[fill={rgb,255:red,213; green,62; blue,79}, draw=black] (1,9) rectangle (2,10);
\node at (1.5,9.5) {$a$};
\filldraw[fill={rgb,255:red,213; green,62; blue,79}, draw=black] (2,9) rectangle (3,10);
\node at (2.5,9.5) {$a$};
\filldraw[fill={rgb,255:red,213; green,62; blue,79}, draw=black] (3,9) rectangle (4,10);
\node at (3.5,9.5) {$a$};
\filldraw[fill={rgb,255:red,213; green,62; blue,79}, draw=black] (4,9) rectangle (5,10);
\node at (4.5,9.5) {$a$};
\filldraw[fill={rgb,255:red,213; green,62; blue,79}, draw=black] (5,9) rectangle (6,10);
\node at (5.5,9.5) {$a$};
\filldraw[fill={rgb,255:red,213; green,62; blue,79}, draw=black] (6,9) rectangle (7,10);
\node at (6.5,9.5) {$a$};
\filldraw[fill={rgb,255:red,213; green,62; blue,79}, draw=black] (7,9) rectangle (8,10);
\node at (7.5,9.5) {$a$};
\filldraw[fill={rgb,255:red,213; green,62; blue,79}, draw=black] (8,9) rectangle (9,10);
\node at (8.5,9.5) {$a$};
\filldraw[fill={rgb,255:red,213; green,62; blue,79}, draw=black] (9,9) rectangle (10,10);
\node at (9.5,9.5) {$a$};

// 2
\filldraw[fill={rgb,255:red,213; green,62; blue,79}, draw=black] (0,8) rectangle (1,9);
\node at (0.5,8.5) {$a$};
\filldraw[fill={rgb,255:red,255; green,255; blue,255}, draw=black] (1,8) rectangle (2,9);
\node at (1.5,8.5) {${\scriptstyle 0}$};
\filldraw[fill={rgb,255:red,244; green,109; blue,67}, draw=black] (2,8) rectangle (3,9);
\node at (2.5,8.5) {$b$};
\filldraw[fill={rgb,255:red,244; green,109; blue,67}, draw=black] (3,8) rectangle (4,9);
\node at (3.5,8.5) {$b$};
\filldraw[fill={rgb,255:red,244; green,109; blue,67}, draw=black] (4,8) rectangle (5,9);
\node at (4.5,8.5) {$b$};
\filldraw[fill={rgb,255:red,244; green,109; blue,67}, draw=black] (5,8) rectangle (6,9);
\node at (5.5,8.5) {$b$};
\filldraw[fill={rgb,255:red,244; green,109; blue,67}, draw=black] (6,8) rectangle (7,9);
\node at (6.5,8.5) {$b$};
\filldraw[fill={rgb,255:red,244; green,109; blue,67}, draw=black] (7,8) rectangle (8,9);
\node at (7.5,8.5) {$b$};
\filldraw[fill={rgb,255:red,244; green,109; blue,67}, draw=black] (8,8) rectangle (9,9);
\node at (8.5,8.5) {$b$};
\filldraw[fill={rgb,255:red,244; green,109; blue,67}, draw=black] (9,8) rectangle (10,9);
\node at (9.5,8.5) {$b$};

// 3
\filldraw[fill={rgb,255:red,213; green,62; blue,79}, draw=black] (0,7) rectangle (1,8);
\node at (0.5,7.5) {$a$};
\filldraw[fill={rgb,255:red,244; green,109; blue,67}, draw=black] (1,7) rectangle (2,8);
\node at (1.5,7.5) {$b$};
\filldraw[fill={rgb,255:red,255; green,255; blue,255}, draw=black] (2,7) rectangle (3,8);
\node at (2.5,7.5) {${\scriptstyle 0}$};
\filldraw[fill={rgb,255:red,253; green,174; blue,97}, draw=black] (3,7) rectangle (4,8);
\node at (3.5,7.5) {$c$};
\filldraw[fill={rgb,255:red,253; green,174; blue,97}, draw=black] (4,7) rectangle (5,8);
\node at (4.5,7.5) {$c$};
\filldraw[fill={rgb,255:red,253; green,174; blue,97}, draw=black] (5,7) rectangle (6,8);
\node at (5.5,7.5) {$c$};
\filldraw[fill={rgb,255:red,253; green,174; blue,97}, draw=black] (6,7) rectangle (7,8);
\node at (6.5,7.5) {$c$};
\filldraw[fill={rgb,255:red,253; green,174; blue,97}, draw=black] (7,7) rectangle (8,8);
\node at (7.5,7.5) {$c$};
\filldraw[fill={rgb,255:red,253; green,174; blue,97}, draw=black] (8,7) rectangle (9,8);
\node at (8.5,7.5) {$c$};
\filldraw[fill={rgb,255:red,253; green,174; blue,97}, draw=black] (9,7) rectangle (10,8);
\node at (9.5,7.5) {$c$};

//4
\filldraw[fill={rgb,255:red,213; green,62; blue,79}, draw=black] (0,6) rectangle (1,7);
\node at (0.5,6.5) {$a$};
\filldraw[fill={rgb,255:red,244; green,109; blue,67}, draw=black] (1,6) rectangle (2,7);
\node at (1.5,6.5) {$b$};
\filldraw[fill={rgb,255:red,253; green,174; blue,97}, draw=black] (2,6) rectangle (3,7);
\node at (2.5,6.5) {$c$};
\filldraw[fill={rgb,255:red,255; green,255; blue,255}, draw=black] (3,6) rectangle (4,7);
\node at (3.5,6.5) {${\scriptstyle 0}$};
\filldraw[fill={rgb,255:red,254; green,224; blue,139}, draw=black] (4,6) rectangle (5,7);
\node at (4.5,6.5) {$d$};
\filldraw[fill={rgb,255:red,254; green,224; blue,139}, draw=black] (5,6) rectangle (6,7);
\node at (5.5,6.5) {$d$};
\filldraw[fill={rgb,255:red,254; green,224; blue,139}, draw=black] (6,6) rectangle (7,7);
\node at (6.5,6.5) {$d$};
\filldraw[fill={rgb,255:red,254; green,224; blue,139}, draw=black] (7,6) rectangle (8,7);
\node at (7.5,6.5) {$d$};
\filldraw[fill={rgb,255:red,254; green,224; blue,139}, draw=black] (8,6) rectangle (9,7);
\node at (8.5,6.5) {$d$};
\filldraw[fill={rgb,255:red,254; green,224; blue,139}, draw=black] (9,6) rectangle (10,7);
\node at (9.5,6.5) {$d$};

// 5
\filldraw[fill={rgb,255:red,213; green,62; blue,79}, draw=black] (0,5) rectangle (1,6);
\node at (0.5,5.5) {$a$};
\filldraw[fill={rgb,255:red,244; green,109; blue,67}, draw=black] (1,5) rectangle (2,6);
\node at (1.5,5.5) {$b$};
\filldraw[fill={rgb,255:red,253; green,174; blue,97}, draw=black] (2,5) rectangle (3,6);
\node at (2.5,5.5) {$c$};
\filldraw[fill={rgb,255:red,254; green,224; blue,139}, draw=black] (3,5) rectangle (4,6);
\node at (3.5,5.5) {$d$};
\filldraw[fill={rgb,255:red,255; green,255; blue,255}, draw=black] (4,5) rectangle (5,6);
\node at (4.5,5.5) {${\scriptstyle 0}$};
\filldraw[fill={rgb,255:red,255; green,255; blue,191}, draw=black] (5,5) rectangle (6,6);
\node at (5.5,5.5) {$e$};
\filldraw[fill={rgb,255:red,255; green,255; blue,191}, draw=black] (6,5) rectangle (7,6);
\node at (6.5,5.5) {$e$};
\filldraw[fill={rgb,255:red,255; green,255; blue,191}, draw=black] (7,5) rectangle (8,6);
\node at (7.5,5.5) {$e$};
\filldraw[fill={rgb,255:red,255; green,255; blue,191}, draw=black] (8,5) rectangle (9,6);
\node at (8.5,5.5) {$e$};
\filldraw[fill={rgb,255:red,255; green,255; blue,191}, draw=black] (9,5) rectangle (10,6);
\node at (9.5,5.5) {$e$};

// 6
\filldraw[fill={rgb,255:red,213; green,62; blue,79}, draw=black] (0,4) rectangle (1,5);
\node at (0.5,4.5) {$a$};
\filldraw[fill={rgb,255:red,244; green,109; blue,67}, draw=black] (1,4) rectangle (2,5);
\node at (1.5,4.5) {$b$};
\filldraw[fill={rgb,255:red,253; green,174; blue,97}, draw=black] (2,4) rectangle (3,5);
\node at (2.5,4.5) {$c$};
\filldraw[fill={rgb,255:red,254; green,224; blue,139}, draw=black] (3,4) rectangle (4,5);
\node at (3.5,4.5) {$d$};
\filldraw[fill={rgb,255:red,255; green,255; blue,191}, draw=black] (4,4) rectangle (5,5);
\node at (4.5,4.5) {$e$};
\filldraw[fill={rgb,255:red,255; green,255; blue,255}, draw=black] (5,4) rectangle (6,5);
\node at (5.5,4.5) {${\scriptstyle 0}$};
\filldraw[fill={rgb,255:red,230; green,245; blue,152}, draw=black] (6,4) rectangle (7,5);
\node at (6.5,4.5) {$f$};
\filldraw[fill={rgb,255:red,230; green,245; blue,152}, draw=black] (7,4) rectangle (8,5);
\node at (7.5,4.5) {$f$};
\filldraw[fill={rgb,255:red,230; green,245; blue,152}, draw=black] (8,4) rectangle (9,5);
\node at (8.5,4.5) {$f$};
\filldraw[fill={rgb,255:red,230; green,245; blue,152}, draw=black] (9,4) rectangle (10,5);
\node at (9.5,4.5) {$f$};

// 7
\filldraw[fill={rgb,255:red,213; green,62; blue,79}, draw=black] (0,3) rectangle (1,4);
\node at (0.5,3.5) {$a$};
\filldraw[fill={rgb,255:red,244; green,109; blue,67}, draw=black] (1,3) rectangle (2,4);
\node at (1.5,3.5) {$b$};
\filldraw[fill={rgb,255:red,253; green,174; blue,97}, draw=black] (2,3) rectangle (3,4);
\node at (2.5,3.5) {$c$};
\filldraw[fill={rgb,255:red,254; green,224; blue,139}, draw=black] (3,3) rectangle (4,4);
\node at (3.5,3.5) {$d$};
\filldraw[fill={rgb,255:red,255; green,255; blue,191}, draw=black] (4,3) rectangle (5,4);
\node at (4.5,3.5) {$e$};
\filldraw[fill={rgb,255:red,230; green,245; blue,152}, draw=black] (5,3) rectangle (6,4);
\node at (5.5,3.5) {$f$};
\filldraw[fill={rgb,255:red,255; green,255; blue,255}, draw=black] (6,3) rectangle (7,4);
\node at (6.5,3.5) {${\scriptstyle 0}$};
\filldraw[fill={rgb,255:red,171; green,221; blue,164}, draw=black] (7,3) rectangle (8,4);
\node at (7.5,3.5) {$g$};
\filldraw[fill={rgb,255:red,171; green,221; blue,164}, draw=black] (8,3) rectangle (9,4);
\node at (8.5,3.5) {$g$};
\filldraw[fill={rgb,255:red,171; green,221; blue,164}, draw=black] (9,3) rectangle (10,4);
\node at (9.5,3.5) {$g$};

// 8
\filldraw[fill={rgb,255:red,213; green,62; blue,79}, draw=black] (0,2) rectangle (1,3);
\node at (0.5,2.5) {$a$};
\filldraw[fill={rgb,255:red,244; green,109; blue,67}, draw=black] (1,2) rectangle (2,3);
\node at (1.5,2.5) {$b$};
\filldraw[fill={rgb,255:red,253; green,174; blue,97}, draw=black] (2,2) rectangle (3,3);
\node at (2.5,2.5) {$c$};
\filldraw[fill={rgb,255:red,254; green,224; blue,139}, draw=black] (3,2) rectangle (4,3);
\node at (3.5,2.5) {$d$};
\filldraw[fill={rgb,255:red,255; green,255; blue,191}, draw=black] (4,2) rectangle (5,3);
\node at (4.5,2.5) {$e$};
\filldraw[fill={rgb,255:red,230; green,245; blue,152}, draw=black] (5,2) rectangle (6,3);
\node at (5.5,2.5) {$f$};
\filldraw[fill={rgb,255:red,171; green,221; blue,164}, draw=black] (6,2) rectangle (7,3);
\node at (6.5,2.5) {$g$};
\filldraw[fill={rgb,255:red,255; green,255; blue,255}, draw=black] (7,2) rectangle (8,3);
\node at (7.5,2.5) {${\scriptstyle 0}$};
\filldraw[fill={rgb,255:red,102; green,194; blue,165}, draw=black] (8,2) rectangle (9,3);
\node at (8.5,2.5) {$h$};
\filldraw[fill={rgb,255:red,102; green,194; blue,165}, draw=black] (9,2) rectangle (10,3);
\node at (9.5,2.5) {$h$};

// 9
\filldraw[fill={rgb,255:red,213; green,62; blue,79}, draw=black] (0,1) rectangle (1,2);
\node at (0.5,1.5) {$a$};
\filldraw[fill={rgb,255:red,244; green,109; blue,67}, draw=black] (1,1) rectangle (2,2);
\node at (1.5,1.5) {$b$};
\filldraw[fill={rgb,255:red,253; green,174; blue,97}, draw=black] (2,1) rectangle (3,2);
\node at (2.5,1.5) {$c$};
\filldraw[fill={rgb,255:red,254; green,224; blue,139}, draw=black] (3,1) rectangle (4,2);
\node at (3.5,1.5) {$d$};
\filldraw[fill={rgb,255:red,255; green,255; blue,191}, draw=black] (4,1) rectangle (5,2);
\node at (4.5,1.5) {$e$};
\filldraw[fill={rgb,255:red,230; green,245; blue,152}, draw=black] (5,1) rectangle (6,2);
\node at (5.5,1.5) {$f$};
\filldraw[fill={rgb,255:red,171; green,221; blue,164}, draw=black] (6,1) rectangle (7,2);
\node at (6.5,1.5) {$g$};
\filldraw[fill={rgb,255:red,102; green,194; blue,165}, draw=black] (7,1) rectangle (8,2);
\node at (7.5,1.5) {$h$};
\filldraw[fill={rgb,255:red,255; green,255; blue,255}, draw=black] (8,1) rectangle (9,2);
\node at (8.5,1.5) {${\scriptstyle 0}$};
\filldraw[fill={rgb,255:red,70; green,156; blue,209}, draw=black] (9,1) rectangle (10,2);
\node at (9.5,1.5) {$i$};

//10
\filldraw[fill={rgb,255:red,213; green,62; blue,79}, draw=black] (0,0) rectangle (1,1);
\node at (0.5,0.5) {$a$};
\filldraw[fill={rgb,255:red,244; green,109; blue,67}, draw=black] (1,0) rectangle (2,1);
\node at (1.5,0.5) {$b$};
\filldraw[fill={rgb,255:red,253; green,174; blue,97}, draw=black] (2,0) rectangle (3,1);
\node at (2.5,0.5) {$c$};
\filldraw[fill={rgb,255:red,254; green,224; blue,139}, draw=black] (3,0) rectangle (4,1);
\node at (3.5,0.5) {$d$};
\filldraw[fill={rgb,255:red,255; green,255; blue,191}, draw=black] (4,0) rectangle (5,1);
\node at (4.5,0.5) {$e$};
\filldraw[fill={rgb,255:red,230; green,245; blue,152}, draw=black] (5,0) rectangle (6,1);
\node at (5.5,0.5) {$f$};
\filldraw[fill={rgb,255:red,171; green,221; blue,164}, draw=black] (6,0) rectangle (7,1);
\node at (6.5,0.5) {$g$};
\filldraw[fill={rgb,255:red,102; green,194; blue,165}, draw=black] (7,0) rectangle (8,1);
\node at (7.5,0.5) {$h$};
\filldraw[fill={rgb,255:red,70; green,156; blue,209}, draw=black] (8,0) rectangle (9,1);
\node at (8.5,0.5) {$i$};
\filldraw[fill={rgb,255:red,255; green,255; blue,255}, draw=black] (9,0) rectangle (10,1);
\node at (9.5,0.5) {${\scriptstyle 0}$};

\end{tikzpicture}
\end{subfigure}%
\caption{The pattern of element values for $\mathbf{M}$ in relation to the inhomogeneous deprivation-structured mixing matrix given by Equation \ref{eq:inhomMixingCD}.
For each $\mathbf{M}$, deprivation reduces monotonically from left to right (and top to bottom) and also element values monotonically increase from $a$ up to $j$.
Left: $\mathbf{M} = \mathbf{U} + \mathbf{U}^\mathrm{T} - \mathrm{diag}(\mathbf{U})$ as per Equation \ref{eq:inhomMixingM}.
Middle: `assortative' $\mathbf{M} = \mathrm{diag}(\mathbf{U})$.
Right: `disassortative' $\mathbf{M} = \mathbf{U} + \mathbf{U}^\mathrm{T} - 2 \, \mathrm{diag}(\mathbf{U})$.}
\label{fig:deprivationmixingmatrix}

\end{figure}

In terms of deprivation the fitted model assumes homogeneous social mixing.
The interest here is the effect on case incidence of an increase in mixing which is inhomogeneous across the deprivation groups: specifically the more advantaged an individual is the greater their mixing.
In the absence of empirical data the deprivation-structured mixing matrix is defined with a parsimonious structure by using an upper triangular matrix $\mathbf{U}$ as follows
\begin{subequations}
    \label{eq:inhomMixing}
    \begin{align}
        \bm{\mathcal{C}}^D(t) &= \mathbf{1}_{J \times J} + \mathbf{M} \,  \varpi(t) \label{eq:inhomMixingCD} \\
        \mathbf{M} &= \mathbf{U} + \mathbf{U}^\mathrm{T} - \mathrm{diag}(\mathbf{U}) \label{eq:inhomMixingM}
    \end{align}
\end{subequations}
with function $\varpi(t)$ and where the elements of $\mathbf{U}$ are constrained according to $u_{i,i} \geq 0$, $u_{i,j}=u_{i,i}$ and $u_{i+1,i+1}>u_{i,i}$ for all $i$.
For convenience the form of the mixing matrix $\mathbf{M}$ is depicted in the left panel of Figure \ref{fig:deprivationmixingmatrix}.
Let $i=1$ refer to the most deprived group and let successive indexes denote successively less deprived groups, as such the least deprived groups have the highest rates of mixing.
A caveat to this approach is that the age-structured mixing matrix remains unchanged therefore some caution is required when interpreting results as the age structure of simulated cases is not expected to reflect the age structure of the observed case data.

The model is fitted to the training data with $\bm{\mathcal{C}}^D = \mathbf{1}_{J \times J}$.
To assess the effect of increased social mixing a forward simulation is run with the deprivation mixing matrix defined using Equation \ref{eq:inhomMixing}. 
As an exemplar, a forward simulation is set up as follows.
First, to represent all educational settings commencing during September to early October let $\varpi(t)$ have a linear increase over time, to this end $\varpi(t) = \omega \, (t-t')$ where $\omega = 0.00085$ for $(t-t')>0$ otherwise $\omega = 0$.
Secondly, let $t'=10$ days account for both the time lag between increased mixing resulting in increased case incidence and the time lag between the start of the simulation (29 August 2021) and the start of the school term (1 September 2021).
Thirdly, as an example let $u_{1,1}=6, u_{2,2}=8, u_{3,3}=11, u_{4,4}=16, u_{5,5}=23, u_{6,6}=35, u_{7,7}=41, u_{8,8}=55, u_{9,9}=75, u_{10,10}=141$, 
where the structure of the mixing matrix is shown in the left panel of Figure \ref{fig:deprivationmixingmatrix}.
Finally, simulate forward $8$ weeks from the end of the training data, as such $t=0,1...,55$.
With this set-up forward simulations show that the case incidence exhibits total deprivation-switching after the 12 October 2021: see solid lines in right panel of Figure \ref{fig:plot_simulated_timeseries_incidence_100k_social_mixing_increase} where the order of IMD deciles has switched between the beginning of the simulation and mid-October 2021.
Furthermore, this simulation is consistent with the dynamics observed in the UKHSA COVID-19 case data, that is the epidemic is growing.
This can be seen by comparing the expected (solid lines) and observed (dotted lines) case incidence in right panel of Figure \ref{fig:plot_simulated_timeseries_incidence_100k_social_mixing_increase} and also Figure S7 where no smoothing is applied.
With this configuration on 13 October 2021 decile one (most deprived) has an increase in social mixing of $1.2$ times and decile ten $5.2$ times.

In the absence of additional data there are too many degrees of freedom to uniquely identify the structure the deprivation-structured mixing matrix.
To emphasize this two further forward simulation experiments are considered where the form of $\mathbf{M}$ differs from that given in Equation \ref{eq:inhomMixingM}.
First, an increase in `assortative' mixing with $\mathbf{M} = \mathrm{diag}(\mathbf{U})$ as depicted in middle panel of Figure \ref{fig:deprivationmixingmatrix}: for example let  $u_{1,1}=65, u_{2,2}=85, u_{3,3}=115, u_{4,4}=150, u_{5,5}=190, u_{6,6}=250, u_{7,7}=270, u_{8,8}=300, u_{9,9}=330, u_{10,10}=370$.
Secondly, an increase in `disassortative' mixing where $\mathbf{M} =  \mathbf{U} + \mathbf{U}^\mathrm{T} - 2 \, \mathrm{diag}(\mathbf{U})$ as depicted in right panel of Figure \ref{fig:deprivationmixingmatrix}: for example let $u_{1,1}=6, u_{2,2}=8, u_{3,3}=11, u_{4,4}=17, u_{5,5}=26, u_{6,6}=40, u_{7,7}=53, u_{8,8}=73, u_{9,9}=169, u_{10,10}=0$.
If in both experiments $\mathbf{U}$ is constructed as described under Equation \ref{eq:inhomMixing}, $\omega = 0.00085$ and $t'=10$ then the forward simulations exhibit deprivation-switching very similar to that shown in Figure \ref{fig:plot_simulated_timeseries_incidence_100k_social_mixing_increase}: conformation of this result may be seen in the Supplementary Material by comparing Figure S7 with Figures S8 and S9.

\textit{\textbf{In summary}}, an increase in social mixing is a plausible explanation for the observed deprivation-switching, with the caveat that only variations in deprivation mixing were considered.  Without additional data the form of the deprivation-structured mixing matrix cannot be uniquely identified as there are too many degrees of freedom.

\subsubsection{Differential changes in behavioural patterns across deprivation deciles}\label{sec:sim_testing_behaviour}

For the current purposes population-level behavioural patterns are interpreted as behaviours that lead to a positive case being reported/unreported as well as those which increase/decrease the risk of disease such as mask wearing, vaccination status, etc.

In this section differential changes in behaviour across the deprivation groups are investigated, via the behavioural adaptation vector $\bm{\chi}$ whose components are given in Equation \ref{eq:chis}.
Each element of $\bm{\tilde{\rho}}$ describes, for a given age stratum, the relative effect across deprivation strata.
The objective is to simulate a change in population-level behaviour, e.g. testing behaviour, in which the higher deprivation strata progressively dominant the case incidence over time.
To this end the model is first fitted to the training data after which a forward simulation is run with a modified $\bm{\chi}=[\bm{\varkappa}_1(t), \bm{\varkappa}_2(t) , ..., \bm{\varkappa}_K(t)]^\mathrm{T}$, whose elements are defined as
\begin{equation}
\bm{\varkappa}_k(t)  = \left (   \phi \left(1  + \tilde{\psi}_k \right)  +  \eta \left( \frac{1}{2} + \tilde{\varrho}_k(t)  f( \mathbf{\tilde{d}}) \right)  \right) \zeta
\end{equation}

where $\tilde{\varrho}_k(t) = \tilde{\rho}_k-\varepsilon \, t$ is clamped to $[-1/2,1/2]$,
$t = 0, 1, 2, ...$ with units of days,
and $\tilde{\rho}_k$ are given by the posterior samples.
As defined previously: $\eta = 2$; $\phi = 2$; $\xi = 0.3$; and $f(\mathbf{\tilde{d}}) = \tanh(-\xi \mathbf{\tilde{d}})$.  The model is fitted to the training data (07/06/2021 - 29/08/2021), as such the expected values of $\bm{\tilde{\rho}}$ are given at the vertical dashed line in Figure \ref{fig:plot_rho_from_multiple_MCMC_runs}.

\begin{figure}[!ht]
  \begin{subfigure}[]{.95\linewidth}
    \centering
    \subcaption{Forecast over 4-week period with depletion of susceptible individuals as described in Section \ref{sec:sim_depletion}.
    For an 8-week forecast CRPS quartiles over all $\mathrm{cprs}_{i,t}$ are $(Q_1,Q_2,Q_3) = (102.4, 198.0, 346.4)$.
    Left: expected daily incidence aggregated by age group. Right: solid lines including insets, expected daily incidence aggregated by IMD decile; and dotted lines, observed case incidence; IMD 1 refers to most deprived.}
    \includegraphics[width=11cm]{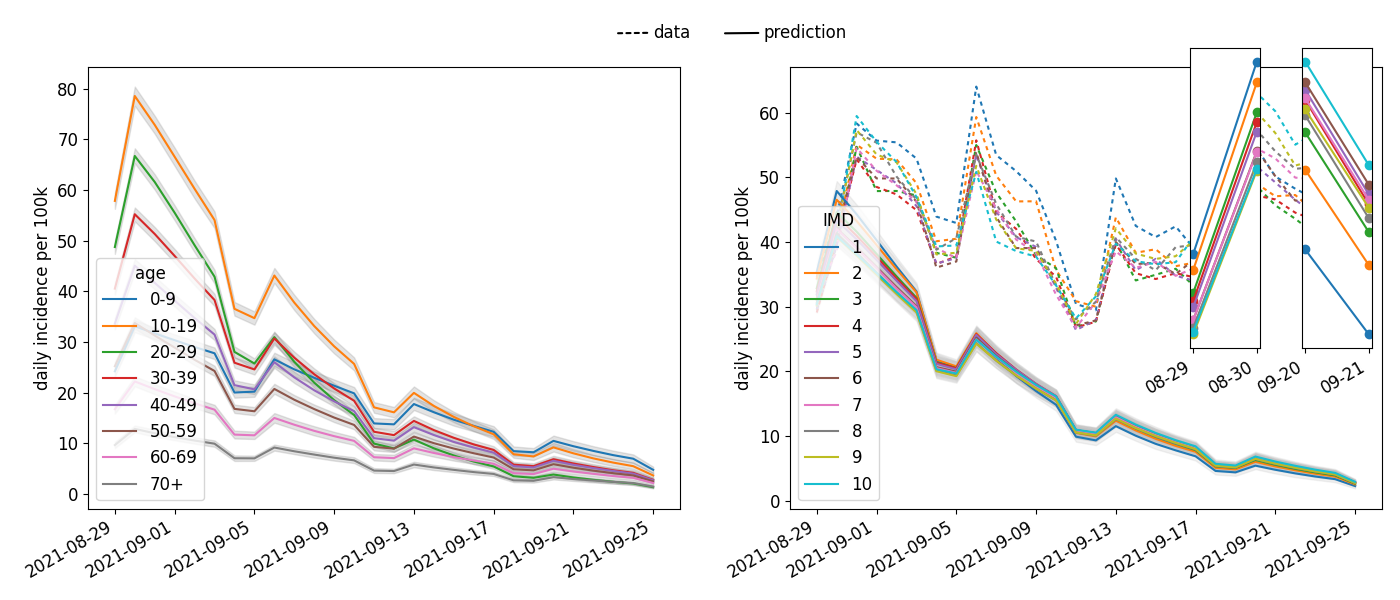}
    \label{fig:plot_simulated_timeseries_incidence_100k_depletion}
  \end{subfigure}
  \begin{subfigure}[]{.95\linewidth}
    \centering
    \subcaption{Forecast over 8-week period with increased social mixing as described in Section \ref{sec:sim_social_mixing}.
    CRPS quartiles over all $\mathrm{cprs}_{i,t}$ are $(Q_1,Q_2,Q_3) = (29.4, 61.9, 148.0)$.
    Left: expected daily incidence aggregated by age group. Right: solid lines, expected daily incidence aggregated by IMD and smoothed with 14-day moving average window; and dotted lines, observed case incidence.}
    \includegraphics[width=11cm]{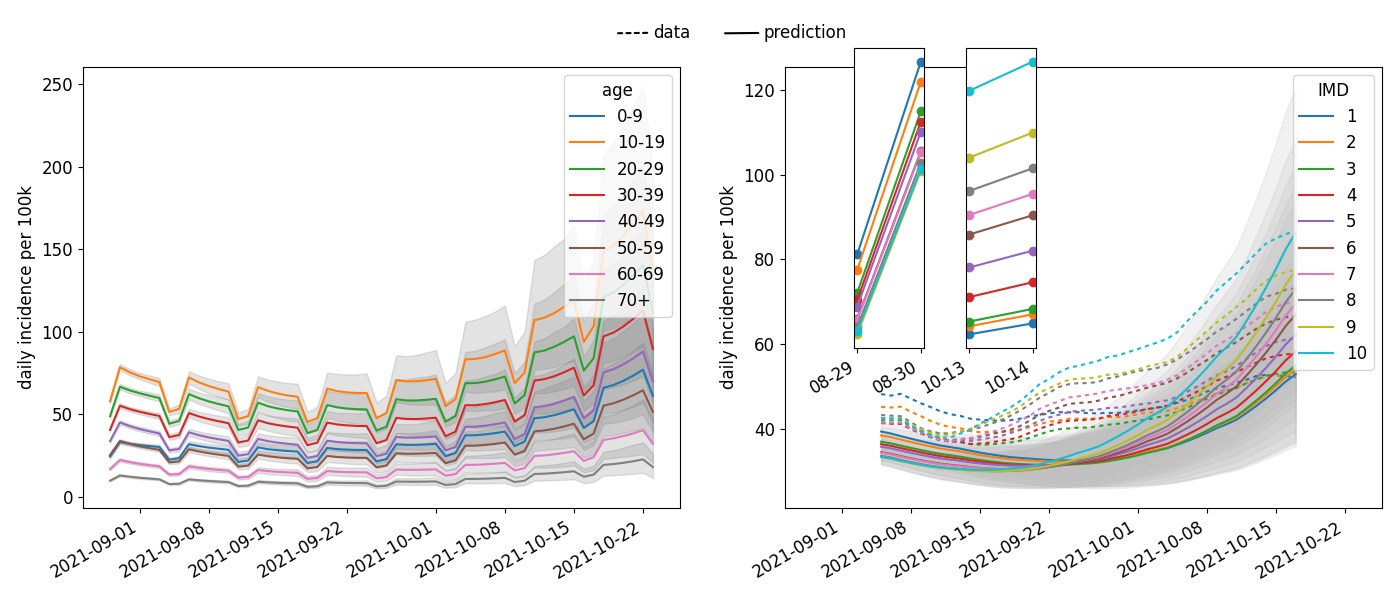}
    \label{fig:plot_simulated_timeseries_incidence_100k_social_mixing_increase}
  \end{subfigure}
  \begin{subfigure}[]{.95\linewidth}
    \centering
    \subcaption{Forecast over 8-week period with perturbation to behavioural adaptation term and increased hazard rate as described in Section \ref{sec:sim_testing_behaviour}: specifically $(\zeta, \varepsilon)=(1.265, 0.017)$.
    CRPS quartiles over all $\mathrm{cprs}_{i,t}$ are $(Q_1,Q_2,Q_3) = (16.9, 42.8, 156.4)$.
    Left: expected daily incidence aggregated by age group. Right: solid lines, expected daily incidence aggregated by IMD and smoothed with 14-day moving average window; and dotted lines, observed case incidence.}
    \includegraphics[width=11cm]{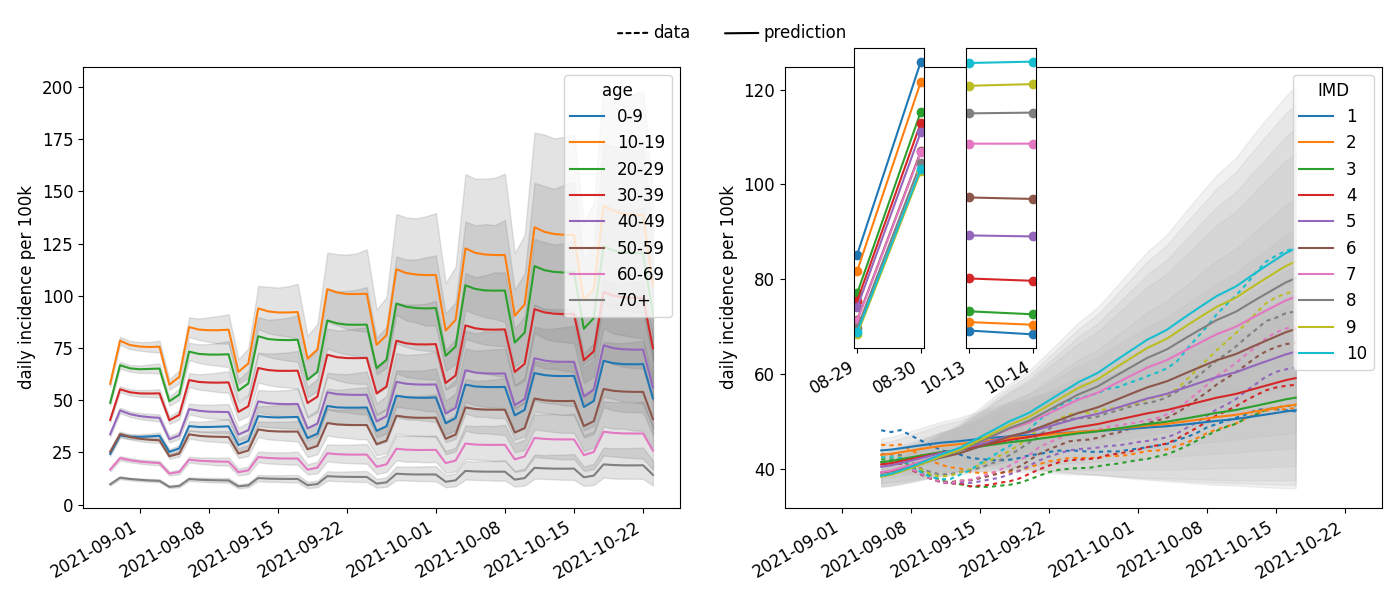}
    \label{fig:plot_simulated_timeseries_incidence_100k_behavioural_chi_with_boost}
  \end{subfigure}  
  \caption{Forecasts of expected daily case incidence (solid lines) with $90\%$ CI, for age groups and deprivation (IMD) deciles, depicting deprivation-switching for model fitted to training data (07/06/2021 - 29/08/2021).}
  \label{fig:simulations}
\end{figure}

If an 8-week forward simulation experiment is run which commences at the end of the training data period and $(\zeta, \varepsilon)=(1.0, 0.017)$ then deprivation-switching occurs but the simulation does not exhibit the correct dynamics in that the simulated epidemic is shrinking, similarly to Figure \ref{fig:fitted_data_ts_incidence_per_100k}, which is contrary to the observations: for completeness the results of this simulation are given in Figure S10.
By construction variations in the values of $\bm{\tilde{\varrho}}(t)=[\tilde{\varrho}_1(t), \tilde{\varrho}_2(t), ..., \tilde{\varrho}_K(t)]$ do not increase the overall hazard rate across all strata (area under $\bm{\chi}$ remains constant), therefore to simulate an epidemic which is growing it is necessary to set $\zeta>1.0$.
To simulate an epidemic where the dynamics are similar to the observed data in so much as the epidemic is  both growing and exhibits deprivation-switching let for example $(\zeta, \varepsilon)=(1.265, 0.017)$.
These results are given in Figure \ref{fig:plot_simulated_timeseries_incidence_100k_behavioural_chi_with_boost}, and also in Figure S11 where the expected values are not smoothed.

A note of caution. This simulation, shown in Figure \ref{fig:plot_simulated_timeseries_incidence_100k_behavioural_chi_with_boost}, has a lower CRPS than the differential social mixing simulation, shown in Figure \ref{fig:plot_simulated_timeseries_incidence_100k_social_mixing_increase}, however it would be a mistake to conclude that in reality it better reflects the unobserved underlying process in the case incidence.
It needs to be appreciated that additional data unavailable to this study is required to disentangle these differential changes in population-level behaviour.

\textit{\textbf{In summary}}, differential changes in behaviour across deprivation groups is a plausible explanation for the observed deprivation-switching.
In the absence of a new more infectious variant and assuming social mixing remains constant over all time, then these simulations would suggest that population-level behavioural changes caused, on average, an increase of approximately $25\%$ in the hazard rate during September and October of 2021.
However without additional data sources the nature and degree of any behavioural changes cannot be identified.

\clearpage
\newpage

\clearpage
\newpage

\section{Discussion}\label{sec:discussion}

We have introduced an infectious disease model whose novelty is that it directly accounts for socioeconomic determinants in terms of deprivation.
This stochastic compartmental model, where population is stratified into deprivation-age groups, was fitted to the 2021 UKHSA English COVID-19 community testing positive case data using a fully Bayesian approach, as such the model learns as the epidemic evolves.
Using the posterior samples from the fitted model a number of metrics were estimated per stratum including reproduction number and case incidence forecasts, from which we can infer the strata driving the epidemic.
Additionally using forward simulation experiments we infer that during the last quarter of 2021 the observed differential case incidence with respect to deprivation cannot be explained by the depletion of susceptible individuals, but could in principle be explained by large-scale differential changes in social mixing patterns and/or population behaviour.

In England, prior to July 2021 the most disadvantaged in society were reporting the most cases of COVID-19 thereby depleting the pool these susceptible individuals.
The more disadvantaged individuals were more likely to have a greater number of contacts per day arising from factors including: overcrowded housing; usage of public transport or car sharing; and lower paid jobs often being public-facing in sectors such as healthcare, retail, transport, manufacturing/factory.
Conversely more privileged sectors in society were more likely to have the flexibility to minimise close contacts due to factors including: less cramped living conditions; personal transport; and ability to work from home \cite{Beale2022, EMG2021}.
During July 2021 all government restrictions were lifted therefore white-collar workers were encouraged to return to their offices and at the same time the hospitality and leisure sectors re-opened.
Consequently, the daily number of contacts increased and perhaps more so among the more privileged in society.
Simultaneously financial support for those in the lower paid jobs, some returning from being furloughed, was not necessarily sufficient to allow them to take sick leave from work therefore it follows that they may be less inclined to test or report a positive LFT.
Consequently it is unsurprising that the UKHSA English COVID-19 data when summarised by deprivation deciles reflects these underlying factors: the most disadvantaged recorded the highest case counts pre-July 2021 switching to the highest case counts from the most privileged post-September 2021.
As stated previously this phenomena is referred to here as \textit{deprivation-switching} but it was unclear which factors at a population level could be responsible.

From an infectious disease modelling perspective there are broadly three mechanisms, not mutually exclusive, by which deprivation-switching could occur: differential depletion of susceptible individuals; large-scale differential changes in social mixing patterns; and large-scale differential changes in population behaviour.
Given posterior samples from fitted model forward simulations did not exhibit deprivation-switching (Figure \ref{fig:fitted_data_ts_incidence_per_100k}) and moreover they predicted the epidemic was shrinking contrary to the observed data.
To investigate this anomaly three forward simulation experiments were constructed.
Posterior samples were computed using the original model fitted to the training data (07/06/2021 - 29/08/2021), then these samples were used in forward simulations where the model was perturbed as follows:
\begin{itemize}
    \item First, forward simulations exhibited deprivation-switching if the depletion of susceptible individuals was five times higher than that recorded in the observed case data, however this results in a shrinking epidemic which is contrary to the observations.
    It is unlikely that the observed case incidence was missing such a large quantity of data but we acknowledge to some extent there may have been an element of depletion in deprived small-scale networks which would not be detected by our model.

    \item Secondly, forward simulations showed if derivation-structured social mixing increases by up to around five times in the most privileged groups (highest deciles) then deprivation-switching occurs with similar dynamics to the observations.
    This would appear not to be too unreasonable in light of the 2011 study by Eames et. al. \cite{Eames2011} of school children mixing patterns, they reported that during the holidays the daily number of recorded encounters approximately halved and that the number of close contact encounters dropped by about one third.
    Furthermore it is known that social mixing significantly increased during the summer and autumn of 2021 as government restrictions were lifted \cite{Smith2022}.
    
    \item Thirdly, forward simulations demonstrated that if there are relative changes in behaviour across deprivation deciles then deprivation-switching occurs.
    Furthermore for the case in question provided the susceptible to exposed hazard rate is increased by about $25\%$ the epidemic grows with similar dynamics to the observations, indicating that there was an underlying process driving up case incidence which was not directly accounted for by the original fitted model.

\end{itemize}

Finally, it is important to appreciate that each of these simulation experiments were conducted separately and that in reality all three underlying processes discussed may have contributed to some degree towards both deprivation-switching and rising case incidence during October and November 2021.
In addition due to the number of degrees of freedom in the model no attempt was made to simultaneously model both deprivation and age-related changes, in this regard the forward simulation experiments assumed that changes in deprivation were independent of age.
To better understand the observed dynamics probed by these three simulation experiments and reduce the number of degrees of freedom in the model additional data not available to this study is required, specifically population-level behavioural data including age-deprivation-structured social mixing data that preferably accounts for temporal factors.
It therefore remains for future work to quantify, with data sources beyond the scope of this work, the degree to which each of these factors were responsible for observed differential dynamics among socioeconomic groups.


Beyond the need for additional data required to determine the unobserved population level social mixing patterns and behavioural patterns there are a number of other general weaknesses relating to this work.
Slowly varying temporal dynamics, such as changes in immunity due to the vaccination programme, are implicitly accounted for by fitting the model to a  12-week window of training data however rapidly changing dynamics are not directly captured (e.g. through time-varying explanatory variables) although they could be detected using exceedance probabilities based on the posterior samples.
The UKHSA English COVID-19 positive case data relied upon voluntary testing as such there will be missing data which is not accounted for in this work, and furthermore the approach outlined is sensitive to fluctuations in case testing rates.
As a result some caution is advised when interpreting results especially in terms of policy decisions.



In the future it would be of interest to consider local spatial effects by implementing a three-way stratification: deprivation, age and space.
This would have the potential to support more nuanced policy decisions at a local level and also allow for the detection of any local depletion of susceptible individuals.
If using a compartmental model the main barrier would be the computational burden of such a high dimensional computation space, hence for a three-way stratification there is a requirement for either more powerful computational approaches or novel mathematical models.
Standard compartmental models, such as our model, assume the population mixes at random (each individual has a small chance of encountering any other individual) however this assumption may be too strong \cite{Keeling20005} given human interactions are network-based, e.g. small-world.
In the context of including additional stratification, such as space, it would therefore be of future interest to consider more complex network-structures where mixing matrices which are not full as this would lighten the computational burden.
There is a balance between the case incidence size and the number of factors a model can realistically support however it would be of interest to consider explanatory variables including ethnicity and also other factors which may relate to the disease transmission, for instance a measure of time spent in communal/public setting e.g. public transport, public facing workplace, and so on.

At the beginning of the COVID-19 pandemic, it proved to be impossible to fit the underlying process observed in the UKHSA data to our prototype models using standard software tools.
This necessitated extensive development of more robust methodology and algorithms,
as detailed in \cite{gemlib} and \cite{imdage}.
In future epidemics with our publicly available software tools \cite{gemlib} and models \cite{imdage} it should be possible to fit our model, or similar models, and thereby more rapidly support policy-makers in determining which groups are at the highest risk or being most adversely affected.

\textbf{In summary}, this work reaches beyond the traditional infectious disease modelling boundaries of spatiotemporal contagion and mortality dynamics to explicitly include socioeconomic characteristics of the underlying population.
With the population stratified by deprivation and age our model has been shown to give insights into the complex dynamics between strata as the epidemic unfolds:  different stratum dominate during different epochs.
In particular observations based on the UKHSA COVID-19 data show that prior to the summer of 2021 the most deprived groups reported the highest case incidence but this reverses during the autumn of 2021.
Evidence presented from the forward simulation experiments suggest that these observed changes in incidence cannot be accounted for by the depletion of susceptible individuals, instead they are likely to be due to substantial shifts in population level behaviour which differ across the deprivation groups such as: differential increases in social mixing; and differential population-level characteristics which may include motivation/ability to test/report a positive test as well as take precautions to mitigate disease risk e.g. mask-wearing, vaccination, etc.
These factors cannot be directly observed in the UKHSA COVID-19 case data therefore it is important to recognise that additional population-level social mixing and behavioural data is required for a deeper understanding beyond the scope of this paper.
If this model along with its corresponding software and sufficient LFT/PCR testing had been in place at the beginning of the COVID-19 pandemic it would have been possible to identify, in real-time, that the elderly and working-age groups from the least privileged groups were bearing the brunt of the pandemic: this identification could have potentially influenced policy decisions thereby reducing fatalities for example in health and social care settings.
Consequently by accounting for the synergy between infectious disease transmission and socioeconomic determinants such models have the potential in future epidemics to identify, in real-time, those who are most adversely impacted and in turn help policy-makers better target their support.

\section*{Competing interests}
    The authors declare no competing interests.

\section*{Author contributions}
    A.C.H. model design, development and implementation.
    C.P.J. MCMC methodology and implementation.
    
    The manuscript was drafted by A.C.H. and revised critically for important intellectual content by all authors.
    All authors gave final approval for publication.

\section*{Acknowledgments}

    A.C.H. and C.P.J. were supported through the Wellcome Trust `GEM: translational software for outbreak analysis' and the UKRI through the JUNIPER modelling consortium (grant number MR/V038613/1).

    The authors would like to thank The High End Computing facility at Lancaster University for providing the facilities required for fitting the models in this paper.

    The views expressed in this paper are those of the authors and not necessarily those of their respective funders or institutions.

\section*{Data availability}
The COVID\nobreakdash-19 data were obtained from the UK Health Security Agency.
These data contain confidential information, with public data deposition non-permissible for socioeconomic reasons. Requests for this data should be made to the UK Health Security Agency.

\bibliography{refs}{}

\begin{thebibliography}{10}

\bibitem{Rasanathan2018}
K.~Rasanathan, ``10 years after the commission on social determinants of
  health: social injustice is still killing on a grand scale,'' {\em The
  Lancet}, vol.~392, no.~10154, pp.~1176--1177, 2018.

\bibitem{Marmot2006}
M.~G. Marmot and R.~G. Wilkinson, {\em Social determinants of health}.
\newblock Oxford: Oxford University Press, 2nd ed.~ed., 2006.

\bibitem{Charlton2013}
J.~Charlton, C.~Rudisill, N.~Bhattarai, and M.~Gulliford, ``Impact of
  deprivation on occurrence, outcomes and health care costs of people with
  multiple morbidity,'' {\em Journal of Health Services Research \& Policy},
  vol.~18, no.~4, pp.~215--223, 2013.

\bibitem{Quinn2014}
S.~C. Quinn and S.~Kumar, ``Health inequalities and infectious disease
  epidemics: A challenge for global health security,'' {\em Biosecurity and
  Bioterrorism: Biodefense Strategy, Practice, and Science}, vol.~12, no.~5,
  pp.~263--273, 2014.

\bibitem{Jensen2021}
N.~Jensen, A.~H. Kelly, and M.~Avendano, ``The covid-19 pandemic underscores
  the need for an equity-focused global health agenda,'' {\em Humanities and
  Social Sciences Communications}, vol.~8, no.~15, 2021.

\bibitem{WHO2020HIE}
{World Health Organization}, {\em Health inequity and the effects of
  COVID‑19: assessing, responding to and mitigating the socioeconomic impact
  on health to build a better future}.
\newblock Regional Office for Europe: World Health Organization, 2020.

\bibitem{PHEGW14472020}
{Public Health England}, {\em Disparities in the risk and outcomes of COVID-19
  (gateway number: GW-1447)}.
\newblock Wellington House, London: {PHE publications}, 2020.

\bibitem{Beale2022}
S.~Beale, I.~Braithwaite, A.~M. Navaratnam, P.~Hardelid, A.~Rodger, A.~Aryee,
  T.~E. Byrne, E.~W.~L. Fong, E.~Fragaszy, C.~Geismar, J.~Kovar, V.~Nguyen,
  P.~Patel, M.~Shrotri, R.~Aldridge, and A.~Hayward, ``Deprivation and exposure
  to public activities during the covid-19 pandemic in england and wales,''
  {\em Journal of Epidemiology \& Community Health}, vol.~76, no.~4,
  pp.~319--326, 2022.

\bibitem{EMG2021}
{Environmental Modelling Group - Transmission Group}, ``Covid-19 risk by
  occupation and workplace,'' 11 February 2021.

\bibitem{GOVIMD2019}
{GOV.UK}, ``English indices of deprivation 2019,'' 26 September 2019.

\bibitem{ONSIMD2019}
{The Office for National Statistics}, ``{Death registrations and populations by
  Index of Multiple Deprivation (IMD) decile, England and Wales, 2019
  (reference number: 12413)},'' 2020.

\bibitem{Bajaj2021}
V.~Bajaj, N.~Gadi, A.~P. Spihlman, S.~C. Wu, C.~H. Choi, and V.~R. Moulton,
  ``Aging, immunity, and covid-19: How age influences the host immune response
  to coronavirus infections?,'' {\em Frontiers in Physiology}, vol.~11, 2021.

\bibitem{Brauer2017}
F.~Brauer, ``Mathematical epidemiology: Past, present, and future,'' {\em
  Infectious Disease Modelling}, vol.~2, no.~2, pp.~113--127, 2017.

\bibitem{Galanis2021}
G.~Galanis and A.~Hanieh, ``Incorporating social determinants of health into
  modelling of covid-19 and other infectious diseases: A baseline
  socio-economic compartmental model,'' {\em Social Science \& Medicine},
  vol.~274, p.~113794, 2021.

\bibitem{Mossong2008}
J.~Mossong, N.~Hens, M.~Jit, P.~Beutels, K.~Auranen, R.~Mikolajczyk,
  M.~Massari, S.~Salmaso, G.~S. Tomba, J.~Wallinga, J.~Heijne,
  M.~Sadkowska-Todys, M.~Rosinska, and W.~J. Edmunds, ``Social contacts and
  mixing patterns relevant to the spread of infectious diseases,'' {\em PLOS
  Medicine}, vol.~5, no.~3, 2008.

\bibitem{Klepac2020}
P.~Klepac, A.~J. Kucharski, A.~J. Conlan, S.~Kissler, M.~L. Tang, H.~Fry, and
  J.~R. Gog, ``Contacts in context: large-scale setting-specific social mixing
  matrices from the {BBC} pandemic project,'' {\em medRxiv}, 2020.

\bibitem{Danon2013}
L.~Danon, J.~M. Read, T.~A. House, M.~C. Vernon, and M.~J. Keeling, ``Social
  encounter networks: characterizing {Great Britain},'' {\em {Proceedings of
  The Royal Society B}}, vol.~280, no.~20131037, 2013.

\bibitem{Davies2020}
{Nicholas G. Davies and Petra Klepac and Yang Liu and Kiesha Prem and Mark Jit
  and CMMID COVID-19 working group and Rosalind M. Eggo}, ``Age-dependent
  effects in the transmission and control of {COVID-19} epidemics,'' {\em
  Nature Medicine}, vol.~26, p.~1205–1211, 2020.

\bibitem{Brauer2019}
F.~Brauer, C.~Castillo-Chavez, and Z.~Feng, {\em Mathematical Models in
  Epidemiology}.
\newblock Springer-Verlag, New York, 1st~ed., 2019.

\bibitem{polymod_data}
J.~Mossong, N.~Hens, M.~Jit, P.~Beutels, K.~Auranen, R.~Mikolajczyk,
  M.~Massari, S.~Salmaso, G.~S. Tomba, J.~Wallinga, J.~Heijne,
  M.~Sadkowska-Todys, M.~Rosinska, and W.~J. Edmunds, ``Polymod social contact
  data,'' 2017.
\newblock Version 1.1.

\bibitem{socialmxr}
S.~Funk, M.~B.-N. Dunbar, C.~A.~B. Pearson, S.~Clifford, C.~Jarvis, and
  A.~Robert, {\em socialmixr}, 2020.
\newblock Available at
  \url{https://https://cran.r-project.org/web/packages/socialmixr/}.

\bibitem{Abbey1952}
H.~Abbey, ``An examination of the reed-frost theory of epidemics,'' {\em Human
  biology}, vol.~24, no.~3, p.~201—233, 1952.

\bibitem{Fine1977}
P.~E.~M. Fine, ``{A commentary on the mechanical analogue to the Reed-Frost
  epidemic model},'' {\em American Journal of Epidemiology}, vol.~106, no.~2,
  pp.~87--100, 1977.

\bibitem{imdage}
A.~C. Hale and C.~P. Jewell, {\em imd-age-covid19uk}, 2022.
\newblock Available at \url{https://gitlab.com/achale/covid19uk-imd-age}.

\bibitem{gemlib}
C.~P. Jewell and A.~C. Hale, {\em gemlib}, 2022.
\newblock Available at \url{https://gitlab.com/gem-epidemics/gemlib}.

\bibitem{Matheson1976}
J.~E. Matheson and R.~L. Winkler, ``Scoring rules for continuous probability
  distributions,'' {\em Management Science}, vol.~22, no.~10, pp.~1087--1096,
  1976.

\bibitem{Eames2011}
K.~T. Eames, N.~L. Tilston, and W.~J. Edmunds, ``The impact of school holidays
  on the social mixing patterns of school children,'' {\em Epidemics}, vol.~3,
  no.~2, pp.~103--108, 2011.

\bibitem{Smith2022}
L.~E. Smith, H.~W.~W. Potts, R.~Aml\^{o}t, N.~T. Fear, S.~Michie, and G.~J.
  Rubin, ``{Patterns of social mixing in England changed in line with
  restrictions during the COVID-19 pandemic (September 2020 to April 2022)},''
  {\em Scientific Reports}, vol.~12, no.~10436, 2022.

\bibitem{Keeling20005}
M.~Keeling and K.~Eames, ``Networks and epidemic models,'' {\em Journal of the
  Royal Society Interface}, vol.~2, pp.~295--307, 10 2005.

\bibitem{Jewell2009}
C.~P. Jewell, M.~J. Keeling, and G.~O. Roberts, ``Predicting undetected
  infections during the 2007 foot-and-mouth disease outbreak,'' {\em Journal of
  The Royal Society Interface}, vol.~6, no.~41, pp.~1145--1151, 2009.

\end{thebibliography}
\bibliographystyle{ieeetr}

\clearpage
\newpage

\appendix
\section*{Appendix}
\renewcommand{\thefigure}{A\arabic{figure}}
\setcounter{figure}{0} 

\subsection*{Visualisation of $\bm{\chi}$}\label{app:visualChi}

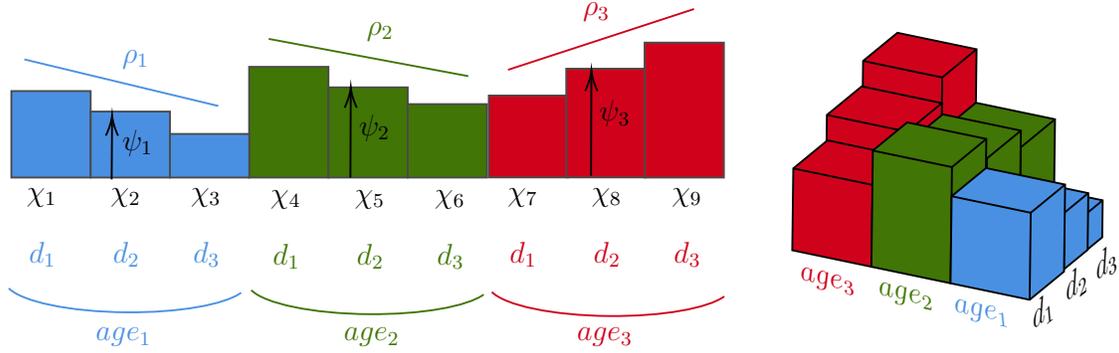
\begin{figure}[ht!]
\centering

\tikzset{every picture/.style={line width=0.75pt}} 

\begin{tikzpicture}[x=0.75pt,y=0.75pt,yscale=-1,xscale=1]

\draw [color={rgb, 255:red, 65; green, 117; blue, 5 }  ,draw opacity=1 ][fill={rgb, 255:red, 208; green, 2; blue, 27 }  ,fill opacity=1 ]   (250.2,105.6) -- (143.2,85.6) ;
\draw  [color={rgb, 255:red, 0; green, 0; blue, 0 }  ,draw opacity=1 ][fill={rgb, 255:red, 74; green, 144; blue, 226 }  ,fill opacity=1 ] (494.7,160.81) -- (502.99,154) -- (590.28,172.22) -- (590.03,192.09) -- (581.74,198.9) -- (494.45,180.68) -- cycle ; \draw  [color={rgb, 255:red, 0; green, 0; blue, 0 }  ,draw opacity=1 ] (590.28,172.22) -- (581.99,179.03) -- (494.7,160.81) ; \draw  [color={rgb, 255:red, 0; green, 0; blue, 0 }  ,draw opacity=1 ] (581.99,179.03) -- (581.74,198.9) ;
\draw  [color={rgb, 255:red, 0; green, 0; blue, 0 }  ,draw opacity=1 ][fill={rgb, 255:red, 208; green, 2; blue, 27 }  ,fill opacity=1 ] (461.84,97.44) -- (480.26,82.32) -- (522.68,91.17) -- (521.75,164.58) -- (503.34,179.71) -- (460.92,170.85) -- cycle ; \draw  [color={rgb, 255:red, 0; green, 0; blue, 0 }  ,draw opacity=1 ] (522.68,91.17) -- (504.26,106.3) -- (461.84,97.44) ; \draw  [color={rgb, 255:red, 0; green, 0; blue, 0 }  ,draw opacity=1 ] (504.26,106.3) -- (503.34,179.71) ;
\draw  [fill={rgb, 255:red, 208; green, 2; blue, 27 }  ,fill opacity=1 ] (442.44,126.4) -- (461.1,111.08) -- (504.09,120.05) -- (503.34,179.71) -- (484.67,195.04) -- (441.69,186.07) -- cycle ; \draw   (504.09,120.05) -- (485.42,135.38) -- (442.44,126.4) ; \draw   (485.42,135.38) -- (484.67,195.04) ;
\draw  [fill={rgb, 255:red, 208; green, 2; blue, 27 }  ,fill opacity=1 ] (424.42,155.09) -- (442.74,140.04) -- (485.25,148.92) -- (484.7,192.83) -- (466.38,207.88) -- (423.87,199.01) -- cycle ; \draw   (485.25,148.92) -- (466.93,163.96) -- (424.42,155.09) ; \draw   (466.93,163.96) -- (466.38,207.88) ;
\draw  [color={rgb, 255:red, 0; green, 0; blue, 0 }  ,draw opacity=1 ][fill={rgb, 255:red, 65; green, 117; blue, 5 }  ,fill opacity=1 ] (503.9,134.99) -- (522.31,119.87) -- (564.73,128.72) -- (564.17,173.44) -- (545.75,188.56) -- (503.34,179.71) -- cycle ; \draw  [color={rgb, 255:red, 0; green, 0; blue, 0 }  ,draw opacity=1 ] (564.73,128.72) -- (546.32,143.85) -- (503.9,134.99) ; \draw  [color={rgb, 255:red, 0; green, 0; blue, 0 }  ,draw opacity=1 ] (546.32,143.85) -- (545.75,188.56) ;
\draw  [fill={rgb, 255:red, 65; green, 117; blue, 5 }  ,fill opacity=1 ] (486.38,141.01) -- (504.56,126.08) -- (546.43,134.82) -- (545.76,187.9) -- (527.58,202.83) -- (485.72,194.09) -- cycle ; \draw   (546.43,134.82) -- (528.25,149.75) -- (486.38,141.01) ; \draw   (528.25,149.75) -- (527.58,202.83) ;
\draw  [fill={rgb, 255:red, 65; green, 117; blue, 5 }  ,fill opacity=1 ] (467.07,145.53) -- (485.49,130.41) -- (527.9,139.26) -- (527.12,201.58) -- (508.7,216.71) -- (466.29,207.86) -- cycle ; \draw   (527.9,139.26) -- (509.49,154.39) -- (467.07,145.53) ; \draw   (509.49,154.39) -- (508.7,216.71) ;
\draw  [fill={rgb, 255:red, 74; green, 144; blue, 226 }  ,fill opacity=1 ] (521.01,168.59) -- (533.71,158.16) -- (582.57,168.36) -- (582.19,198.82) -- (569.49,209.25) -- (520.62,199.05) -- cycle ; \draw   (582.57,168.36) -- (569.87,178.79) -- (521.01,168.59) ; \draw   (569.87,178.79) -- (569.49,209.25) ;
\draw  [fill={rgb, 255:red, 74; green, 144; blue, 226 }  ,fill opacity=1 ] (509.3,170.03) -- (527.72,154.91) -- (570.14,163.76) -- (569.55,210.45) -- (551.13,225.57) -- (508.72,216.72) -- cycle ; \draw   (570.14,163.76) -- (551.72,178.89) -- (509.3,170.03) ; \draw   (551.72,178.89) -- (551.13,225.57) ;
\draw  [color={rgb, 255:red, 74; green, 74; blue, 74 }  ,draw opacity=1 ][fill={rgb, 255:red, 208; green, 2; blue, 27 }  ,fill opacity=1 ] (261.2,116) -- (303.71,116) -- (303.71,159.93) -- (261.2,159.93) -- cycle ;
\draw  [color={rgb, 255:red, 74; green, 74; blue, 74 }  ,draw opacity=1 ][fill={rgb, 255:red, 208; green, 2; blue, 27 }  ,fill opacity=1 ] (302.71,101.6) -- (345.22,101.6) -- (345.22,159.93) -- (302.71,159.93) -- cycle ;
\draw  [color={rgb, 255:red, 74; green, 74; blue, 74 }  ,draw opacity=1 ][fill={rgb, 255:red, 208; green, 2; blue, 27 }  ,fill opacity=1 ] (344.71,87.6) -- (387.22,87.6) -- (387.22,160) -- (344.71,160) -- cycle ;
\draw  [color={rgb, 255:red, 74; green, 74; blue, 74 }  ,draw opacity=1 ][fill={rgb, 255:red, 65; green, 117; blue, 5 }  ,fill opacity=1 ] (132.71,100.6) -- (175.22,100.6) -- (175.22,160) -- (132.71,160) -- cycle ;
\draw  [color={rgb, 255:red, 74; green, 74; blue, 74 }  ,draw opacity=1 ][fill={rgb, 255:red, 65; green, 117; blue, 5 }  ,fill opacity=1 ] (175.22,111.6) -- (217.72,111.6) -- (217.72,160) -- (175.22,160) -- cycle ;
\draw  [color={rgb, 255:red, 74; green, 74; blue, 74 }  ,draw opacity=1 ][fill={rgb, 255:red, 65; green, 117; blue, 5 }  ,fill opacity=1 ] (217.72,120.6) -- (260.23,120.6) -- (260.23,160) -- (217.72,160) -- cycle ;
\draw  [color={rgb, 255:red, 74; green, 74; blue, 74 }  ,draw opacity=1 ][fill={rgb, 255:red, 74; green, 144; blue, 226 }  ,fill opacity=1 ] (5.23,113.6) -- (47.74,113.6) -- (47.74,160) -- (5.23,160) -- cycle ;
\draw  [color={rgb, 255:red, 74; green, 74; blue, 74 }  ,draw opacity=1 ][fill={rgb, 255:red, 74; green, 144; blue, 226 }  ,fill opacity=1 ] (47.74,124.6) -- (90.25,124.6) -- (90.25,160) -- (47.74,160) -- cycle ;
\draw  [color={rgb, 255:red, 74; green, 74; blue, 74 }  ,draw opacity=1 ][fill={rgb, 255:red, 74; green, 144; blue, 226 }  ,fill opacity=1 ] (90.25,136.6) -- (132.76,136.6) -- (132.76,160) -- (90.25,160) -- cycle ;
\draw [color={rgb, 255:red, 0; green, 0; blue, 0 }  ,draw opacity=1 ][fill={rgb, 255:red, 0; green, 0; blue, 0 }  ,fill opacity=1 ]   (187,160) -- (187.19,113.6) ;
\draw [shift={(187.2,111.6)}, rotate = 90.24] [color={rgb, 255:red, 0; green, 0; blue, 0 }  ,draw opacity=1 ][line width=0.75]    (10.93,-3.29) .. controls (6.95,-1.4) and (3.31,-0.3) .. (0,0) .. controls (3.31,0.3) and (6.95,1.4) .. (10.93,3.29)   ;
\draw [color={rgb, 255:red, 0; green, 0; blue, 0 }  ,draw opacity=1 ][fill={rgb, 255:red, 0; green, 0; blue, 0 }  ,fill opacity=1 ]   (59,161) -- (59.23,128.6) ;
\draw [shift={(59.24,126.6)}, rotate = 90.4] [color={rgb, 255:red, 0; green, 0; blue, 0 }  ,draw opacity=1 ][line width=0.75]    (10.93,-3.29) .. controls (6.95,-1.4) and (3.31,-0.3) .. (0,0) .. controls (3.31,0.3) and (6.95,1.4) .. (10.93,3.29)   ;
\draw [color={rgb, 255:red, 0; green, 0; blue, 0 }  ,draw opacity=1 ][fill={rgb, 255:red, 0; green, 0; blue, 0 }  ,fill opacity=1 ]   (316,159) -- (316.19,104.6) ;
\draw [shift={(316.2,102.6)}, rotate = 90.2] [color={rgb, 255:red, 0; green, 0; blue, 0 }  ,draw opacity=1 ][line width=0.75]    (10.93,-3.29) .. controls (6.95,-1.4) and (3.31,-0.3) .. (0,0) .. controls (3.31,0.3) and (6.95,1.4) .. (10.93,3.29)   ;
\draw [color={rgb, 255:red, 74; green, 144; blue, 226 }  ,draw opacity=1 ][fill={rgb, 255:red, 208; green, 2; blue, 27 }  ,fill opacity=1 ]   (116.2,121.6) -- (12.2,96.6) ;
\draw [color={rgb, 255:red, 208; green, 2; blue, 27 }  ,draw opacity=1 ][fill={rgb, 255:red, 208; green, 2; blue, 27 }  ,fill opacity=1 ]   (371.2,69.6) -- (271.53,102.19) ;
\draw  [draw opacity=0] (128.44,221.77) .. controls (119.57,228.12) and (95.32,232.53) .. (66.83,232.29) .. controls (36.42,232.04) and (10.93,226.57) .. (3.85,219.4) -- (66.97,215.88) -- cycle ; \draw  [color={rgb, 255:red, 74; green, 144; blue, 226 }  ,draw opacity=1 ] (128.44,221.77) .. controls (119.57,228.12) and (95.32,232.53) .. (66.83,232.29) .. controls (36.42,232.04) and (10.93,226.57) .. (3.85,219.4) ;  
\draw  [draw opacity=0] (258.44,222.77) .. controls (249.57,229.12) and (225.32,233.53) .. (196.83,233.29) .. controls (166.42,233.04) and (140.93,227.57) .. (133.85,220.4) -- (196.97,216.88) -- cycle ; \draw  [color={rgb, 255:red, 65; green, 117; blue, 5 }  ,draw opacity=1 ] (258.44,222.77) .. controls (249.57,229.12) and (225.32,233.53) .. (196.83,233.29) .. controls (166.42,233.04) and (140.93,227.57) .. (133.85,220.4) ;  
\draw  [draw opacity=0] (387.44,223.77) .. controls (378.57,230.12) and (354.32,234.53) .. (325.83,234.29) .. controls (295.42,234.04) and (269.93,228.57) .. (262.85,221.4) -- (325.97,217.88) -- cycle ; \draw  [color={rgb, 255:red, 208; green, 2; blue, 27 }  ,draw opacity=1 ] (387.44,223.77) .. controls (378.57,230.12) and (354.32,234.53) .. (325.83,234.29) .. controls (295.42,234.04) and (269.93,228.57) .. (262.85,221.4) ;  

\draw (319,117) node [anchor=north west][inner sep=0.75pt]  [color={rgb, 255:red, 0; green, 0; blue, 0 }  ,opacity=1 ]  {$\psi _{3}$};
\draw (63.2,132.6) node [anchor=north west][inner sep=0.75pt]  [color={rgb, 255:red, 0; green, 0; blue, 0 }  ,opacity=1 ]  {$\psi _{1}$};
\draw (310.2,64.6) node [anchor=north west][inner sep=0.75pt]  [color={rgb, 255:red, 208; green, 2; blue, 27 }  ,opacity=1 ]  {$\rho _{3}$};
\draw (190.2,124.6) node [anchor=north west][inner sep=0.75pt]  [color={rgb, 255:red, 0; green, 0; blue, 0 }  ,opacity=1 ]  {$\psi _{2}$};
\draw (194.2,75) node [anchor=north west][inner sep=0.75pt]  [color={rgb, 255:red, 65; green, 117; blue, 5 }  ,opacity=1 ]  {$\rho _{2}$};
\draw (63.2,90) node [anchor=north west][inner sep=0.75pt]  [color={rgb, 255:red, 74; green, 144; blue, 226 }  ,opacity=1 ]  {$\rho _{1}$};
\draw (13.2,193) node [anchor=north west][inner sep=0.75pt]  [color={rgb, 255:red, 74; green, 144; blue, 226 }  ,opacity=1 ]  {$d_{1}$};
\draw (58.2,193) node [anchor=north west][inner sep=0.75pt]  [color={rgb, 255:red, 74; green, 144; blue, 226 }  ,opacity=1 ]  {$d_{2}$};
\draw (101.2,193) node [anchor=north west][inner sep=0.75pt]  [color={rgb, 255:red, 74; green, 144; blue, 226 }  ,opacity=1 ]  {$d_{3}$};
\draw (144,194) node [anchor=north west][inner sep=0.75pt]  [color={rgb, 255:red, 65; green, 117; blue, 5 }  ,opacity=1 ]  {$d_{1}$};
\draw (189,194) node [anchor=north west][inner sep=0.75pt]  [color={rgb, 255:red, 65; green, 117; blue, 5 }  ,opacity=1 ]  {$d_{2}$};
\draw (232,194) node [anchor=north west][inner sep=0.75pt]  [color={rgb, 255:red, 65; green, 117; blue, 5 }  ,opacity=1 ]  {$d_{3}$};
\draw (271,193) node [anchor=north west][inner sep=0.75pt]  [color={rgb, 255:red, 208; green, 2; blue, 27 }  ,opacity=1 ]  {$d_{1}$};
\draw (316,193) node [anchor=north west][inner sep=0.75pt]  [color={rgb, 255:red, 208; green, 2; blue, 27 }  ,opacity=1 ]  {$d_{2}$};
\draw (359,193) node [anchor=north west][inner sep=0.75pt]  [color={rgb, 255:red, 208; green, 2; blue, 27 }  ,opacity=1 ]  {$d_{3}$};
\draw (182,239) node [anchor=north west][inner sep=0.75pt]  [color={rgb, 255:red, 65; green, 117; blue, 5 }  ,opacity=1 ]  {$age_{2}$};
\draw (307,239) node [anchor=north west][inner sep=0.75pt]  [color={rgb, 255:red, 208; green, 2; blue, 27 }  ,opacity=1 ]  {$age_{3}$};
\draw (49,239) node [anchor=north west][inner sep=0.75pt]  [color={rgb, 255:red, 74; green, 144; blue, 226 }  ,opacity=1 ]  {$age_{1}$};
\draw (12,164) node [anchor=north west][inner sep=0.75pt]  [color={rgb, 255:red, 0; green, 0; blue, 0 }  ,opacity=1 ]  {$\chi _{1}$};
\draw (57,164) node [anchor=north west][inner sep=0.75pt]  [color={rgb, 255:red, 0; green, 0; blue, 0 }  ,opacity=1 ]  {$\chi _{2}$};
\draw (100,164) node [anchor=north west][inner sep=0.75pt]  [color={rgb, 255:red, 0; green, 0; blue, 0 }  ,opacity=1 ]  {$\chi _{3}$};
\draw (142.8,165) node [anchor=north west][inner sep=0.75pt]  [color={rgb, 255:red, 0; green, 0; blue, 0 }  ,opacity=1 ]  {$\chi _{4}$};
\draw (187.8,165) node [anchor=north west][inner sep=0.75pt]  [color={rgb, 255:red, 0; green, 0; blue, 0 }  ,opacity=1 ]  {$\chi _{5}$};
\draw (230.8,165) node [anchor=north west][inner sep=0.75pt]  [color={rgb, 255:red, 0; green, 0; blue, 0 }  ,opacity=1 ]  {$\chi _{6}$};
\draw (269.8,164) node [anchor=north west][inner sep=0.75pt]  [color={rgb, 255:red, 0; green, 0; blue, 0 }  ,opacity=1 ]  {$\chi _{7}$};
\draw (314.8,164) node [anchor=north west][inner sep=0.75pt]  [color={rgb, 255:red, 0; green, 0; blue, 0 }  ,opacity=1 ]  {$\chi _{8}$};
\draw (357.8,164) node [anchor=north west][inner sep=0.75pt]  [color={rgb, 255:red, 0; green, 0; blue, 0 }  ,opacity=1 ]  {$\chi _{9}$};
\draw (468.15,213.78) node [anchor=north west][inner sep=0.75pt]  [color={rgb, 255:red, 65; green, 117; blue, 5 }  ,opacity=1 ,rotate=-11.86,xslant=-0.32]  {$age_{2}$};
\draw (508.7,222.71) node [anchor=north west][inner sep=0.75pt]  [color={rgb, 255:red, 74; green, 144; blue, 226 }  ,opacity=1 ,rotate=-11.86,xslant=-0.32]  {$age_{1}$};
\draw (426.15,205.78) node [anchor=north west][inner sep=0.75pt]  [color={rgb, 255:red, 208; green, 2; blue, 27 }  ,opacity=1 ,rotate=-11.86,xslant=-0.32]  {$age_{3}$};
\draw (568.49,212.25) node [anchor=north west][inner sep=0.75pt]  [color={rgb, 255:red, 0; green, 0; blue, 0 }  ,opacity=1 ,rotate=-321.81,xslant=0.65]  {$d_{2}$};
\draw (584.6,201.25) node [anchor=north west][inner sep=0.75pt]  [color={rgb, 255:red, 0; green, 0; blue, 0 }  ,opacity=1 ,rotate=-321.81,xslant=0.65]  {$d_{3}$};
\draw (550.6,227.25) node [anchor=north west][inner sep=0.75pt]  [color={rgb, 255:red, 0; green, 0; blue, 0 }  ,opacity=1 ,rotate=-321.81,xslant=0.65]  {$d_{1}$};

\end{tikzpicture}

\caption{Visualisation of $\bm{\chi}$ given 9 strata, that is 3 age strata each and 3 deprivation strata.  Each age group has its own colour. On both diagrams the vertical height of each box is proportional to the value of an element of $\bm{\chi}$, age groups are denoted by $age_{1,2,3}$ and deprivations groups $d_{1,2,3}$.
Left: structure of the elements of $\bm{\chi}$ alongside its relationship to $\bm{\psi}$ whose elements determine the intercept per age group and $\bm{\rho}$ whose elements determine the slope per age group.
Right: structure of the elements of $\bm{\chi}$ after it is reshaped, in the main text this configuration is used to depict the values of the elements of $\bm{\chi}$ as a heatmap e.g. Figure \ref{fig:fitted_data_chi}.
}
\label{fig:visualofChi}
\end{figure}

\subsection*{Outline of discrete-time Markov chain methodology}\label{app:mcmcoutline}

Let the number of individuals in each state $q \in \{\mathrm{S},\mathrm{E},\mathrm{I},\mathrm{R}\}$ at time $t$ and in stratum $i$ be denoted by $x^q_{i,t}$.
The number of events occurring between each pair of states at timestep $t$ with length $\delta_t$ is
\begin{equation*}
y^{qr}_{i,t} \sim \mathrm{Binomial}(x^q_{i,t}, p^{qr}_{i}(t))
\end{equation*}
where the transitions from $q$ to $r$ are $(qr) \in \{ (\mathrm{S}\mathrm{E}), (\mathrm{E}\mathrm{I}), (\mathrm{I}\mathrm{R}) \}$ and
the transition probability is the CDF of the exponential distribution 
\begin{equation*}
p^{qr}_{i}(t) = 1 - \mathrm{exp}(-h^{qr}_i(t) \, \delta_t)
\end{equation*}
with the rate of transition $(qr)$ denoted by $h^{qr}_i$.
With this arrangement and with reference to Equation \ref{eq:odes} then states at successive time points are computed as follows:
\begin{subequations}
\label{app:eq:states}
\begin{align}
    x^\mathrm{S}_{i,t+1} &= x^\mathrm{S}_{i,t} - y^{\mathrm{S}\mathrm{E}}_{i,t} \notag \\
    x^\mathrm{E}_{i,t+1} &= x^\mathrm{E}_{i,t} + y^{\mathrm{S}\mathrm{E}}_{i,t} - y^{\mathrm{E}\mathrm{I}}_{i,t} \notag \\
    x^\mathrm{I}_{i,t+1} &= x^\mathrm{I}_{i,t} + y^{\mathrm{E}\mathrm{I}}_{i,t} - y^{\mathrm{I}\mathrm{R}}_{i,t} \notag \\
    x^\mathrm{R}_{i,t+1} &= x^\mathrm{R}_{i,t} + y^{\mathrm{I}\mathrm{R}}_{i,t} \notag 
\end{align}
\end{subequations}
To fit the model to the stratified time series of COVID-19 data we assume that the aggregated daily case incidence at each $i$ and $t$ is equivalent to observing events $y^{\mathrm{I}\mathrm{R}}_{i,t}$.
Consequently there is censoring of events in so much as $y^{\mathrm{S}\mathrm{E}}_{i,t}$ and $y^{\mathrm{E}\mathrm{I}}_{i,t}$ are unobserved, therefore
to distinguish between observed and unobserved events let $z^{\mathrm{S}\mathrm{E}}_{i,t} = y^{\mathrm{S}\mathrm{E}}_{i,t}$ and $z^{\mathrm{E}\mathrm{I}}_{i,t} = y^{\mathrm{E}\mathrm{I}}_{i,t}$.
Conditional on a set of initial states $\mathbf{X}_0$ at $t=0$ and vector of model parameters $\bm{\theta}$, then the log likelihood of observed transitions $\mathbf{Y}$, unobserved transitions $\mathbf{Z}$, and states $\mathbf{X}$ is
\begin{equation*}
\begin{aligned}
    \ell(\mathbf{Y}, \mathbf{Z}, \mathbf{X} | \mathbf{X}_0, \bm{\theta}) \propto \sum^T_{t=1} \sum^L_{i=1} [
    & z^{\mathrm{S}\mathrm{E}}_{i,t} \, \mathrm{log} \, p^{\mathrm{S}\mathrm{E}}_{i,t} + (x^\mathrm{S}_{i,t} - z^{\mathrm{S}\mathrm{E}}_{i,t}) \, \mathrm{log} \, (1-p^{\mathrm{S}\mathrm{E}}_{i,t}) \, + \\
    & z^{\mathrm{E}\mathrm{I}}_{i,t} \, \mathrm{log} \, p^{\mathrm{E}\mathrm{I}}_{i,t} + (x^\mathrm{E}_{i,t} - z^{\mathrm{E}\mathrm{I}}_{i,t}) \, \mathrm{log} \, (1-p^{\mathrm{E}\mathrm{I}}_{i,t}) \, + \\
    & y^{\mathrm{I}\mathrm{R}}_{i,t} \, \mathrm{log} \, p^{\mathrm{I}\mathrm{R}}_{i,t} + (x^\mathrm{I}_{i,t} - y^{\mathrm{I}\mathrm{R}}_{i,t}) \, \mathrm{log} \, (1-p^{\mathrm{I}\mathrm{R}}_{i,t}) \, ]
\end{aligned}
\end{equation*}
with $\mathbf{X}_0 = [ \bm{x}^\mathrm{S}_{0}, \bm{x}^\mathrm{E}_{0}, \bm{x}^\mathrm{I}_{0}, \bm{x}^\mathrm{R}_{0} ]^{\mathrm{T}}$, $\bm{\theta} = [\bm{\psi}, \bm{\rho}, \gamma_1, \alpha_0, \bm{\alpha_t} ]^{\mathrm{T}}$ and given training data (i.e. $y^{\mathrm{I}\mathrm{R}}_{i,t} \; \forall \; i,t$) with and a total of $T$ time points and $L$ strata.
Note that methodology for estimating the unobserved events is beyond the scope of this appendix however it is based on a MCMC algorithm within a data-augmentation framework, for reference see \cite{Jewell2009}.

\end{document}


\maketitle

\clearpage
\newpage

\begin{landscape}
    \begin{figure}
        \centering
        \captionsetup{width=0.85\linewidth}
        \includegraphics[width=0.90\linewidth]{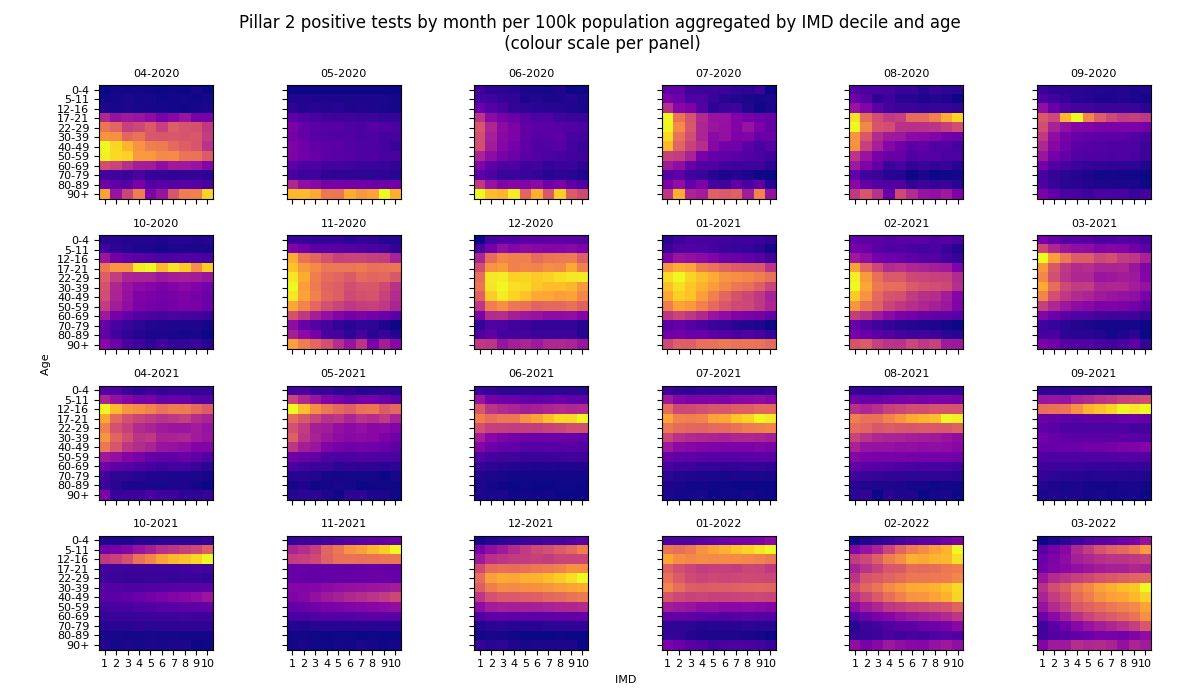}
        \caption{UKHSA COVID-19 case data from April 2020 to March 2022: specifically community testing positive case data, known as `pillar 2'. Age is split into 12 bands. Deprivation (IMD) is divided into 10 deciles (1 most deprived, 10 least deprived). Each panel relates to a given month and each pixel represents the case count normalised by population size. For visualisation proposes each panel has its own colour scale: yellow=high, orange=medium, blue=low. This figure tells the story of the pandemic from a deprivation-age perspective, for instance the effectiveness of the vaccination programme is visible during the first 9 months of 2021 in so much as cases are progressively focused into lower age groups.}
        \label{fig:55positive_tests_p2_grid_over_imds_ages_per_month_per_panel_normalised_scale_timeseries}
    \end{figure}
\end{landscape}

\begin{landscape}
    \begin{figure}
        \centering
        \captionsetup{width=0.98\linewidth}
        \includegraphics[width=0.98\linewidth]{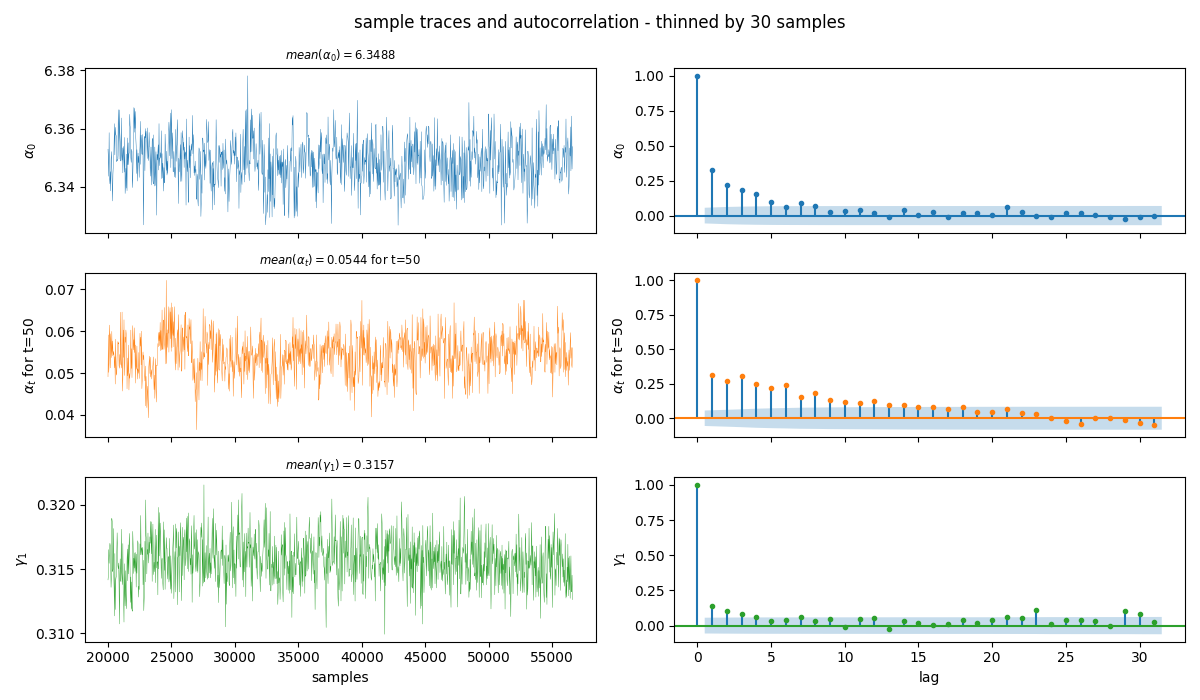}
        \caption{MCMC posterior samples for model fitted to the training data, showing samples for $\alpha_0$, $\alpha_t$ with $t=50$ and $\gamma_1$. The burn-in of $20000$ samples is omitted.}
        \label{fig:diagnostic_traces_burnin20000}
    \end{figure}
\end{landscape}

\begin{landscape}
    \begin{figure}
        \centering
        \captionsetup{width=0.98\linewidth}
        \includegraphics[width=0.98\linewidth]{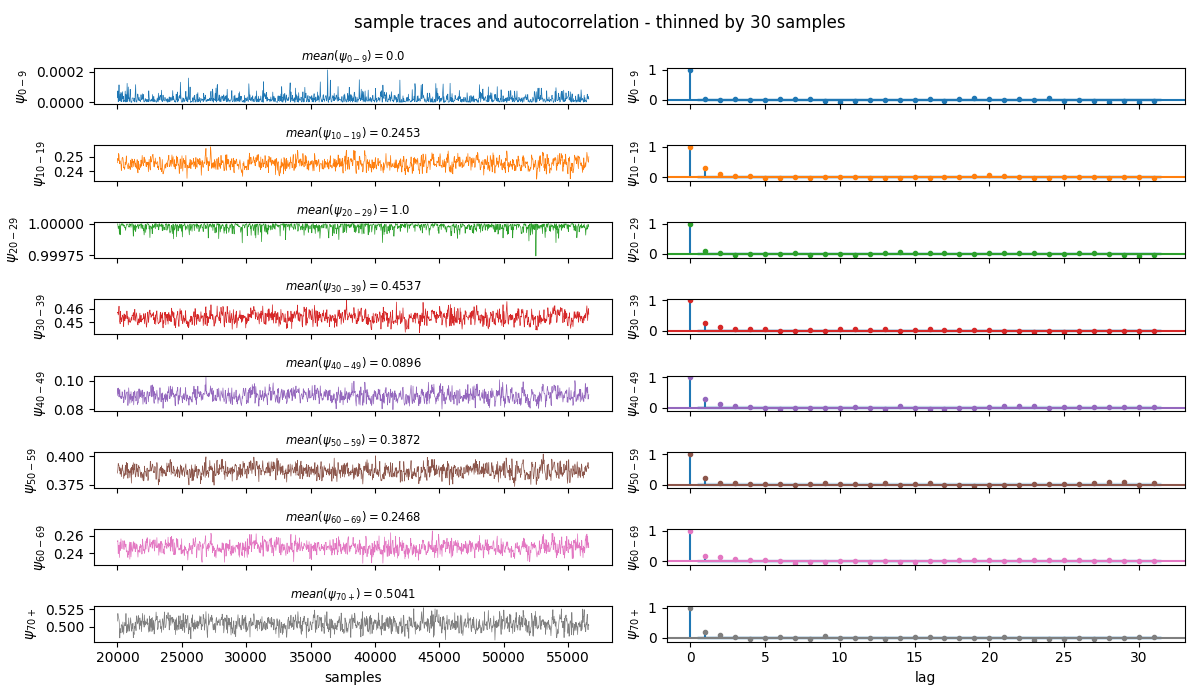}
        \caption{MCMC posterior samples for the model fitted to the training data showing samples relating to the elements of $\bm{\psi}$: subscripts denote age group.
        The burn-in of $20000$ samples is omitted.}
        \label{fig:diagnostic_traces_psi_burnin_20000}
    \end{figure}
\end{landscape}

\begin{landscape}
    \begin{figure}
        \centering
        \captionsetup{width=0.98\linewidth}
        \includegraphics[width=0.98\linewidth]{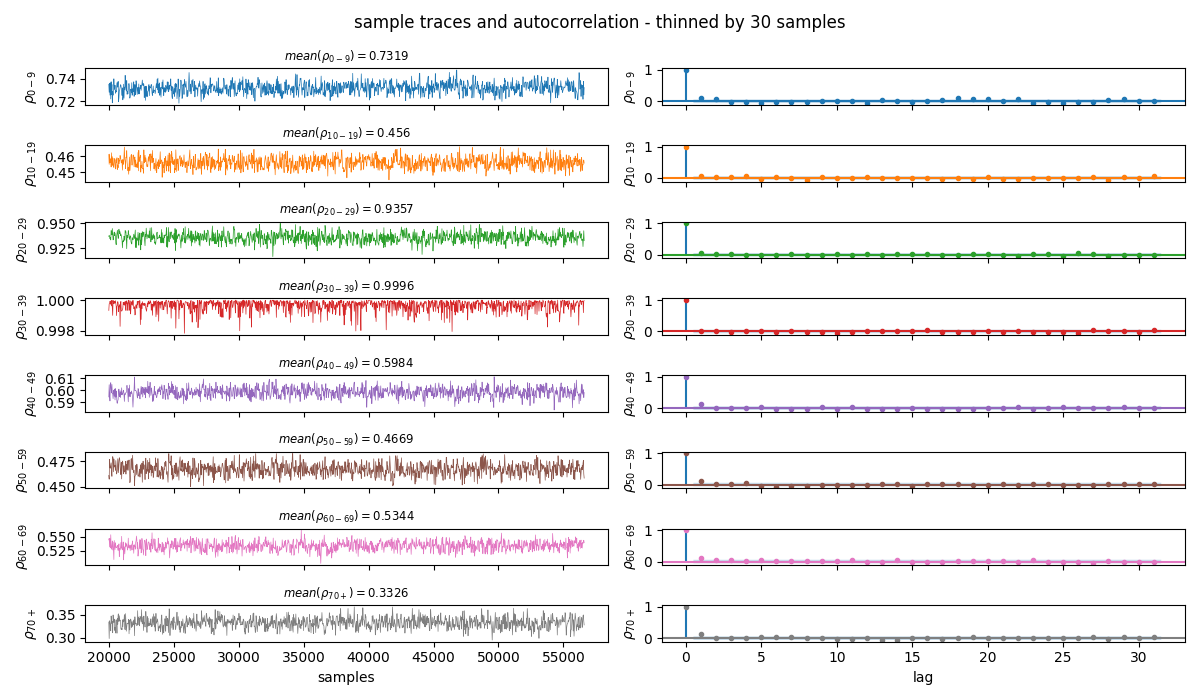}
        \caption{MCMC posterior samples for the model fitted to the training data showing samples relating to the elements of $\bm{\rho}$: subscripts denote age group.
        The burn-in of $20000$ samples is omitted.}
        \label{fig:diagnostic_traces_rho_burnin_20000}
    \end{figure}
\end{landscape}

\begin{landscape}
    \begin{figure}
        \centering
        \captionsetup{width=0.98\linewidth}
        \includegraphics[width=0.98\linewidth]{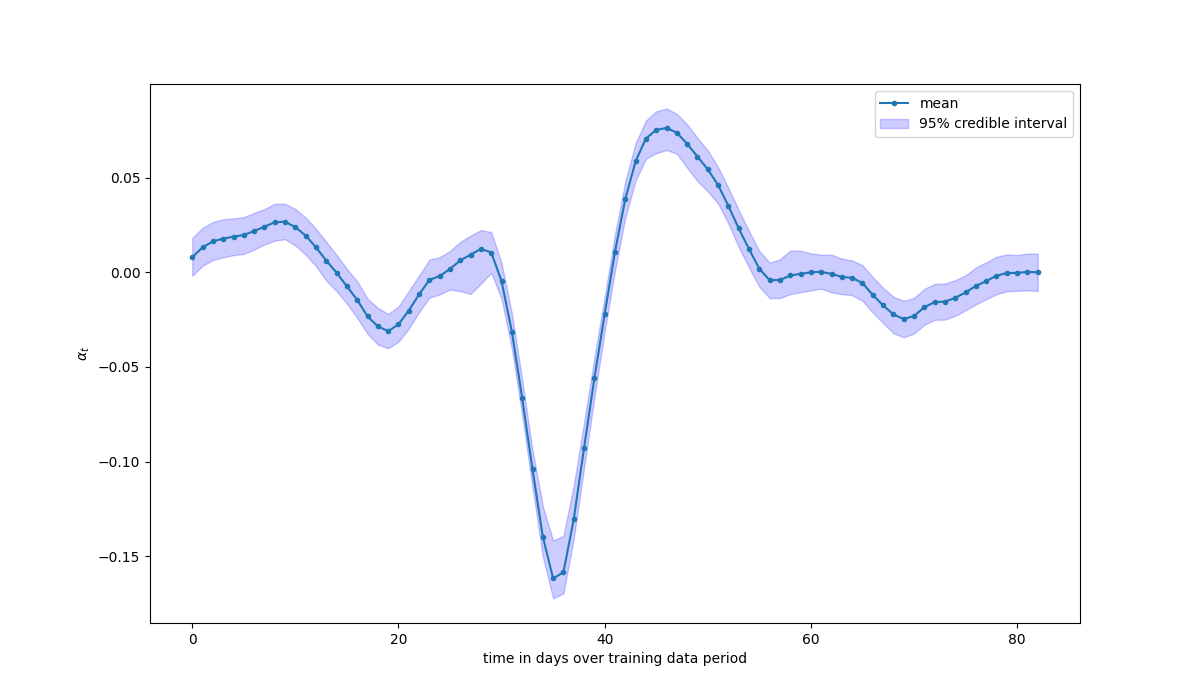}
        \caption{MCMC posterior samples for the model fitted to the training data in relation to $\bm{\alpha_t}$.}
        \label{fig:diagnostic_alpha_t_burnin20000}
    \end{figure}
\end{landscape}

\begin{landscape}
    \begin{figure}
        \centering
        \captionsetup{width=0.98\linewidth}
        \includegraphics[width=0.98\linewidth]{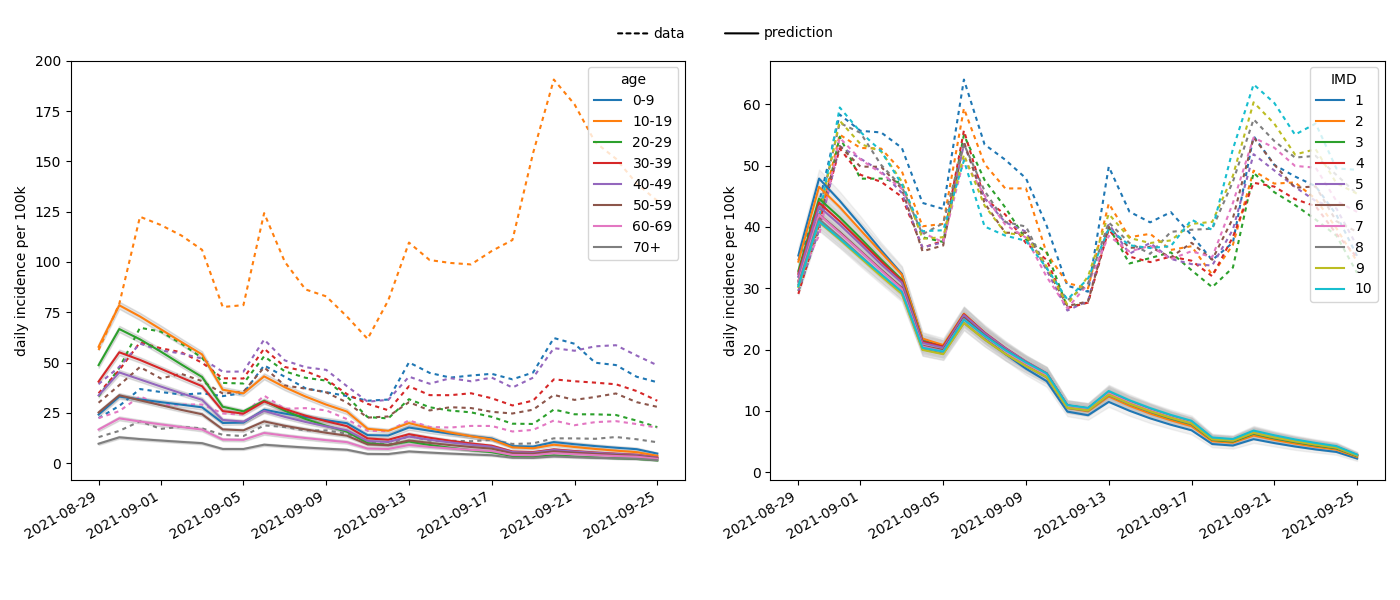}
        \caption{Forecast over 4-week period when there is a depletion of susceptible individuals as described in Section 4.4.1.
        The model is fitted to training data from 07/06/2021 to 29/08/2021.
        Forecasts of expected daily case incidence are shown alongside their respective 90\% credible intervals. 
        Left: solid line, expected daily incidence aggregated by age group; and dotted line, observed case incidence aggregated by age group.
        Right: solid line, expected daily incidence aggregated by IMD (deprivation) decile; and dotted line, observed case incidence aggregated by IMD decile; IMD 1 refers to the most deprived and 10 the least.}
        \label{fig:plot_simulated_timeseries_incidence_100k_depletion_supp_mat}
    \end{figure}
\end{landscape}

\begin{landscape}
    \begin{figure}
        \centering
        \captionsetup{width=0.98\linewidth}
        \includegraphics[width=0.98\linewidth]{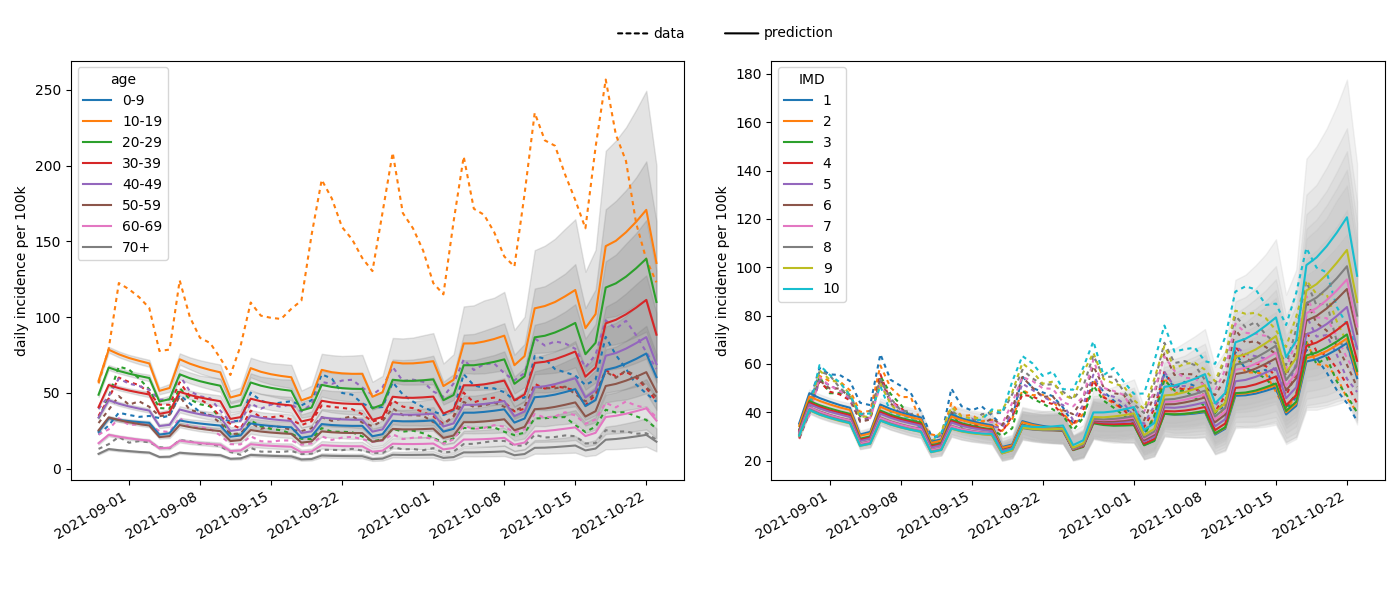}
        \caption{Forecast over 8-week period for the simulation experiment where there is an increase in deprivation-structured mixing in accordance with
        $\bm{\mathcal{C}}^D(t)  =  \mathbf{1}_{J \times J} + \mathbf{M} \, \varpi(t)$ where $\mathbf{M} = \mathbf{U} + \mathbf{U}^\mathrm{T} - \mathrm{diag}(\mathbf{U})$.
        The model is fitted to training data from 07/06/2021 to 29/08/2021.
        Forecasts of expected daily case incidence are shown alongside their respective 90\% credible intervals. 
        Left: solid line, expected daily incidence aggregated by age group; and dotted line, observed case incidence aggregated by age group.
        Right: solid line, expected daily incidence aggregated by IMD (deprivation) decile; and dotted line, observed case incidence aggregated by IMD decile.}
        \label{fig:plot_simulated_timeseries_incidence_100k_social_mixing_increase_supp_mat}
    \end{figure}
\end{landscape}

\begin{landscape}
    \begin{figure}
        \centering
        \captionsetup{width=0.98\linewidth}
        \includegraphics[width=0.98\linewidth]{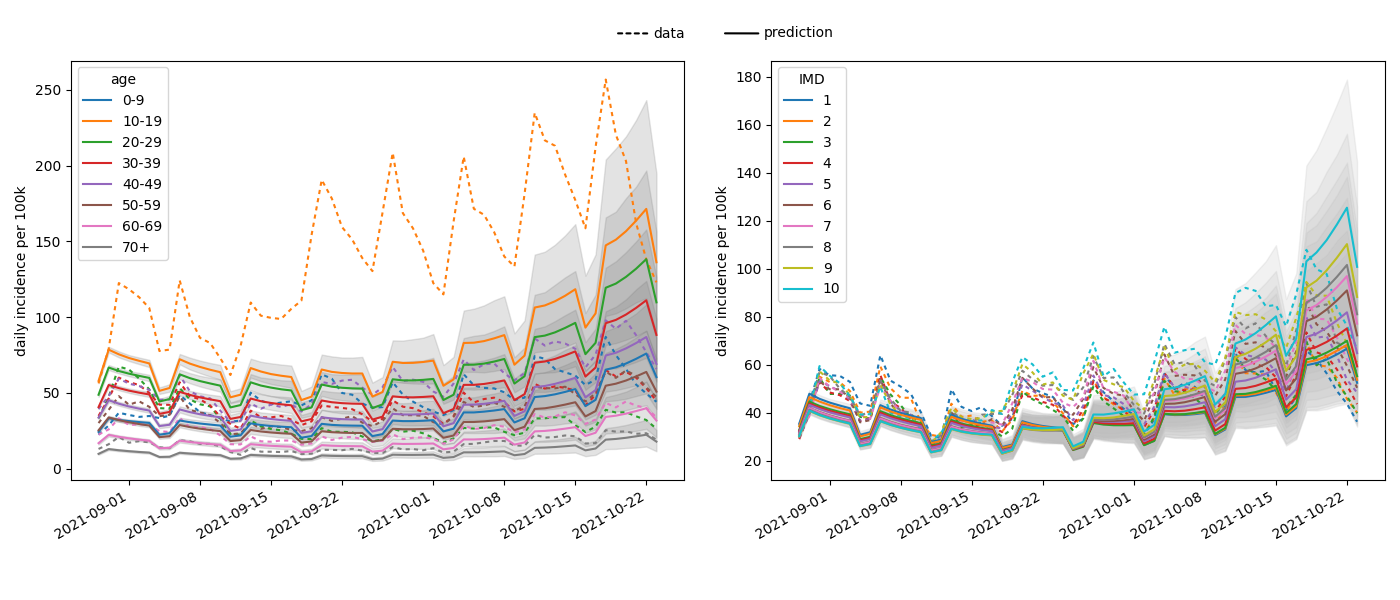}
        \caption{Forecast over 8-week period for the `assortative' simulation experiment where there is an increase in deprivation-structured mixing in accordance with $\bm{\mathcal{C}}^D(t)  = \mathbf{1}_{J \times J} + \mathbf{M} \,  \varpi(t)$ where $\mathbf{M} = \mathrm{diag}(\mathbf{U})$.
        The model is fitted to training data from 07/06/2021 to 29/08/2021.
        Forecasts of expected daily case incidence are shown alongside their respective 90\% credible intervals. 
        Left: solid line, expected daily incidence aggregated by age group; and dotted line, observed case incidence aggregated by age group.
        Right: solid line, expected daily incidence aggregated by IMD (deprivation) decile; and dotted line, observed case incidence aggregated by IMD decile.}
        \label{fig:plot_simulated_timeseries_incidence_100k_social_mixing_assortative_supp_mat}
    \end{figure}
\end{landscape}

\begin{landscape}
    \begin{figure}
        \centering
        \captionsetup{width=0.98\linewidth}
        \includegraphics[width=0.98\linewidth]{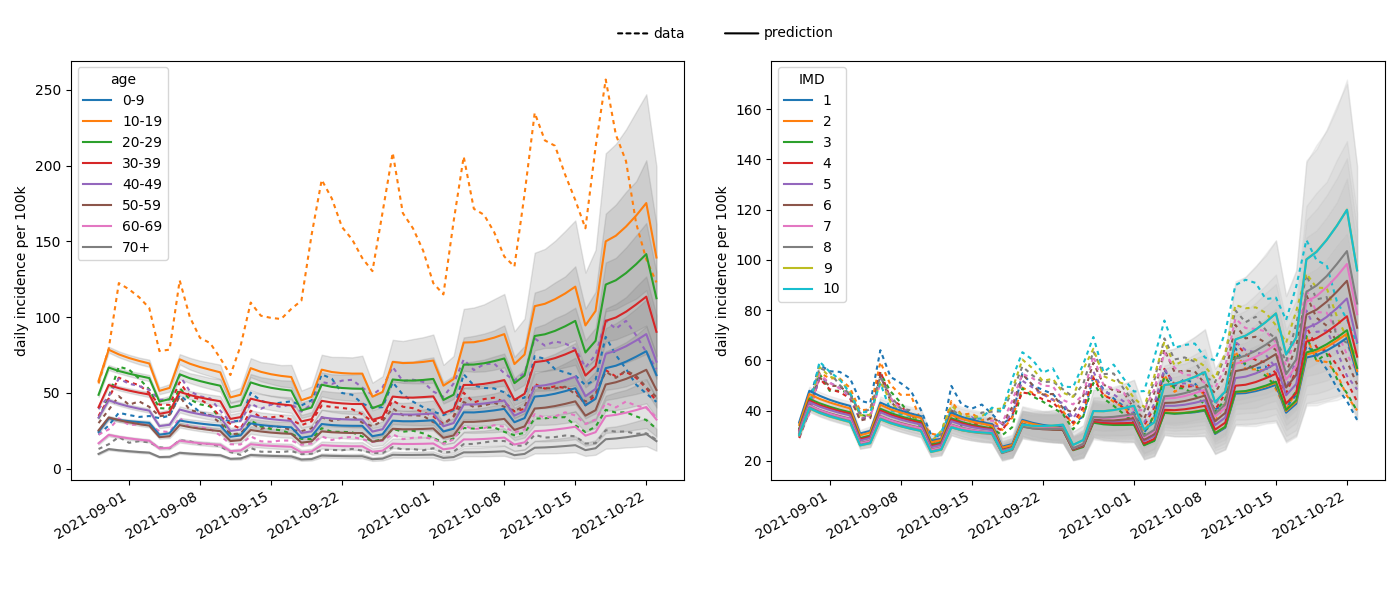}
        \caption{Forecast over 8-week period for the `disassortative' simulation experiment where there is an increase in deprivation-structured mixing in accordance with $\bm{\mathcal{C}}^D(t)  = \mathbf{1}_{J \times J} + \mathbf{M} \varpi(t)$ where $\mathbf{M} = \mathbf{U} + \mathbf{U}^\mathrm{T} - 2 \, \mathrm{diag}(\mathbf{U})$.
        The model is fitted to training data from 07/06/2021 to 29/08/2021.
        Forecasts of expected daily case incidence are shown alongside their respective 90\% credible intervals. 
        Left: solid line, expected daily incidence aggregated by age group; and dotted line, observed case incidence aggregated by age group.
        Right: solid line, expected daily incidence aggregated by IMD (deprivation) decile; and dotted line, observed case incidence aggregated by IMD decile.}
        \label{fig:plot_simulated_timeseries_incidence_100k_social_mixing_not_assortative_supp_mat}
    \end{figure}
\end{landscape}

\begin{landscape}
    \begin{figure}
        \centering
        \captionsetup{width=0.98\linewidth}
        \includegraphics[width=0.98\linewidth]{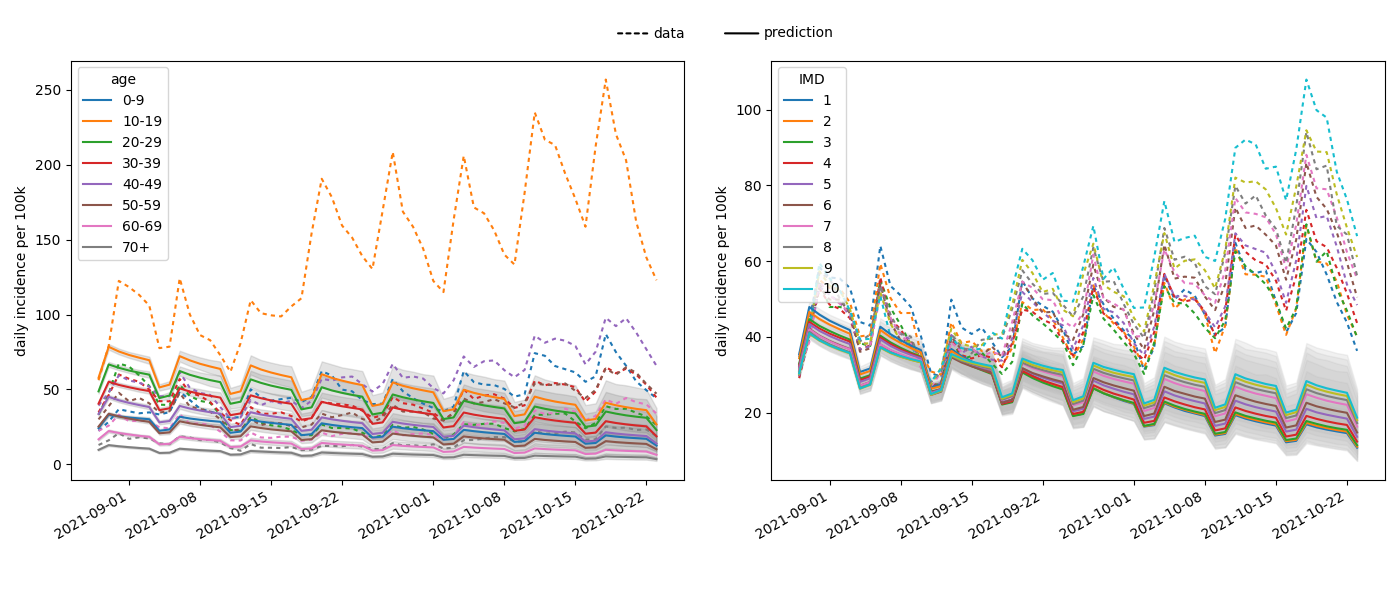}
        \caption{
        Forecast over 8-week period with perturbation to behavioural ascertainment term and no increase in hazard rate: specifically $(\zeta, \varepsilon) = (1.0, 0.017)$. 
        The model is fitted to training data from 07/06/2021 to 29/08/2021.
        Forecasts of expected daily case incidence are shown alongside their respective 90\% credible intervals.
        Left: solid line, expected daily incidence aggregated by age group; and dotted line, observed case incidence aggregated by age group.
        Right: solid line, expected daily incidence aggregated by IMD (deprivation) decile; and dotted line, observed case incidence aggregated by IMD decile.}
        \label{fig:plot_simulated_timeseries_incidence_100k_behavioural_chi_no_boost_supp_mat}
    \end{figure}
\end{landscape}

\begin{landscape}
    \begin{figure}
        \centering
        \captionsetup{width=0.98\linewidth}
        \includegraphics[width=0.98\linewidth]{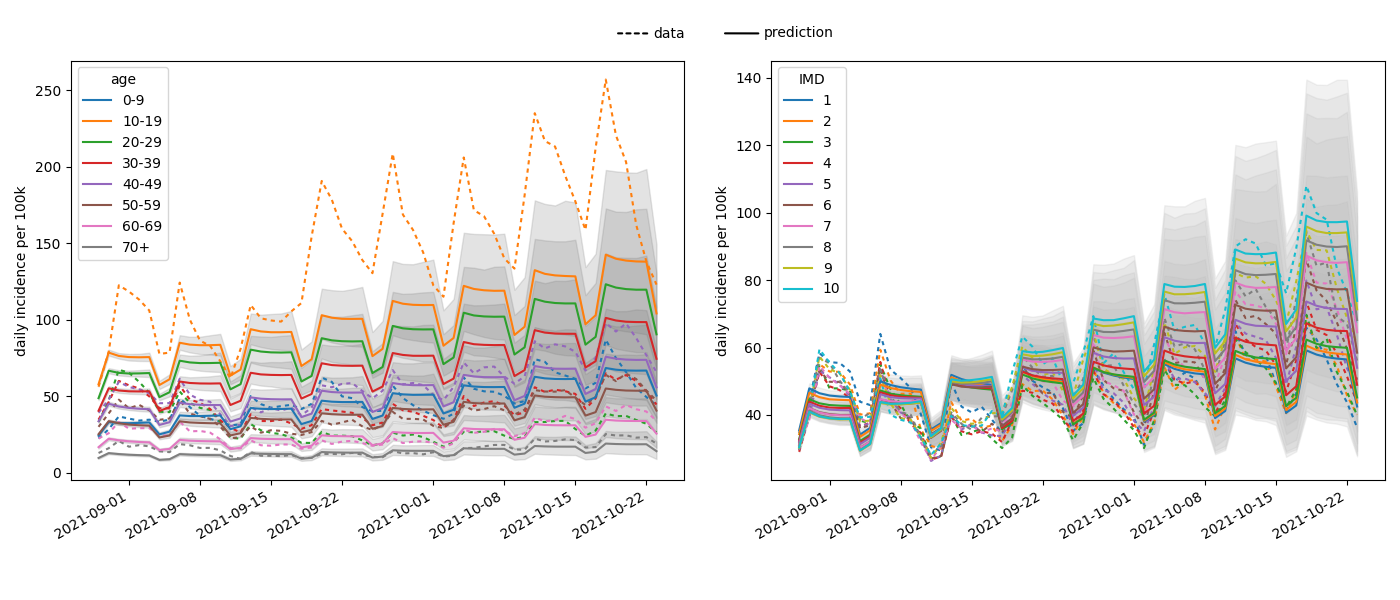}
        \caption{
        Forecast over 8-week period with perturbation to behavioural ascertainment term and increased hazard rate: specifically $(\zeta, \varepsilon) = (1.265, 0.017)$. 
        The model is fitted to training data from 07/06/2021 to 29/08/2021.
        Forecasts of expected daily case incidence are shown alongside their respective 90\% credible intervals.
        Left: solid line, expected daily incidence aggregated by age group; and dotted line, observed case incidence aggregated by age group.
        Right: solid line, expected daily incidence aggregated by IMD (deprivation) decile; and dotted line, observed case incidence aggregated by IMD decile.}
        \label{fig:plot_simulated_timeseries_incidence_100k_behavioural_chi_with_boost_supp_mat}
    \end{figure}
\end{landscape}

\begin{landscape}
    \begin{figure}
        \centering
        \captionsetup{width=0.98\linewidth}
        \includegraphics[width=0.98\linewidth]{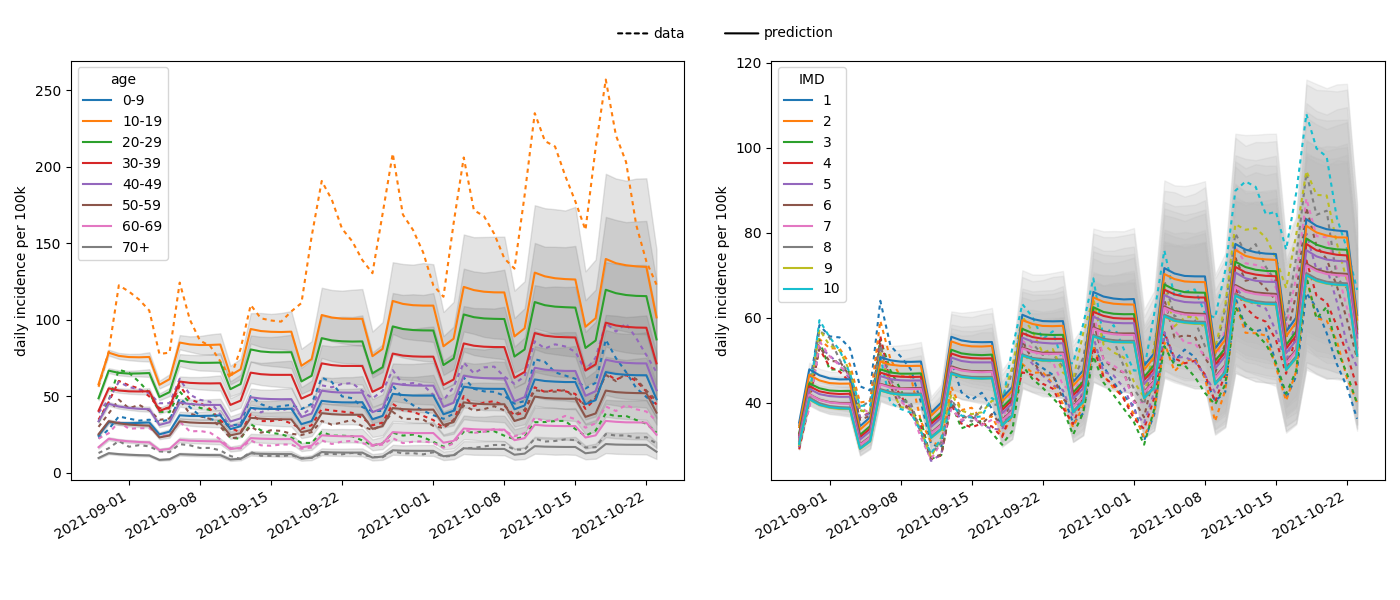}
        \caption{
        Forecast over 8-week period for the model described by Equations 2d and 4a (in the main paper) but where the susceptible to exposed hazard rate in Equation 4a is multiplied by a factor of $1.265$: this is interpreted as a $26.5\%$ increase in the hazard rate.
        This multiplying factor is the only difference between this model and the original model plotted in Figure 3c in the main paper.
        Both models are fitted to training data from 07/06/2021 to 29/08/2021 and neither model shows deprivation-switching.
        For this model, the 8-week forecast CRPS quartiles over all $\mathcal{R}_{i,t}$ are $(Q_1,Q_2,Q_3) = (18.8,  52.8, 165.3)$: note that this is substantially lower than original model in Figure 3c.
        Forecasts of expected daily case incidence are shown alongside their respective 90\% credible intervals.
        Left: solid line, expected daily incidence aggregated by age group; and dotted line, observed case incidence aggregated by age group.
        Right: solid line, expected daily incidence aggregated by IMD (deprivation) decile; and dotted line, observed case incidence aggregated by IMD decile.
        Note that deprivation-switching is not observed.}
        \label{fig:plot_simulated_timeseries_incidence_100k_no_effects_but_with_boost}
    \end{figure}
\end{landscape}

\begin{landscape}
    \begin{figure}
        \centering
        \captionsetup{width=0.96\linewidth}
        \includegraphics[width=0.96\linewidth]{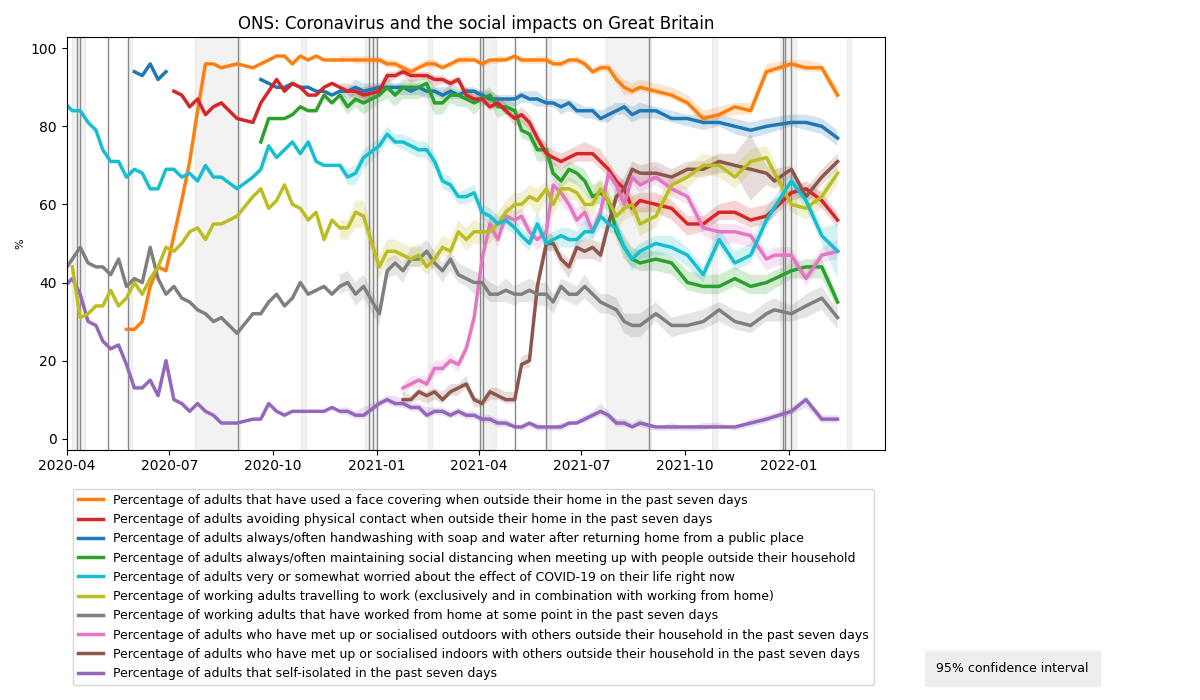}
        \caption{
        Data from the \textit{Coronavirus and the social impacts on Great Britain}$ ^{\dagger}$ study by the Office for National Statistics. Using a subset of these data this figure depicts trends in social behaviour between April 2020 and February 2022.  Note that vertical lines depict public and school holidays.\\ $^{\dagger}$\scriptsize{https://www.ons.gov.uk/peoplepopulationandcommunity/healthandsocialcare/healthandwellbeing/datasets/coronavirusandthesocialimpactsongreatbritaindata}}
        \label{fig:plot_social_impacts_2022-03-28_supp_matt}
    \end{figure}
\end{landscape}

\begin{landscape}
\begin{figure}
    \centering
    \captionsetup{width=0.98\linewidth}
    \includegraphics[width=0.98\linewidth]{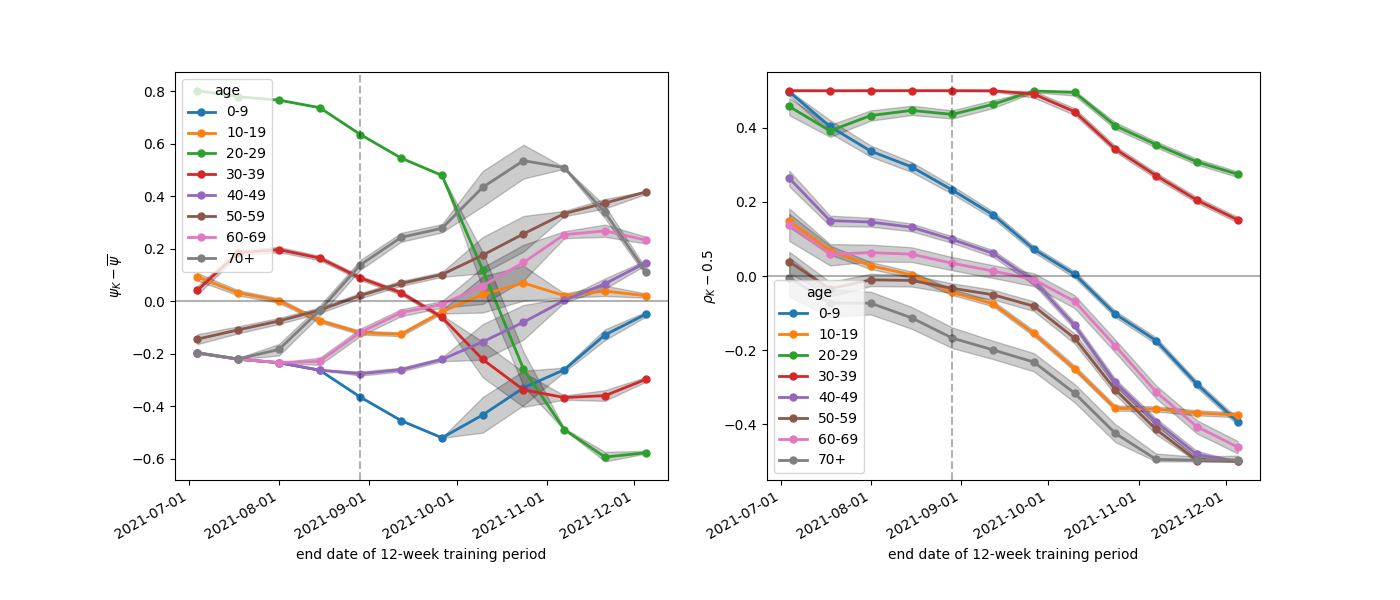}
    \caption{Expected value and 98\% credible interval of posterior samples for model parameters $\bm{\tilde{\psi}} = \bm{\psi} - \hat{\psi}$ (intercept term in $\bm{\chi}$) and $\bm{\tilde{\rho}} = \bm{\rho} - 0.5$ (slope term in $\bm{\chi}$)  where the model is fitted to a number of different training datasets, each with a 12-week window.  The circles are at fortnightly intervals and located at the final date of the estimate's respective training data period.  The vertical dashed line relates to posterior estimates from the model fitted to the training data, 07/06/2021 - 29/08/2021.
    Left: the relative contribution of each the age group $k$ to the intercept term in $\bm{\chi}$, for example the 20-29 group (green) contribute strongly at beginning of July 2021 but weakly by December 2021. Right: given age group $k$, then if $(\rho_k-0.5) >0$ the most deprived groups report the most cases, $(\rho_k-0.5) <0$ the least deprived groups report the most cases, otherwise there is no difference between the groups or the trend is non-linear and symmetric e.g. a second-degree polynomial.}
    \label{fig:plot_rho_and_psi_from_multiple_MCMC_runs_supp_mat}
\end{figure}
\end{landscape}